\documentclass[12pt]{emulateapj}
\usepackage{epsfig}
\usepackage{graphicx}
\usepackage{amssymb}
\usepackage{lscape}
\usepackage{longtable}
\usepackage{natbib}
\usepackage{lscape}
\bibpunct{(} {)} {;} {a} {} {,}
\newcommand*{\DISKBB}{\texttt{DISKBB}}
\newcommand*{\DISKPN}{\texttt{DISKPN}}
\newcommand*{\EQPAIR}{\texttt{EQPAIR}}
\newcommand*{\THCOMP}{\texttt{THCOMP}}
\newcommand*{\COMPTT}{\texttt{COMPTT}}
\newcommand*{\DBBTT}{\texttt{DBBTT}}
\newcommand*{\DBBEQPAIR }{\texttt{DBBEQPAIR}}
\newcommand*{\DBBTTPEG }{\texttt{DBBTTPEG}}
\newcommand*{\PEGPWRLW}{\texttt{PEGPWRLW}}
\def\sco{Scorpius X-1}
\def\msun{$M_{\odot}$}
\def\mdot{$\dot{M}$}

\shorttitle{Spectral Evolution of \sco}
\shortauthors{A. D' A{\'\i} et al.}

\begin{document}

\title{Broad-band Spectral Evolution of Scorpius X-1 along its Color-Color Diagram}
\author{A. D'A\'i\altaffilmark{1}, P. $\dot{Z}$ycki\altaffilmark{2}, 
T. Di Salvo\altaffilmark{1},R. Iaria\altaffilmark{1}, 
G. Lavagetto\altaffilmark{1},  N.R. Robba\altaffilmark{1}}

\altaffiltext{1}{Dipartimento di Scienze Fisiche ed Astronomiche, 
Universit\'a di Palermo, Archirafi 36, 90123 Palermo, Italy; email: dai@fisica.unipa.it}
\altaffiltext{2}{Nicolaus Copernicus Astronomical Center, Bartycka 18, 00-716 Warsaw, Poland}

\begin{abstract}
  We analyze a  large collection of RXTE archive  data from April 1997
  to August 2003  of the bright X-ray source Scorpius  X-1 in order to
  study the  broadband spectral evolution of the  source for different
  values  of  the inferred  mass  accretion  rate  by studying  energy
  spectra  from selected  regions in  the Z-track  of  its Color-Color
  Diagram.   A  two-component  model,  consisting of  a  soft  thermal
  component interpreted as thermal emission from an accretion disk and
  a  thermal Comptonization  component,  is unable  to  fit the  whole
  3--200 keV energy spectrum  at low accretion rates. Strong residuals
  in the highest energy band of the spectrum require the addition of a
  third component that can be  fitted with a power-law component, that
  could represent  a second thermal Comptonization from  a much hotter
  plasma,  or a hybrid  thermal/non-thermal Comptonization,  where the
  electrons in  the Comptonizing cloud have  a Maxwellian distribution
  of velocities with a power-law hard tail.  The presence of this hard
  emission in  \sco~ has been previously reported,  however, without a
  clear  relation with  the accretion  rate.  We  show, for  the first
  time, that there exists a  common trend in the spectral evolution of
  the source, where the spectral parameters change in correlation with
  the position of the source in the CD.  In particular, using a hybrid
  thermal/non-thermal Comptonization model (EQPAIR code), we show that
  the power supplied to the non-thermal distribution can be as high as
  half  of the  total  hard  power injected  in  heating the  electron
  distribution.  We  also found that  a small sample of  spectra, when
  the source resides  at the top of the FB can  also show intense hard
  X-ray emission.   We discuss the physical  implications derived from
  the  results of  our analysis,  with  a particular  emphasis on  the
  hardest part of the X-ray emission and its possible origins.
\end{abstract}
\maketitle
\section{Introduction}
\subsection{Scorpius X-1}
\sco~  is the brightest  persistent X-ray  source in  the sky  and the
first  identified X-ray  extra-solar  source \citep{giacconi62}.   The
X-ray source  is an old,  low magnetized neutron star  (NS), accreting
matter  transferred  through   Roche-lobe  overflow  from  a  low-mass
companion \citep[recently identified as an  M class star of $\sim$ 0.4
\msun,~][]{steeghs02}.  \sco~  is a  prototype  of  the  class of  the
Low-Mass X-ray Binaries  (LMXBs), and assuming a distance  of $2.8 \pm
0.3$ kpc \citep{bradshaw99}, the source emits close to the theoretical
Eddington limit for a  1.4 \msun~ NS ($L_{Edd} \sim 2~\times~10^{38}~$
erg s$^{-1}$).\\
Based on the timing behavior of LMXBs in correlation with the position
of   a  given   source  in   the  X-ray   color-color   diagram  (CD),
\citet{hasinger89} grouped  these sources  into two categories:  the Z
sources  and the Atoll  sources.  The  former are  brighter, radiating
close to $L_{Edd}$, the latter  are less bright, emitting at 0.01--0.1
$L_{Edd}$.   Z  sources exhibit  the  classical three-shaped  branches
describing a  Z pattern in the  CD: the Horizontal Branch  (HB) at the
top of the Z track, followed by the Normal Branch (NB) and the Flaring
Branch  (FB)  at  the  bottom   of  the  pattern.   There  are  strong
indications \citep[e.g.][]{vrtilek91} suggesting  that what drives the
changing  in the  spectral and  temporal  properties of  LMXBs is  the
instantaneous  accretion rate  ($\dot{M}$), which,  for Z  sources, is
believed to increase  monotonically from the HB to  the FB. Similarly,
Atoll sources, display two  different spectral/timing states: the soft
and luminous  banana state, associated to higher  accretion rates, and
the hard  and less luminous  island state, associated to  lower \mdot.
This early and straightforward classification  has now moved to a more
complex picture, as the patterns  that Atoll and Z sources describe in
their CDs,  when displayed on long  timescales, appear to  be more and
more similar to each  other \citep{muno02, gierlinski02}, although some
observational  facts still  hold: Z sources  move on the  CDs on
shorter timescales, have higher luminosities and present a timing
phenomenology different from that of the Atoll sources.\\
\sco, as  all the sources  of its class, shows  several quasi-periodic
oscillations (QPOs) along  all the branches of its  CD: the horizontal
branch oscillations (HBOs), the  normal branch oscillations (NBOs) and
the  flaring branch  oscillations (FBOs).   In \sco~  NBOs,  with peak
frequencies in the  range 4.5--7 Hz, and FBOs,  in the frequency range
6--25 Hz,  seem to be physically  related to each other  since the NBO
peak frequency  smoothly joins  the FBO peak  frequency as  the source
moves from  the NB to the  FB \citep{casella05}.  Van der  Klis et al.
(1996) reported the first observation of  HBOs in \sco~ at 45 Hz, with
an inferred  harmonic near 90 Hz;  the power spectrum can  also show a
pair of QPOs, whose frequencies  are in the range 800-1100 Hz, denoted
kHzQPOs,  that  shift  simultaneously  in frequency,  with  an  almost
constant  or  weakly  frequency-dependent peak  separation  \citep[see
e.g.][]{zhang06}.   The   physical  interpretation  of   these  timing
features is not unique and the related scientific debate is still open
\citep[see][~for a review]{klis04}.\\
Spectral  studies of  \sco~  have been  so  far not  so extensive  and
detailed  as  timing studies.   This  is in  part  due  to the  strong
brightness of  the source that actually prevents  its observation with
the  most  sophisticated  and  high-resolution X-ray  satellites  like
BeppoSAX and  ASCA in  the recent past  and Chandra and  XMM-Newton at
present.   Three  articles have  investigated  so  far the  broad-band
spectral  behavior of  the source,  using  data from  the Rossi  X-ray
Timing  Explorer (RXTE).   \citet{barnard03} used  data from  both the
Proportional  Counter Array  (PCA) and  the High  Energy  X-ray Timing
Experiment  (HEXTE), showing that  the spectrum,  in the  energy range
2.5--50  keV,  can  be  fitted  by a  simple  two-component  model,  a
black-body  soft  component  and  a  cutoff power-law,  plus  a  broad
Gaussian line, and  interpreted these results in the  framework of the
so  called Birmingham model  \citep{church01}. In  this interpretation
the spectral emission consists of black-body emission from the NS plus
Comptonized   emission  from  an   extended  accretion   disk  corona.
\citet{bradshaw03} used  only PCA data  (in the 2.5--18.2  keV range),
adopting  a model  consisting of  a black  body emission  plus  a bulk
motion Comptonization component and a reflection component in the form
of a  broad Gaussian line.  \citet{damico01} studied  HEXTE data above
20 keV, in order to test the presence of high energy X-ray emission in
the spectra of  the source.  Data were fitted,  using a bremsstrahlung
component,  to mimic  the  effects  of the  thermal  component, and  a
power-law component,  whenever the  statistics and clear  residuals in
the fit  required it.  In this  way no apparent  correlation was found
between the  presence of the hard  tail and position of  the source in
the CD, contrarily to what had been reported for other Z sources.\\
More recently,  \citet{disalvo06}, using INTEGRAL data  in the 20--200
keV energy band, detected a  hard X-ray tail, with photon index values
between  2 and  3.4, whose  intensity  decreased as  the source  moved
towards  higher accretion  rates.  Data  did  not show  evidence of  a
high-energy  cutoff up  to $\sim$  200 keV,  suggesting  a non-thermal
origin for the hard tail.\\
\sco~   is   also   an   interesting   source   of   radio   emission.
\citet{formalont01},  through an  extensive VLBA  monitoring campaign,
have shown that  the radio emission is composed  of a point-like radio
core emission at the position of  the X-ray source and of two opposite
radio  lobes, moving  through the  ISM with  relativistic  speeds $v/c
=0.45 \pm 0.03$, and with an angle of 46$^{\circ} \pm$6$^{\circ}$ with
respect to our line of sight.  It is also evident a connection between
the  phenomenon of  radio flare  at  the core  of the  source and  the
following flaring of  the lobes, so that it is  argued that the energy
production  of the radio  emission is  confined near  the NS  and only
afterward transported to the lobes via the working surface of a jet
\citep[see e.g.   the magnetohydrodynamical simulations of][]{kato04}.\\
These  results  stimulated  the  search for  the  possible  connection
between inflow  and outflow mechanisms working in  the violent regimes
of accretion onto a NS or  a black hole (BH). Energetic mass outflows,
collimated sometimes  in a  typical jet pattern,  can be  produced not
only  in  extragalactic X-ray  sources,  such  as  quasars and  active
galactic  nuclei (AGN),  but  also in  galactic  X-ray binary  systems
\citep[see~][~for  a  review]{fender02}.  It  has  been observed  that
galactic BH  binary systems  are able to  produce strong  jet emission
(sources  for which  a jet  has  been already  spatially resolved  are
usually   denoted  \textit{microquasars}),   with   radio-loud  states
associated  to  hard/low  X-ray   states,  alternated  to  periods  of
radio-quenching  during  soft/high   X-ray  states.   For  NS  systems
radio-loud episodes are generally  interpreted as a jet signature.  In
this context, all the Z sources seem to owe, under certain conditions,
a  radio-jet  nature.  Detections  of  radio-loud  states are  usually
associated with spectral  states of low \mdot, i.e.\ on  the HB of the
Z-track  \citep{fender00}, while  the  few radio  detections in  atoll
systems found  the sources residing in  the island state  of their CDs
\citep{migliari03}.
\subsection{Spectral properties of Z sources class} \label{introduction}
We rapidly  review in  this section the  spectral properties of  the Z
sources.  The class of the  Z sources display a homogeneous pattern of
spectral properties; their energy spectra can be usually decomposed as
follows:\\
\begin{itemize}
\item a  soft thermal  component, with characteristic  temperatures in
  the 0.5--1  keV range, interpreted  as thermal emission from  the NS
  surface or an optically thick, geometrically thin accretion disk;
\item  a  thermal  Comptonized  component, where  the  electron  cloud
  distribution has  a temperature  in the 2.5--3  keV range,  the soft
  seed photon  temperature is around 1  keV and the  optical depth can
  have  rather large  values  ($\tau \geq$  3);  this relatively  cold
  thermal  Comptonization is  thought to  take place  close to  the NS
  surface;  the soft  seed  photon temperature  generally exceeds  the
  highest  temperature reached  in  the accretion  disk,  so that  the
  source of  the soft  emission is probably  confined in  the boundary
  layer  between the  inner  edge of  the  accretion disk  and the  NS
  surface.   The high  optical depth  values associated  to  the cloud
  saturates   the  seed   photon  spectrum   close  to   the  electron
  temperature, so  that a quasi  Wien spectrum results.  A theoretical
  interpretation  of this  spectral  decomposition has  been given  by
  \citet{inogamov99}, where the boundary layer emission is expected to
  be radiation pressure  supported and its emission is  locally at the
  Eddington rate.  Change in luminosity  is attributed to a  change in
  the  emitting area rather  than in  its emissivity.   Phase resolved
  spectroscopy of  bright LMXBs \citep{revnivtsev05}  further supports
  this interpretation,  establishing in all  the analyzed Z  sources a
  common  cut-off, due  to  the boundary  layer  thermal emission,  at
  $\simeq$ 2.4 keV.
\item  a reflection  component,  often simply  modelled  with a  broad
  Gaussian line, in  the 6.4-6.9 keV range. In all  the Z sources this
  line  has  been  always  observed;  spectra from  BeppoSAX  or  RXTE
  observations  were not  able to  constraint the  shape of  the line,
  while  the  relative  width  was  in  the  range  0.1-1  keV.   High
  resolution spectra  in this energy  range have been obtained  so far
  only with  the BBRXT  \citep{smale93} and the  XMM-Newton satellites
  \citep{costantini02} for  Cygnus X-2; in both  observations the line
  was associated to highly  ionized iron (centroid energy $\simeq$ 6.7
  keV), while the line appears  to be intrinsically broad (FWHM $\sim$
  1 keV).  \citet{brandt94} pointed  out that the determination of the
  line  profile can be  a primary  diagnostic tool,  but, at  the same
  time,  kinematic and  relativistic effects  can greatly  distort the
  line from  the simple Gaussian  profile. Moreover, if  reflection is
  caused  by hard  X-rays  reflected  by a  cold  accretion disk,  the
  contribution of the Compton reflected continuum should be taken into
  account, when the iron  line emission is a considerable contribution
  to the total energy flux;
\item a power-law hard tail, which can contribute up to few percent to
  the total energy flux,  whose strenght usually varies in correlation
  with  the  position  of the  source  in  the  CD, namely  being  the
  strongest on  the HB,  gradually decreasing as  the source  moves to
  higher accretion rates and totally  fading in the FB \citep[e.g.  GX
  17+2,][]{disalvo00}.   The   photon  index  of   the  power-law  was
  generally  found in  the range  1.5--3 with  no evident  high energy
  cutoff up to energies of about 200 keV.
\end{itemize}
In this  article we  report a complete  investigation of  the spectral
properties  of \sco~  through an  extensive analysis  of  RXTE archive
data, indicating a clear connection  between position of the source on
the CD and spectral behavior.
\section{Data reduction and analysis}
The scientific payload of RXTE consists of three instruments, the PCA,
the HEXTE,  and the All Sky  Monitor (ASM).  The PCA  consists of five
co-aligned  Xenon  proportional  counter  units (PCUs)  with  a  total
effective area  of about 6500  cm$^2$. The instrument is  sensitive in
the energy range from 2 keV  up to 60 keV \citep{jahoda}, although the
response matrix is best calibrated in the energy range 3-22 keV.  Data
can be processed using  several different configuration modes; for our
analysis  we exclusively use  the \textit{Standard2}  mode, with  16 s
time resolution and 128 energy channels in the 2--60 keV energy range.
The   HEXTE  consists   of  two   cluster  of   four  NaI/CsI-phoswich
scintillation counters  that are sensitive from  15 keV up  to 220 keV
\citep{rothschild}.  We use the \textit{Standard Mode}, with 64 energy
channels, for the reduction and analysis of the HEXTE data. Background
subtraction  is done  by using  the source-background  rocking  of the
collimators.  We use HEXTE response matrices of 1999 August.\\
We  collected a  large amount  of RXTE  archive data,  discarding only
minor shorter  observations, from  April 1997 up  to August  2003.  We
present in Table~\ref{tab1} the  datasets we used for our analysis,
indicating  the associated  proposal number,  the starting  and ending
times of  each observation in Terrestrial Time,  and the corresponding
exposure times.\\
Data  have  been  processed  using the  standard  selection  criteria,
discarding data  taken during Earth occultations  and passages through
the South Atlantic  Anomaly.  We only used data from  PCU2 for the PCA
and  data from  Cluster A  for  the HEXTE  instrument. We  constructed
color-color   diagrams    (CDs)   of   the    source   by   extracting
energy-dependent   lightcurves  using   PCA  energy   channels  5--10,
11--16,17--22 and 23--44,  with a 64 s bintime.   These channel ranges
correspond  to  the  energy  ranges 1.94--4.05  keV,  4.05--6.18  keV,
6.18--8.32  keV, 8.32--16.26  keV  respectively.  We  define the  Soft
Color (hereafter SC) as the ratio of count rates in the 4.05--6.18 keV
and 1.94--4.05  energy bands, while  the Hard Color (hereafter  HC) as
the ratio  of count  rates in the  8.32--16.26 keV and  6.18--8.32 keV
energy ranges.   However, the channel  to energy conversion  depend on
the  period  of activity  of  the  satellite  (there are  5  different
instrumental Epochs), most of our data belong to Epoch 3 for which the
given  energy bands  are appropriate;  for  the Epoch  5 datasets  the
energy  ranges  are  slightly  shifted,  so that  this  results  in  a
consequent shift  on the HC  and on the  SC axes.  Because we  are not
interested in the  secular shifts of the track in the  CD, nor we want
to  accumulate spectra  taken at  different Epochs,  but to  perform a
statistical study  of the broad-band  spectral behavior of  the source
for different CD  positions in different datasets, we  did not perform
the energy dependent corrections to the colors.\\
We selected regions in the  CDs for each observation dataset, in order
to cover  a homogeneous part  of the Z-track  and to have at  the same
time a suitable statistics.  From these selections we derived the good
time intervals  (GTI) which we  used for extracting  the corresponding
spectra for the PCA and HEXTE data.\\
Given the high luminosity of the  source all the PCA spectra have been
dead-time corrected. Some observations  were performed with the use of
an offset pointing; in these cases we extracted the responses matrixes
of both the  HEXTE and the PCA instruments,  following the indications
given in the on-line  Data Analysis documentation pages.  We processed
and analyzed  data using version 6.0  of the FTOOLS  package suite and
version    11.3.1     of    Xspec.   
\section{The Data}
\begin{deluxetable*}{l l l r}
\centering
\tablewidth{0pt}
\tablecolumns{4}
\small
\tablecaption{LOG OF THE RXTE OBSERVATIONS USED IN THE ANALYSIS \label{tab1}}
\tablecolumns{4}
\tablehead{
\colhead{Proposal} & \colhead{Starting time (TT)}  & \colhead{Ending time (TT)}     
& \colhead{Effective Time}\\
\colhead{Number (CD)} &  &    
& \colhead{ks}\\} 

\startdata
20035A                       &  1997-04-18 15:20:32   &  1997-04-24 19:02:40 & 75.6 \\
20035B                       &  1998-01-03 21:17:52   &  1998-01-08 13:19:44 & 60.0  \\
30036                        &  1998-01-07 13:29:52   & 1998-01-08 22:16:48  & 30.4 \\
30035A                       &  1998-05-30 00:28:00   & 1998-06-02 07:01:36  & 47.8 \\
30035B                       &  1998-07-02 06:26:40   & 1998-07-05 10:32:00  & 32.2 \\
40020                        &  1999-01-06 05:25:20   & 1999-01-16 23:34:40  & 122.1 \\
70014                        &  2002-03-15 01:02:56   & 2002-05-16 23:20:32  & 139.0 \\
70015                        &  2002-05-19 21:55:44   &  2002-05-19 23:54:40 & 2.24  \\
80021                        &  2003-07-30 03:00:48   & 2003-08-13 16:43:44  & 146.9 \\
\enddata
\tablecomments{ 
  Observation log  of the RXTE  datasets used for our  analysis.  Some
  datasets, although  having the same proposal  number contain pointed
  observation  separated by long  time intervals.   In these  cases we
  collected all  the pointing observations  close in time in  just one
  CD,  and distinguish  the relative  CDs by  a letter,  following the
  proposal  number.   Starting  and  ending  times  are  expressed  in
  Terrestrial Time.  Effective exposures times are filtered by our GTI
  screening criteria.
}
\end{deluxetable*}
Hereafter  we shall refer  to a  collection of  close in  time pointed
observations  used to  extract one  CD by  using its  proposal number.
From datasets 20035 (and 30035) we have obtained two different CDs; we
will label  them, respectively, 20035A and 20035B  (30035A and 30035B,
see  the  {\textit Spectrum}  column  in Table~\ref{tab2}).   Datasets
20035A, 20035B,  30036, 30035A,  30035B and 40020  belong to  Epoch 3,
while datasets  70014, 70015  and 80021 belong  to Epoch  5.  Datasets
70014 and 70015 have been merged in one CD, because these observations
were close in time and the source pointing was the same.\\
The eight CDs that we have  extracted present the source mostly on the
NB/FB, as  these are the  states where the  source spends most  of the
time.   In  \sco~ the  HB  is  not  \textit{horizontal} but  rather  a
vertical continuation  of the NB  (a characteristic shared also  by GX
17+2  and  GX 349+2,  the  so-called  Sco-like  sources).  For  visual
clarity  we will  refer to  the \sco~  CD pattern,  mainly  making the
distinction between the  left (or FB) track and the  right (NB and HB)
track,  and between  the top  (harder) parts  and the  bottom (softer)
parts of the tracks.\\
The  count  rates  associated   with  the  extracted  spectra  show  a
significant  trend, generally correlated  with the  HC; HEXTE  and PCA
count rates  are the highest  at the top  of the FB  and monotonically
decrease  towards the FB/NB  apex; following  the direction  bottom NB
$\rightarrow$  HB, the  trend is  inverted.  This  behavior  is mostly
stressed in  the HEXTE data (compare  the countrate in  spectrum 07 in
the top FB with spectrum 10  in the transition region FB/NB, for which
the rate is reduced by two thirds in the HEXTE, but only by one third
in the PCA).\\
\begin{figure*}
\centering
\begin{tabular}{l l l l}
\includegraphics[height=3.5cm, width=2.5cm ]{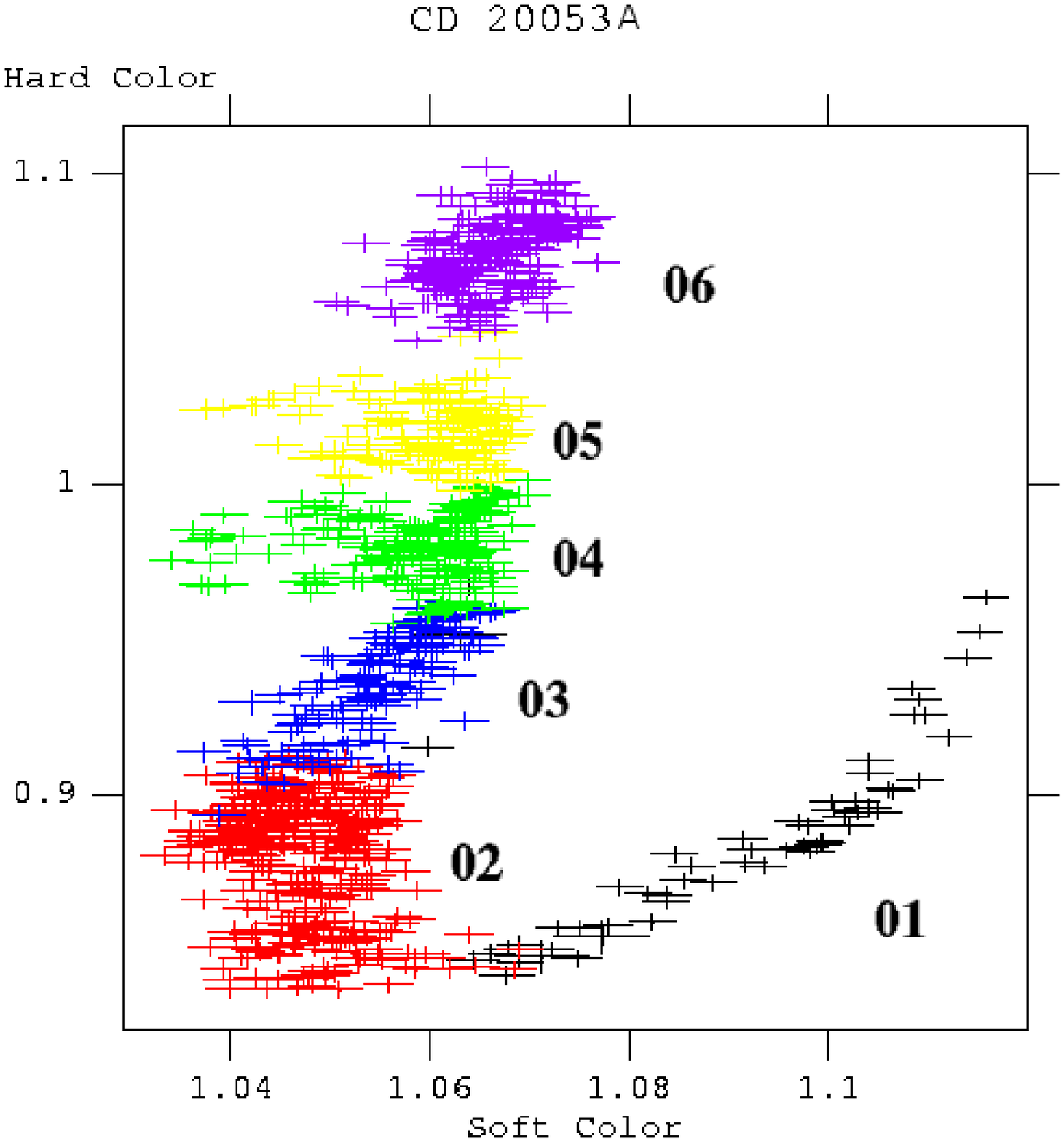} &
\includegraphics[height=3.5cm, width=2.5cm ]{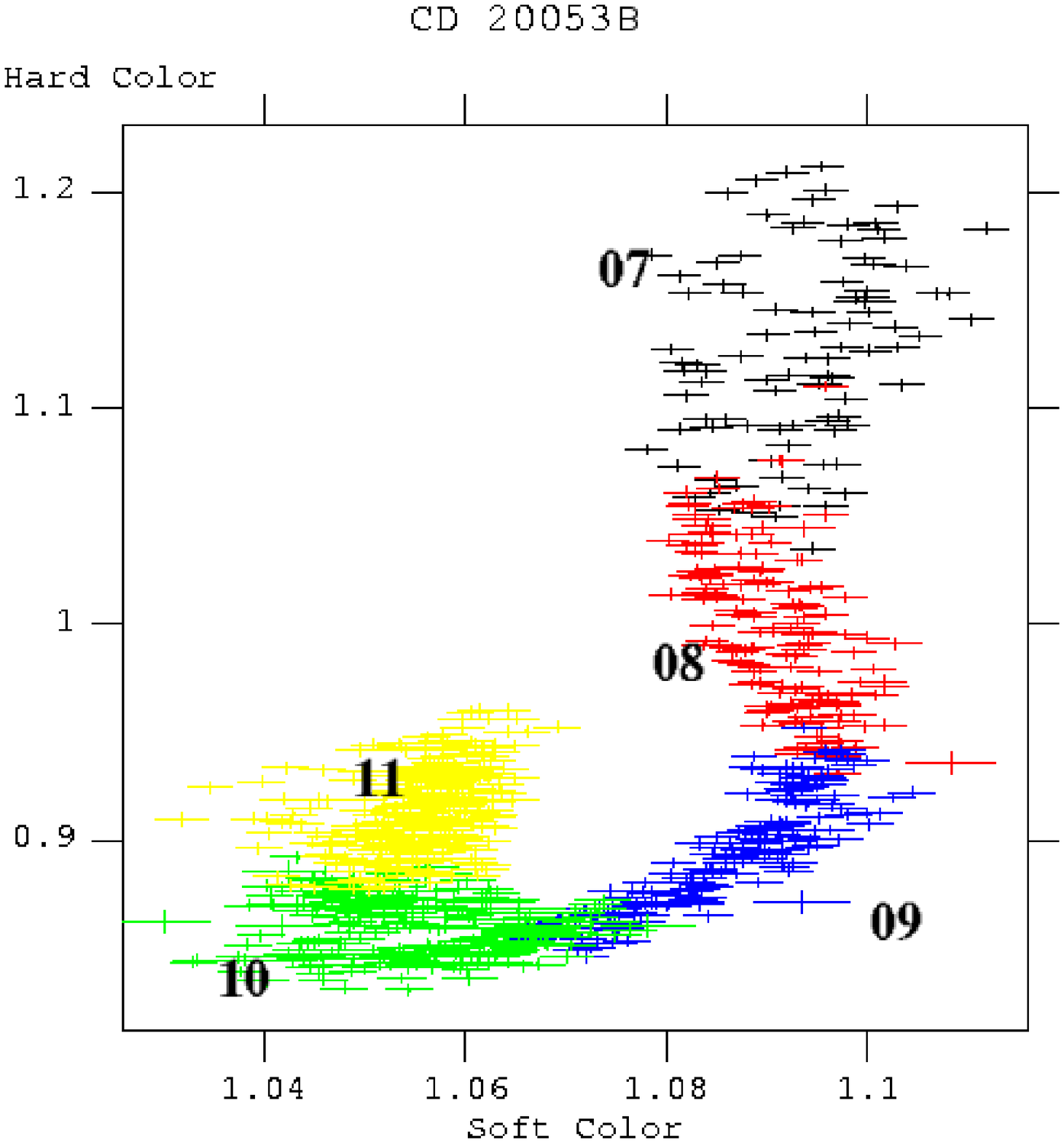} &
\includegraphics[height=3.5cm, width=2.5cm ]{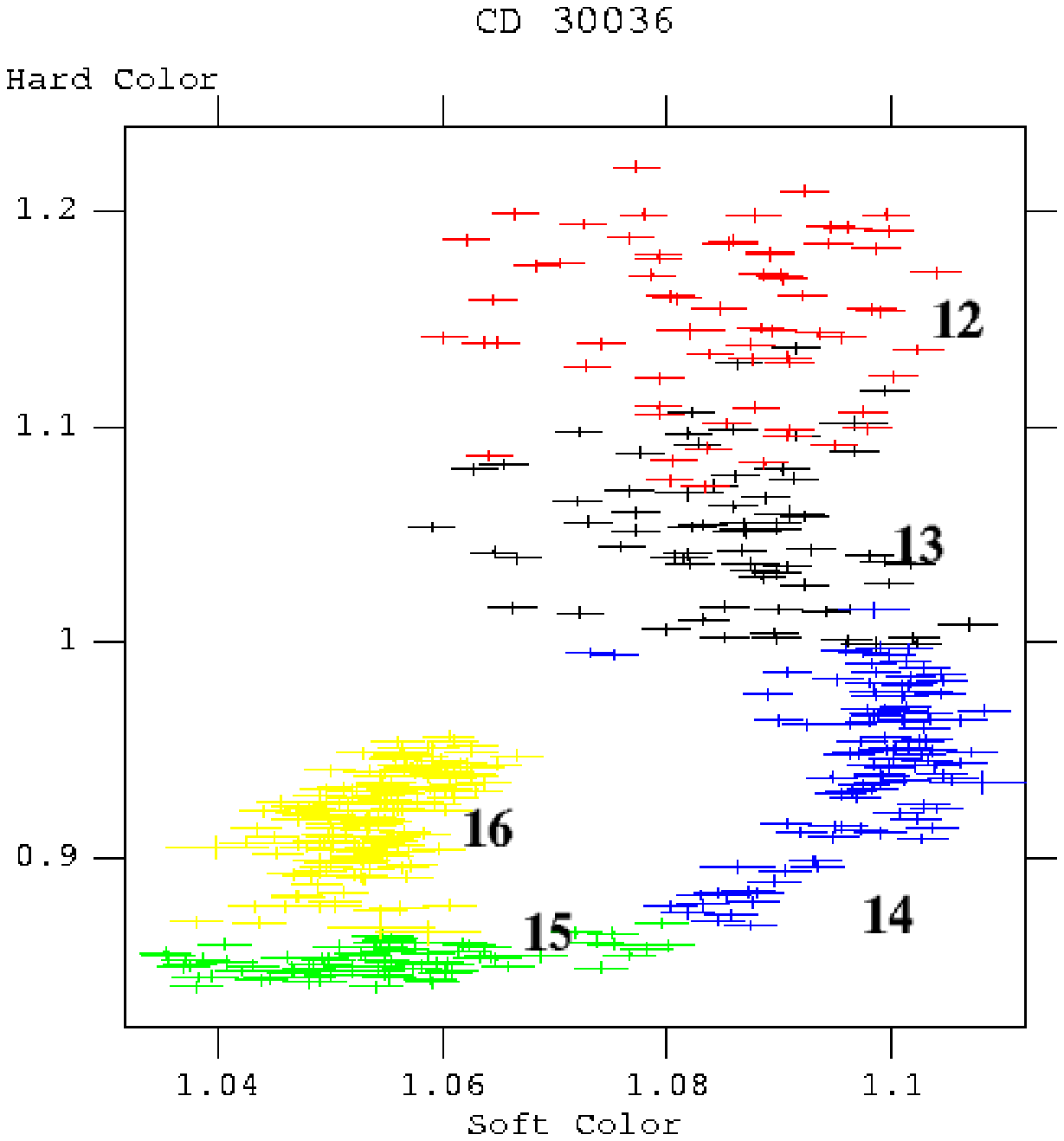} &
\includegraphics[height=3.5cm, width=2.5cm ]{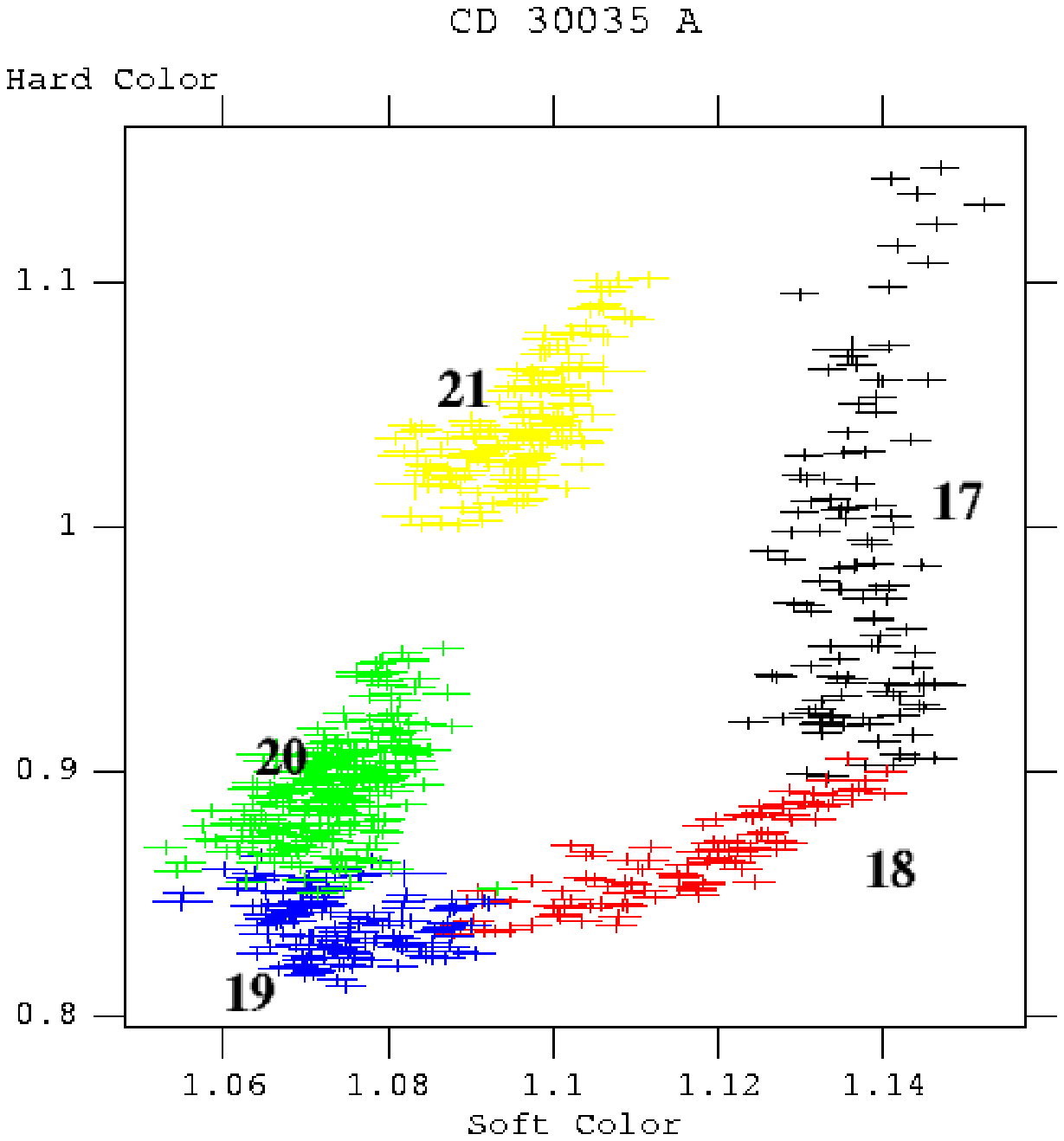} \\
\includegraphics[height=3.5cm, width=2.5cm ]{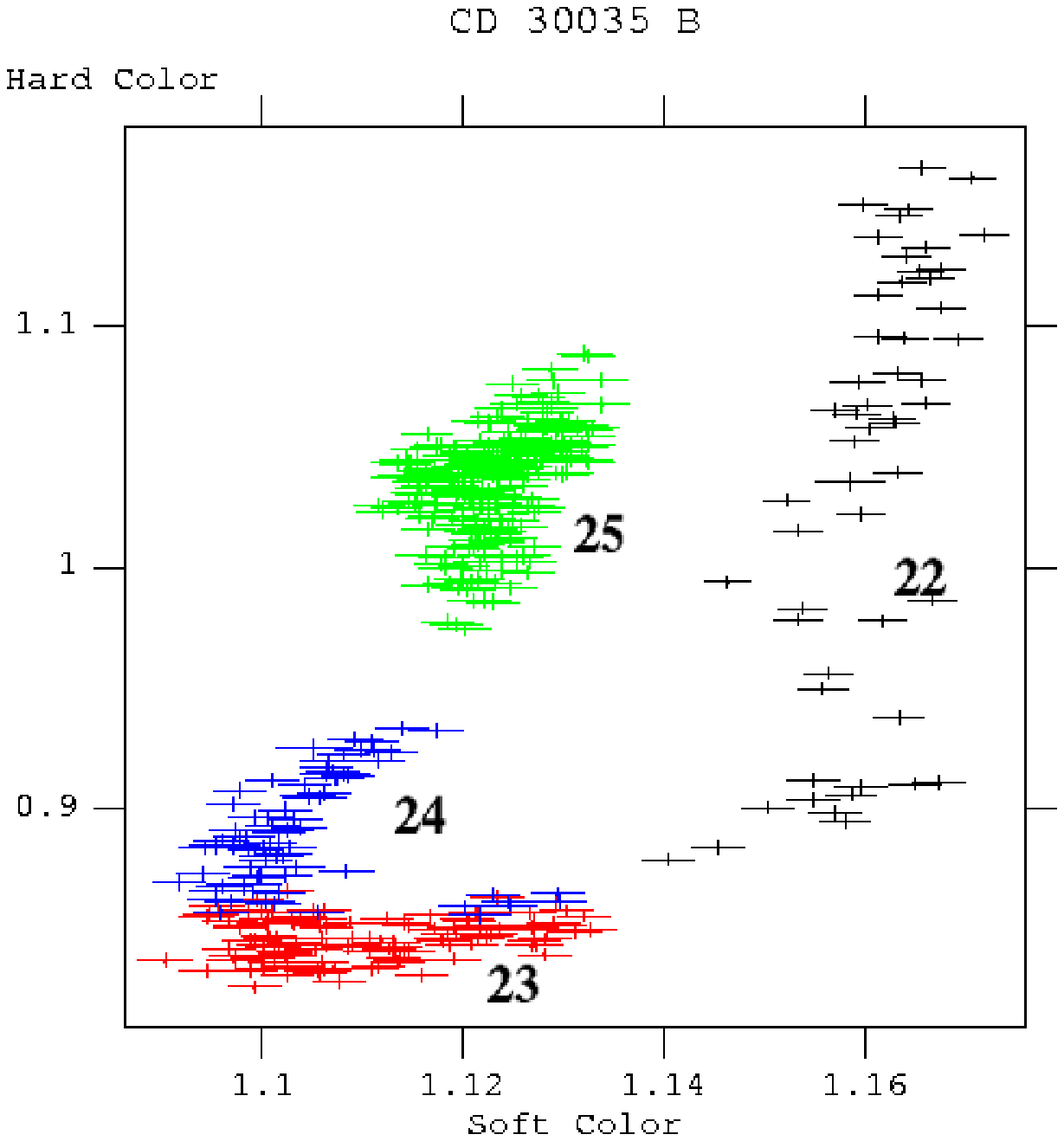} &
\includegraphics[height=3.5cm, width=2.5cm ]{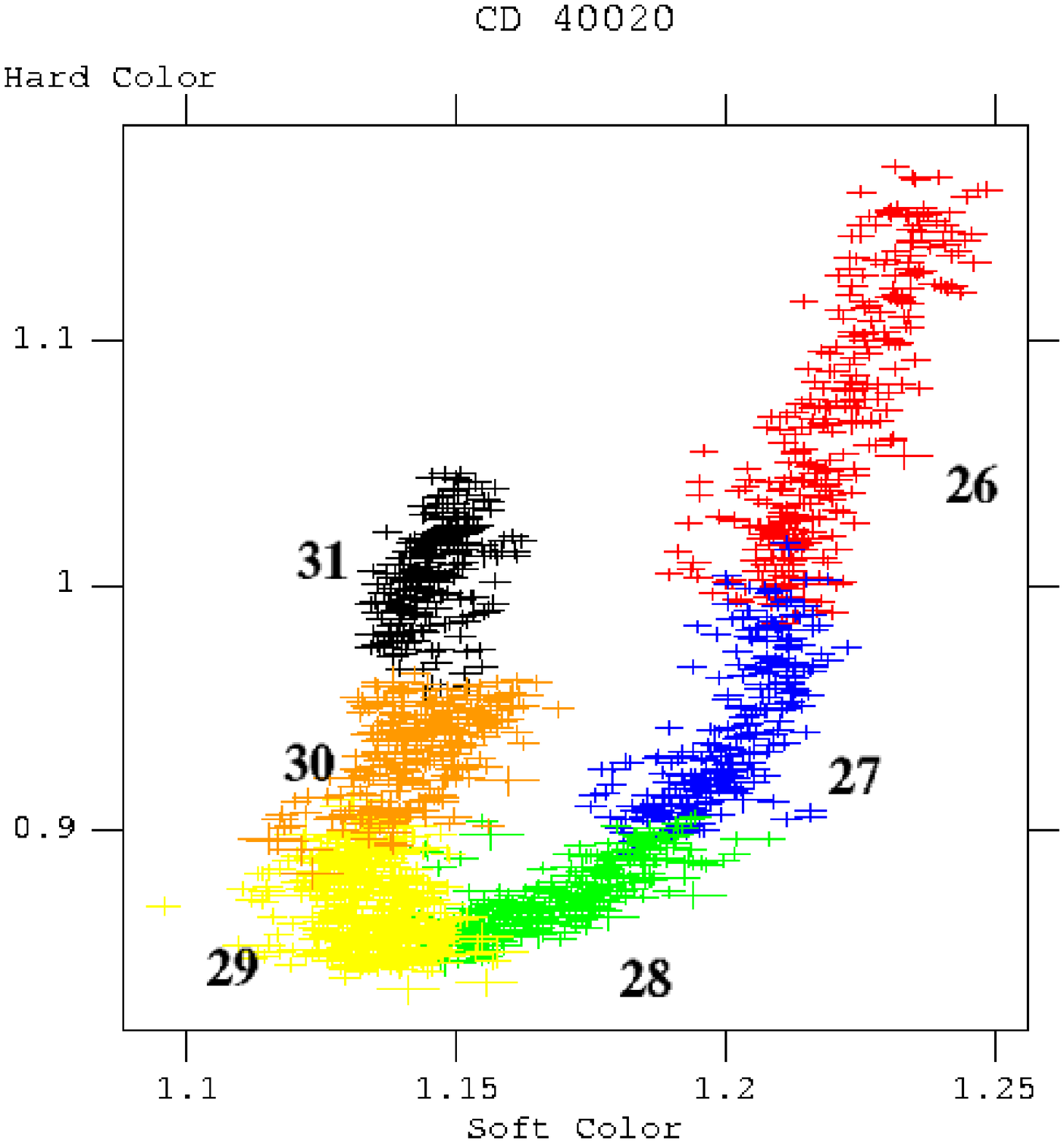} &
\includegraphics[height=3.5cm, width=2.5cm ]{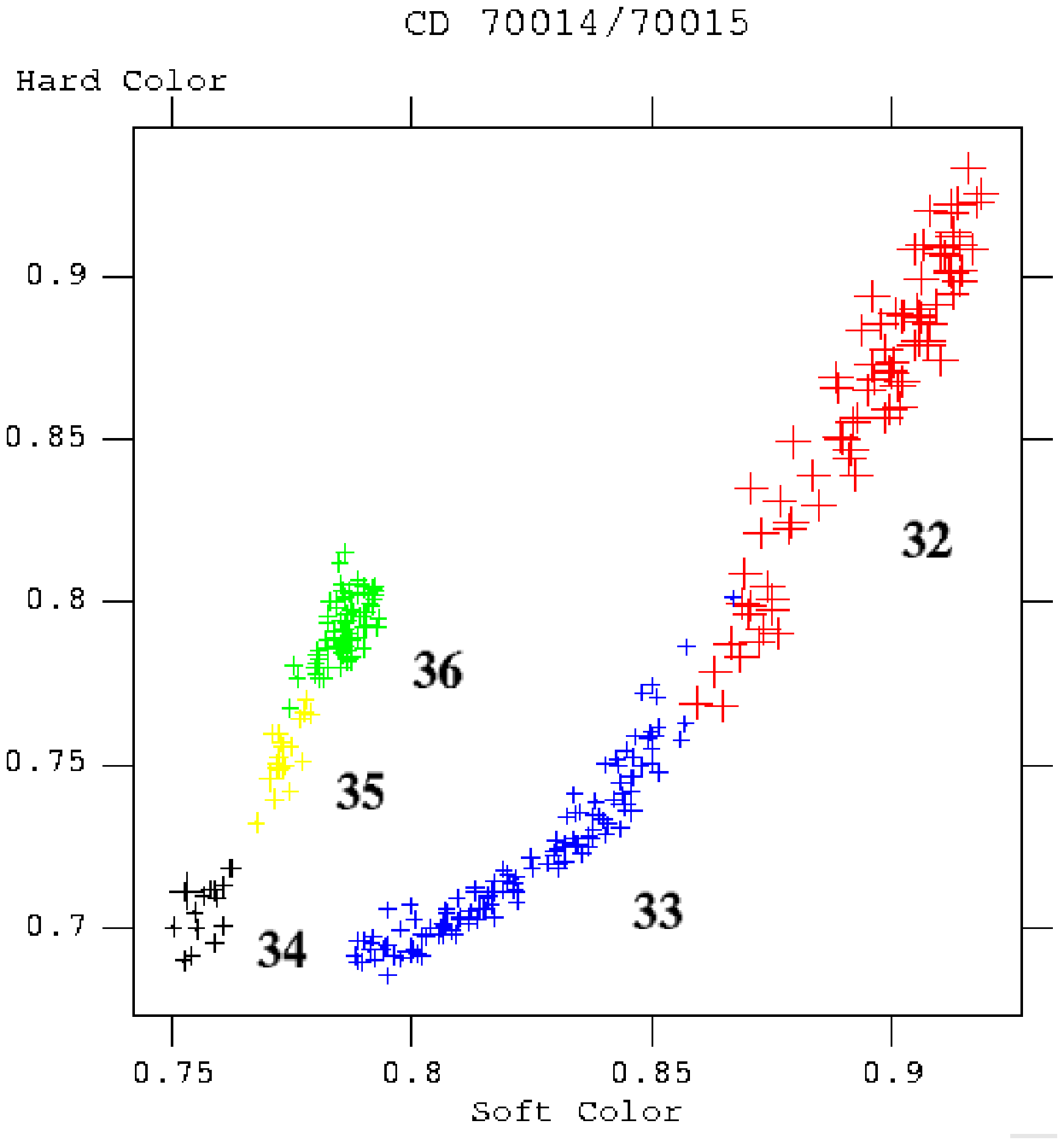} &
\includegraphics[height=3.5cm, width=2.5cm ]{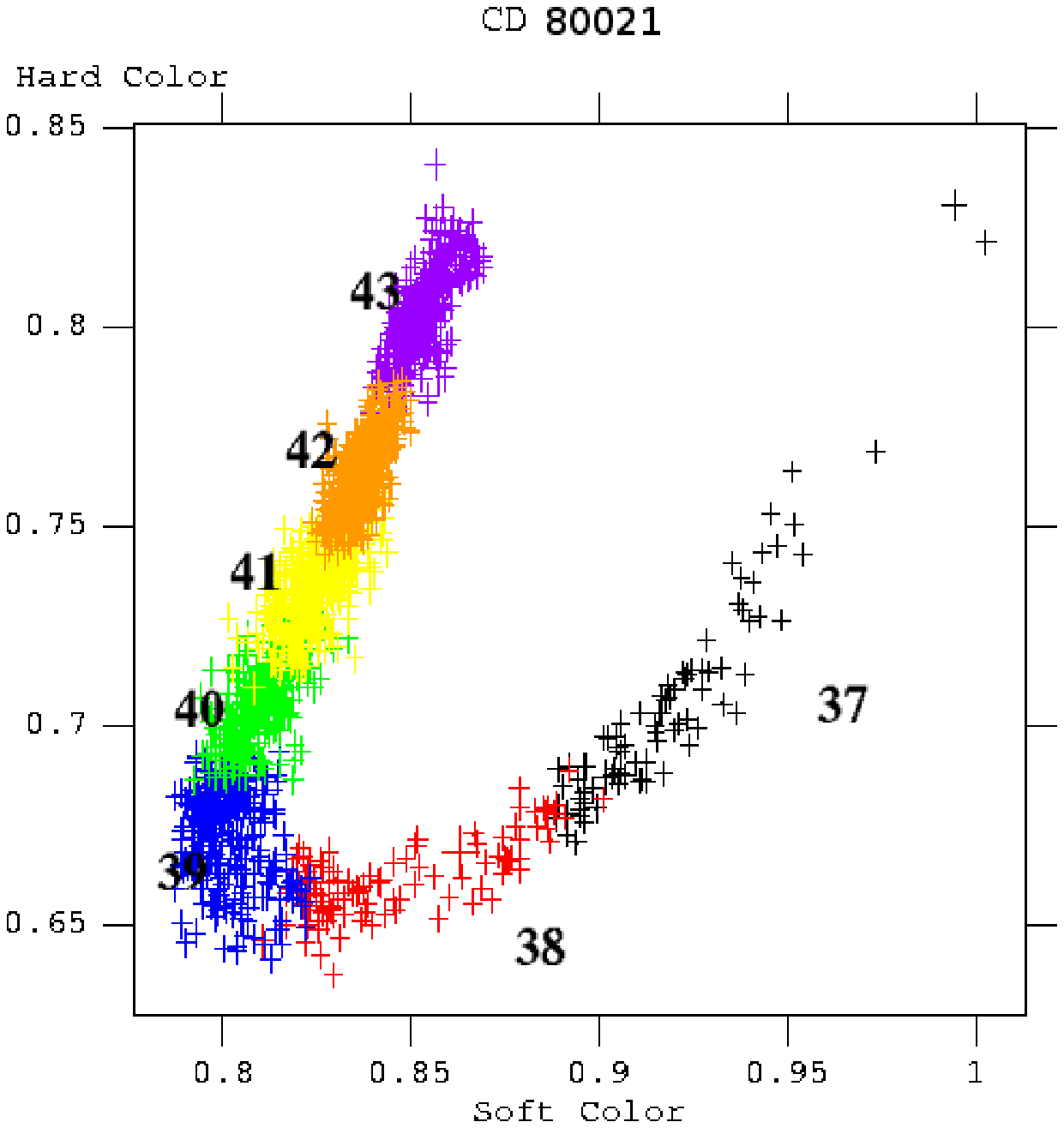} \\
\end{tabular}
\caption{  \small \linespread{1}
CDs extracted from RXTE datasets used in our spectral analysis. 
Different colors indicate different selected regions from which the CD-resolved
spectra were produced.
We label the spectra through a progressive numbering as shown in the CDs.}
\label{fig1}
\end{figure*}
In  four datasets  (20035A,  20035B, 40020  and  30036) the  satellite
pointed the source on axis;  in three datasets (30035A, 30035B, 80021)
an offset  pointing has  been used.  This  influenced the  high energy
statistics, where in  the same position on the  CD, an offset pointing
produced an HEXTE count rate lowered by almost one order of magnitude.
Spectra  extracted  from  datasets  70014  and  70015  have  also  low
statistics  in the  HEXTE band,  as  the integration  time is  shorter
compared with the other spectra.\\
We generally used a 3.0--22 keV  energy band for the PCA spectra and a
20--200 keV energy band for the HEXTE spectra.  HEXTE channels from 28
to 63 have been rebinned, grouping four channels into one.  In some of
the  spectra that  we extracted  we found,  however, model-independent
mismatches in the overlapping energy  region between the HEXTE and the
PCA;  because   these  mismatches   are  of  instrumental   origin  we
occasionally restrict  these energy bands until these  features did no
longer impact on  our fit results.  In Table~\ref{tab2}  we report the
exact energy  bands used for  the PCA and  the HEXTE datasets  for the
selected spectra.\\
The choose to assign an appropriated systematic error for the PCA data
is an essential step in the analysis of the spectra. This operation is
not straightforward,  as the commonly used standard  candle, the Crab,
provides  in this  case only  a  partial calibration  tool.  The  Crab
countrate for  every energy channel below  $\sim$ 20 keV  is about one
order  of magnitude  less than  the \sco~  countrate, thus  making the
extrapolation  of  the Crab  based  systematic  error  to the  \sco~
spectra riskful.  The  mostly adopted view to assign  a 1\% systematic
error for  all the energy channels  has a strong impact  on the softer
part of the spectrum, where there are the mostly weighted channels; we
found that  in this case the  fit is strongly driven  by residuals in
the 3-6 keV range, leading to a modelization of the spectrum that does
not seem  physically plausible. The overall  effect is a  shift of any
soft  component outside  the RXTE  energy band,  and  unrealistic high
values for the extrapolated flux.\\
The only way to avoid the  strong driving effect of the first channels
is to arbitrarily  raise the systematics in this  range; we found that
assigning a  3\% systematic error for  channels between 3  and 3.9 keV
(namely the  first two PCA  channels available in our  spectra), while
leaving a 1 \% systematic error  for all the other channels, allows to
derive a  physically plausible scenario for the  spectral evolution of
the source, does not result  in extreme lower/higher values of reduced
$\chi^2$  for  every fit  performed,  allows  us  to exploit  all  the
available  energy band  and to  generally constrain  all  the spectral
components of the adopted models.  We occasionally still found in some
fits systematical residuals in the soft part of the spectrum, that are
more stressed  in the softer  spectra, whose origin can  be presumedly
instrumental and related to the xenon L-edge near 5 keV.  We choose to
ignore these  effects, in  order to not  overparameterize the  fit, as
they do not influence the determination of the spectral parameters and
do not impact the $\chi^2$ values of the fits.\\
No  systematic  error  has  been  associated to  the  HEXTE  data.   A
normalization constant is left free to vary between the PCU2 and HEXTE
spectra to take into account residual flux calibration uncertainties.
\section{Spectral  models} \label{spectralmodels}  As  pointed out  in
Section 1,  spectra of LMXBs  are usually described  as the sum  of at
least  two spectral components.   Because the  PCA energy  band starts
from  3  keV, it  is  not  possible to  constrain  the  effect of  the
photoelectric interstellar  absorption on the  source flux.  Following
\citet{christian97},  we  fixed  its  value to  3  $\times$  10$^{21}$
cm$^{-2}$, for each  fit performed.  To fit the  CDs selected spectra,
we tried  at first a  series of models  given by the combination  of a
soft  thermal  component,  such  as  a blackbody  or  multicolor  disk
blackbody  \citep[\DISKBB~  in  Xspec,~][]{mitsuda84}, and  a  thermal
Comptonized component.   For the  latter we, alternatively,  tried the
\texttt{COMPPS}          \citep{poutanen96},          {\texttt{COMPTT}
  \citep{titarchuck94}  and   \THCOMP~ \citep[not  included   in  the
  standard XSPEC  package, see~][]{zdziarski96} Comptonization models.
  For any model adopted, large residuals in the 6--10 keV energy range
  indicated  the presence of  iron features,  that we  simply modelled
  with a broad Gaussian line in the 6--7 keV energy range.  We noted a
  general tendency for any adopted model to have a very broad Gaussian
  line (with $\sigma \geq 0.8$  keV) and, generally, with centroids at
  energies $\leq  6.4$ keV.  In  order to avoid  unrealistically large
  Gaussian lines,  absorbing the underneath  continuum, we constrained
  the line to  have centroid energies only in  the 6.4--7.0 keV energy
  range, with a line width less than 0.8 keV, while the normalization
  of the line was a completely free to vary parameter.\\
  We found an adequate description  of the data, for energies below 30
  keV, using  a two-component  model given by  the combination  of the
  \DISKBB~  and  the  \COMPTT  components, a  modelization  that  was
  firstly introduced  by \citet{mitsuda84}, and  that is known  as the
  {\it   Eastern  model}.   Other   Comptonization  models   can  also
  statistically  equally well  represent the  spectral sample,  but we
  preferred to illustrate  our results using this Compton  model as it
  is one of the mostly used  in the LMXBs spectral analysis and allows
  an easy way  to compare our results with  results obtained for other
  sources.\\
Using   the  other   often  used   modelization,  the   Western  model
\citep{white86}  that predicts  a  $\sim$ 1--2  keV blackbody  thermal
emission and a Comptonization  component with disk soft seed-photon in
the 0.4--0.8  keV energy range \citep[see  e.g][]{barnard03}, we found
higher  $\chi^2$   values  for  each  spectrum   and  an  extrapolated
super-Eddington luminosity along each  part of the Z-track, that would
be dramatic for  some FB spectra which would overcome  by more that an
order of  magnitude the Eddington  limit, thus making  this assumption
not  realistic.  Moreover,  the power  dissipated in  the Comptonizing
disk corona would be constantly much greater than the power dissipated
as thermal emission on the NS-surface and this breaks our expectations
that the power  dissipated near the compact object  should be at least
of the same order of the  power dissipated along the disk, if the disk
is  very  close to  the  NS  \citep[see  also][~for a  more  extensive
discussion]{done02}.\\
The \DISKBB~ has two free parameters, the temperature expressed in keV
at the inner  disk radius (kT$_{DB}$) and a  normalization factor that
depends on the  inner radius of the accretion  disk, the distance from
the source and the inclination angle between the line of sight and the
normal to  the disk; the \COMPTT~  has four free  parameters: the soft
seed  photon temperature,  kT$_{0}$, the  electron temperature  of the
Comptonizing  cloud,   kT$_e$,  the  optical  depth,   $\tau$,  and  a
normalization constant.  We assumed  that the Comptonizing geometry is
spherical.  We will  refer to this two-component model  as the \DBBTT~
model.\\
As it is evident from Table ~\ref{tab2} the \DBBTT~ model is, however,
gradually unable  to satisfactorily fit  the whole energy  spectrum as
the  source resides  in zones  of  low inferred  accretion rate.   The
residuals, with respect to this  model, clearly indicated an excess of
flux at energies greater than 30 keV.  There is also a second group of
spectra  (namely spectra  7,8,12,13 and  26)  that gave  for the  same
reason an unsatisfactory high $\chi^2$ value. These spectra lie at the
top of the  FB in the datasets that have  the higher HEXTE statistics.
We will refer to this particular set as the top-FB spectra.\\
Because the mechanisms producing this hard excess in the Z sources are
not clear yet, in the following we will adopt two working hypothesises
to fit the broadband continuum of the source: a) as in the case of the
quasi saturated boundary layer  Comptonization, this component is also
the result of a thermal Comptonization process, but from a much hotter
plasma; b)  a broad band spectrum  is the result of  a hybrid electron
velocity  distribution, i.e.  a  thermal electron  distribution, which
produces  the   optically  thick  Comptonization,   with  a  power-law
non-thermal  tail which  is  responsible for  the high-energy  excess.
Because, for case a), the hot  plasma can only reside in regions close
to  the  central  compact  object, where  magnetic  and  gravitational
effects are the strongest, we assume that the boundary layer radiation
field is the primary source  of soft-seed photons also for this second
thermal Comptonization.   We note that this assumption  is also needed
in order to avoid an  overestimation of the contribution of steep hard
tail  at low  energies.  The  use of  an unbroken  power-law  at steep
photon indeces can give a major contribution in the lowest energy band
of the  spectrum so  that the fit  is driven by  low-energy residuals,
possibly   resulting   in   an   erroneous   continuum   modelization.
Thereafter, a low-energy cutoff at the seed-photon temperature must be
taken into account as the expected photon field spectrum is within the
instrumental range.\\
We added, thereafter, to our basic two-component model, a power-law in
order to mimic this  second Comptonization, multiplied by a low-energy
cut-off.  We set  the value of the low-energy  cut-off anchored to the
kT$_0$ value of the \COMPTT~  component, while we choose the power-law
to have a pegged normalization  expressed as flux of this component in
the 20.0--200.0  keV energy range  (\PEGPWRLW, in XSPEC).   We checked
for each  spectrum the  $\chi^2$ improvement by  adding a  high energy
cutoff, noting  that we could obtain  only lower limits  on the cutoff
energy,  generally in the  30--100 keV  range without  any significant
improvement in the  $\chi^2$. Thereafter we did not  use a high energy
cutoff.   We  will  refer  at  this  model  simply  as  \DBBTTPEG~  or
\textit{two Comptonizations} model.\\
To model the  possibility given in case b) we  also fitted the spectra
showing a  strong hard  excess (namely spectra  for which  the \DBBTT~
model gave a  null hypotesis probability less than  0.05 (it generally
corresponded to a reduced $\chi^2$ value of 1.4) with a model given by
the  sum of  a  thermal  disk component  plus  the recently  developed
thermal/nonthermal   hybrid   Comptonization   model  named   \EQPAIR~
\citep[see][~for  a full  description of  the model]{coppi99,coppi00}.
It  embodies   Compton  scattering,  $e^{\pm}$   pair  production  and
annihilation, $pe^{\pm}$ and  $e^{\pm}e^{\pm}$ thermal and non-thermal
bremsstrahlung,  and energy exchange  between thermal  and non-thermal
parts of the $e^{\pm}$  distribution via Coulomb scattering.  Selected
electrons  are accelerated  to suprathermal  energies and  the thermal
part of  the $e^{\pm}$ distribution can be  additionally heated.  This
model assumes a spherical  plasma cloud with isotropic and homogeneous
distribution of photons and  $e^{\pm}$, and soft seed photons produced
uniformly within the plasma,  with a thermal temperature of $kT_{BB}$.
The  properties of  the plasma  depend  on its  compactness, ${\it  l}
\equiv   \mathcal{L}   \sigma_T  /   (\mathcal{R}   m_e  c^2)$   where
$\mathcal{L}$ is the  total power in the source,  $\mathcal{R}$ is the
radius of the sphere and $\sigma_T$ is the Thomson cross section.  The
compactness  is divided  in  a hard  compactness,  ${\it l_h}$,  which
corresponds  to  the power  supplied  to  the  electrons, and  a  soft
compactness, ${\it  l_s}$, corresponding to the power  supplied in the
form of soft photons.  The compactnesses corresponding to the electron
acceleration and to the additional  heating of the thermal part of the
$e^{\pm}$  distribution  are  denoted  as ${\it  l_{nth}}$  and  ${\it
  l_{th}}$, respectively,  and ${\it l_h} \equiv {\it  l_{nth}} + {\it
  l_{th}}$.  The  non-thermal energy distribution of  the electrons in
the  plasma is  assumed to  be a  power law  replacing  the Maxwellian
exponential tail from  a gamma Lorentz factor $\gamma_{min}  = 1.3$ up
to a $\gamma_{max} = 1000$  with an energy index $G_{inj}$.  The total
Thomson optical depth  and the electron temperature of  the plasma are
computed self-consistently from the  assumed optical depth $\tau_p$ of
the background electron-proton plasma.\\
We  choose to  model the  thermal disk  emission using  as in  the hot
Comptonization  model the  \DISKBB~  component for  spectra in  HB/NB,
while    we     adopted    the    post-Newtonian     model    \DISKPN~
\citep{gierlinski99} in  the case  of the topFB  spectra, as,  in this
accretion state,  we expect the disk  to reach its  minimum inner disk
radius, close  to the  NS; the \DISKPN~  model takes into  account the
effective gravitational  potential in the neighborhood  of the compact
object  by   computing  the  emergent  disk  spectrum   using  a  more
appropriate  post-Newtonian  potential.   We  assumed  that  the  disk
reaches the last stable orbit at  6 R$_g$ and fixed this parameter for
the  fits performed,  while  we left  free  to vary  the maximum  disk
temperature (T$_{max}$)  and a normalization factor (K  = $M^2 cos(i)/
D^2 \beta^4 $, where $M$ is the central mass expressed in solar units,
$i$ is the inclination of the  disk, $D$ is the distance to the source
in kpc,  $\beta$ the color/effective temperature).  We  have left free
to  vary   for  the  \EQPAIR~  component   the  following  parameters:
$kT_{BB}$, $l_h/l_s$, $l_{nth}/l_{h}$,  $\tau_p$ and the normalization
factor.   The $G_{inj}$  was not  very sensible  to variations  in the
range between 1.5 and 3 for  spectra in the HB/NB, while for the topFB
spectra had  to be left as  a free to  vary parameter, as this  gave a
significant  decrease in  the $\chi^2$  when left  as a  free  to vary
parameter.   The  value  of  the  soft  compactness  ${\it  l_s}$  was
unconstrained by the  fit (it is mainly driven  by the pair production
rate, which  manifests itself with  the annihilation line  at energies
$\simeq$ 500 keV, obviously far beyond our energy limits), and we kept
it frozen  at the value  of 10 \citep[see  also][]{gierlinski99}.  All
the other parameters of the model were frozen at the default values.
We will refer to this modelization as the \DBBEQPAIR~ model, or simply,
the \textit{hybrid Comptonization} model. 
\section{Results and discussion}

\begin{deluxetable*}{l r r r  c c c c c}
\centering
\tablecolumns{9}
\small
\tablewidth{0pt}
\tablecaption{COLOR-DIAGRAM RESOLVED SPECTRA AND FITS RESULTS \label{tab2}}
\tablecolumns{9}
\tablehead{
\colhead{Spectrum} & 
\colhead{Exposure} & 
\multicolumn{2}{c}{Countrates (cts/s)}   
& \multicolumn{2}{c}{Energy range (keV)}         
& \colhead{\DBBTT}                          
& \colhead{\DBBTTPEG} 
& \colhead{\DBBEQPAIR}\\
&  \colhead{s}      & \colhead{PCA}           & \colhead{HEXTE} & \colhead{PCA}       & \colhead{HEXTE}
&  \colhead{$\chi^2_{red}$  (d.o.f.)}      & \colhead{$\chi^2_{red}$  (d.o.f.)}     & \colhead{$\chi^2_{red}$ (d.o.f.)}        \\}
\startdata
01 20053A        &  5728    & 20920 &  621    & 3--22   & 20--200   & 0.63 (62)        &     0.59 (61)                 &                \\
02 20053A        & 22400    & 18600 &  521    & 3--22   & 20--200   & 1.46 (62)        &     0.80 (60)       &   0.85 (61)    \\
03 20053A        &  9163    & 19760 &  651    & 3--22   & 20--200   & 1.07 (62)        &     0.62 (61)                &         \\
04 20053A        & 11040    & 20550 &  736    & 3--22   & 20--200   & 2.53 (62)        &     0.55 (60)       &   0.55 (61)      \\
05 20053A        &  8576    & 21060 &  848    & 3--22   & 20--200   & 2.84 (62)        &     1.14 (60)       &   1.14 (61)       \\
06 20053A        & 13472    & 21740 & 1015    & 3--22   & 20--200   & 5.62 (62)        &     0.64 (60)       &   0.63 (61)       \\
\hline
07 20053B   & 5698     & 27960 & 1675 & 3--22   & 24--200 & 2.12 (59)       & 1.21 (58)      &  1.27 (57)  \\
08 20053B   & 9680     & 25820 & 1054 & 3--22   & 20--200 & 1.89 (62)       & 0.99 (60)      &  1.04 (60)  \\
09 20053B   & 11472    & 21870 &  642 & 3--22   & 20--200 & 1.03 (62)       & 0.93 (60)      &   \\
10 20053B   & 13920    & 19320 &  512 & 3--22   & 20--200 & 1.35 (62)       & 0.54 (60)      &   \\
11 20053B   & 18720    & 19930 &  613 & 3--22   & 20--200 & 1.65 (62)       & 0.68 (60)      & 0.55 (61)  \\
\hline
12 30036     & 4208 & 28590 & 1840     & 3--22   & 24--200                 &  2.39 (59)  & 1.41 (58)  &  1.40 (57)   \\
13 30036     & 4336 & 26930 & 1271     & 3--22   & 22--200                 &  1.91 (61)  & 1.32 (60)  &  1.42 (60)             \\
14 30036     & 7072 & 23480 & 799      & 3--22   & 20--200                 &  0.85 (62)  & 0.86 (61)  &                       \\
15 30036     & 4736 & 19130 & 492      & 3--22   & 20--200                 &  0.60 (62)  & 0.59 (61)  &                       \\
16 30036     & 10320 & 20000 & 633     & 3--22   & 20--200                 &  2.36 (62)  & 0.65 (60)  &  0.67 (61)  \\
\hline
17 30035A & 6880   & 23680 & 84     & 3--22   & 24--200                 &  1.11 (59)  & 1.07 (58)  &                   \\
18 30035A & 6608   & 19260 & 43     & 3--22   & 24--200                 &  0.69 (59)  & 0.68 (58)  &                 \\
19 30035A & 8704   & 17380 & 39     & 3--22   & 24--200                 &  1.17 (59)  & 0.87 (58)  &                  \\
20 30035A & 15456  & 18020 & 50     & 3--22   & 24--200                 &  2.62 (59)  & 0.71 (57)  & 0.72 (58)      \\
21 30035A & 10064  & 20040 & 92     & 3--22   & 24--200                 &  2.84 (59)  & 0.50 (57)  & 0.53 (58)     \\
\hline
22 30035B   & 3344  & 24310  & 100    & 3--22   & 24--200                 &  1.03 (59)       & 1.04 (58)      &  \\
23 30035B   & 7344  & 16770  & 39     & 3--22   & 22--200                 &  1.06 (61)       & 1.08 (60)     &  \\
24 30035B   & 4976  & 16520  & 43     & 3--22   & 22--200                 &  1.12 (61)       & 0.99 (60)     &  \\
25 30035B   & 16144 & 18900  & 76     & 3--22   & 24--200                 &  2.56 (59)       & 0.49 (57)     & 0.44 (58) \\
\hline
26 40020     & 3401  & 23380 &  708     & 3--22   & 24--200                 & 1.86 (59)   & 0.83 (57)   &  0.86 (58)   \\ 
27 40020     & 3642  & 20360 &  544     & 3--22   & 20--200                 & 1.26 (62)   & 0.81 (60)   &    \\
28 40020     & 18493 & 15810 &  300     & 3--22   & 20--200                 & 1.14 (62)   & 0.69 (60)   &    \\
29 40020     & 11967 & 14920 &  286     & 3--22   & 20--200                 & 1.29 (62)   & 1.18 (61)   &    \\ 
30 40020     & 5735  & 16100 &  383     & 3--22   & 20--200                 & 0.85 (62)   & 0.67 (61)   &    \\
31 40020     & 7537  & 17100 &  470     & 3--22   & 20--200                 & 3.59 (62)   & 0.70 (60)   &  0.66 (61)  \\
\hline
32  70014  & 822  & 24440 & 948     & 3--22   & 22--200                 &  1.16 (58)      & 0.91 (56)      &    \\
33  70014  & 5668 & 18740 & 368     & 3--22   & 20--200                 &  1.24 (58)      & 0.72 (56)     &     \\ 
34  70015  & 976  & 15250 & 280     & 3--22   & 20--200                 &  0.63 (58)      & 0.63 (57)     &    \\
35  70015  & 1248 & 16300 & 363     & 3--22   & 20--200                 &  0.75 (58)      & 0.69 (57)     &    \\  
36  70014  & 3804 & 17570 & 447     & 3--22   & 22--200  &  1.42 (58)   & 0.73 (56)    & 0.72 (57)   \\
\hline
37 80021  & 5344  & 13030 &  12.8 & 3--22   & 28--200  &  0.73 (53)      & 0.59 (52)     &    \\ 
38 80021  & 6752  & 9900  &  2.0  & 3--22   & 28--200  &  0.60 (53)      & 0.46 (52)     &    \\
39 80021  & 17312 & 8828  & 7.6   & 3--22   & 28--200  &  0.54 (53)      & 0.51 (52)     &    \\
40 80021  & 14624 & 9071  & 27.0  & 3--22   & 28--200  &  0.64 (53)      & 0.57 (52)     &    \\
41 80021  & 29504 & 8990  & 11.7  & 3--22   & 28--200  &  1.27 (53)      & 0.78 (52)     &    \\
42 80021  & 30384 & 10320 & 6.6   & 3--22   & 28--200  &  1.37 (53)      &  0.67 (53)    & 0.70 (52)   \\
43 80021  & 18656 & 10730 & 11.0  & 3--22   & 28--200  &  1.57 (53)      &  0.74 (53)    & 0.69 (52)    \\
\enddata
\tablecomments{ \scriptsize We  report in this table,  from the left  to the right
  column, the  chosen identifying number  of the CDs  selected spectra
  (as  shown   in  Figure~\ref{fig1})  together   with  the  associated
  observational dataset,  its exposure times,  the relative background
  subtracted  countrates  for  the  PCA  and HEXTE,  the  energy  band
  considered in the fits and the reduced $\chi^2$ values obtained from
  the fits,  adopting the \DBBTT,  the \DBBTTPEG~ and  the \DBBEQPAIR~
  model;  the degrees  of  freedom (dof)  for  each fit  are shown  in
  parenthesis.}
\end{deluxetable*}

We have examined 43 energy spectra of \sco~ in the 3.0--200 keV energy
band using RXTE  pointed observations from April 1997  to August 2003;
the spectra  have been extracted  from selected regions chosen  in the
X-ray CD. We produced a CD for  each close in time dataset in order to
avoid  shifts of instrumental  origin, and  repeated our  analysis for
each  CD  obtained  in  this   way,  thus  having  at  hand  a  robust
representation  of the  spectral evolution  of the  source in  all its
accretion states.\\
To fit the  spectra we firstly adopted a  two-component model given by
the  sum  of a  soft  thermal disk  emission  and  a hard  Comptonized
emission.  We checked  for the  presence of  harder X-ray  emission by
adding a high-energy power-law to  our basic model, while for a subset
of spectra  we adopted a hybrid  thermal/non-thermal Comptonization as
described in the previous section.\\
We find that  the basic two-component model \DBBTT~  can fit the 3-200
keV energy band  spectrum (fits for which we  obtained a $\chi^2_{red}
\leq 1.4  $) in  25 out of  a total  43 spectra; these  spectra mostly
belong to the FB and to the  NB. Hard tails, in this group of spectra,
are not significantly detected in  14 spectra (these are spectra 1, 9,
14, 15, 17, 18, 22, 23,  24, 29, 34, 35, 39, 40, see Table~\ref{tab2})
when the  source lies in the  FB/NB apex of  its CD. For the  other 11
spectra, although  the addition  of a new  component is  not strictly
required by  the fit, we found  a general improvement  in the $\chi^2$
value.\\
On the contrary, we find that this model fails to describe the spectra
in  18 of  the  43 selected  spectra  (fits for  which  we obtained  a
$\chi^2_{red} \geq 1.4 $); the  fits generally getting worse and worse
from the  bottom of the FB to  the upper NB.  From  the residuals with
respect to this  model, it is clearly seen an  excess of hard emission
at  energies  above  30  keV.   We report  in  Figure~\ref{fig9}  four
representative deconvolved  spectra for the case in  which the \DBBTT~
model does not give an adequate description to the broad-band spectral
behavior, choosing three  spectra from the HB/NB zone  and one for the
topFB  zone: spectrum  06 (at  the top  of the  left track  of dataset
20035A), spectrum 16 (at the top  of the left track of dataset 30036),
spectrum  31 (at  the top  of  the left  track of  dataset 40020)  and
spectrum 07 (at the top of the right track of dataset 20035B).\\
In the following  we will group the spectra  according to the relative
position in  their CDs, in: topFB  spectra (spectra 7, 8,  12, 13, 17,
22, 26, 32, 37), FB spectra (1, 9, 14, 18, 27, 28, 33, 38), NB spectra
(2, 3, 10, 15, 19, 20, 23, 24,  29, 30, 34, 35, 39, 40) and HB spectra
(4, 5, 6, 11, 16, 21, 25,  31, 36, 41, 42, 43).  We caution the reader
that this  distinction is  based only on  the results of  our spectral
analysis for spectra taken in particular zones of the Z-track, without
any reference to  the temporal behavior of the  source. Although it is
clear,  looking just  at the  lightcurves, to  distinguish  within one
dataset the FB from the NB,  the distinction between the NB and the HB
is rather subtle, as the two  track mostly overlap.  For this reason a
HB  spectrum,  for this  analysis,  is to  be  considered  only as  a
simple label  for a  spectrum extracted from  the top part  of left
track of each CD.\\

\section{The hot Comptonization model}  
\subsection{Energetics}
In the following we will discuss  the fit results based on the adopted
model \DBBTTPEG. Although in some spectra this model over-parameterizes
the fit, the spectral behavior of the source and our conclusions would
not  change,  if  we  used  the best-fitting  values  of  the  \DBBTT~
modelization (we show in Table~\ref{tab3a} and~\ref{tab3b} the best-fit values of the
spectral  parameter, together  with associated  errors quoted  at 90\%
confidence level).  In some cases  we had to freeze some parameters in
order  to  make the  fit  physically  consistent  and avoid  solutions
without physical  meaning. In  spectra 7,  8, 12, 13,  17, 37,  38 the
electron temperature  had to  be fixed to  10 keV (a  value consistent
with what we found in other CDs  for the same CD zone; if free the fit
would give  a best-fit with kT$_0  >$ kT$_e$ which  is clearly wrong),
the  photon index  of the  \PEGPWRLW~ component  in spectra  where this
component  is  not  significantly  detected,  or  when  its  value  is
unconstrained by  the fit,  is frozen  to a reference  value of  2; the
\DISKBB~  parameters of  spectra 42  and  43 had  to be  fixed to  the
best-values as obtained  by the best-fit model \DBBTT~  (if free, this
component vanishes, while the  \COMPTT~ and \PEGPWRLW~ have comparable
flux values).\\
In this spectral decomposition we calculated
the total luminosity as follows:
\begin{equation}
L_x = 4 \pi D^2 (F_{TT} + F_{PEG} + a \xi N_{DB} T_{DB}^4)
\end{equation}
where $F_{TT}$  is the \COMPTT~  flux and $F_{PEG}$ is  the \PEGPWRLW~
flux  extrapolated in  the 0.1--200  keV energy  range; $\xi  = 1/(f^4
\kappa^2)$,   where    $f$   is   the    spectral   hardening   factor
\citep{shimura95},  is assumed  equal to  1.7, and  $\kappa  \sim 0.4$
corrects for the fact that the  radius at which is reached the maximum
temperature    is    greater   than    the    inner   disk    distance
\citep{vierdayanti06}.  $a$ is a constant dependent on the distance of
the source, which we assumed to be 2.8 kpc, and the inclination of the
disk, which we assumed  45$^{\circ}$. In Figure~\ref{fig1} we show the
contribution of the disk and of the Comptonized component to the total
flux.  The luminosities on both axis  are in Eddington units for a 1.4
\msun.\\
The  source span  a luminosity  range  between 0.9  L$_{Edd}$ and  2.4
L$_{Edd}$, thus being in  a super-Eddington state almost independently
of the position of the source on the CD.  We note that the only way to
drastically reduce  this range  is to suppose  a much  massive compact
object (the possibility of a massive  NS hosted in the \sco~ system is
also   discussed  in   \citet{stella99}  in   the  framework   of  the
relativistic  precession  model used  to  explain  the frequencies  of
quasi-periodic oscillations in NSs and BH systems) given that both the
distance and the inclination angle  are known with good accuracy (10\%
and 13\% relative error, respectively).\\
The disk flux  and the hard flux correlate as  expected with the total
flux,  although the disk  emission presents  a more  pronounced linear
relation,  while the hard  flux seems  to saturate  as the  total flux
reaches the  highest values. If we  assume a linear  dependency of the
two contributions  versus the total  emission we derive a  constant of
proportionality  equals to 0.56  for the  disk flux  and 0.44  for the
Comptonized flux.  When we look  at the ratio of the two contributions
along the Z-track  (see Figure~\ref{fig2}}, a 5\% error  on this ratio
is assumed, after  having tested the corresponding hard  and soft flux
values  for  different parameter  values  of  the spectral  components
inside the uncertainties errorbars as derived from the fits), there is
a general trend  to have a higher hard/soft ratio  as the source moves
from higher to  lower accretion states; this is  in agreement with our
expectation that, at lower accretion  rate the disk could be truncated
and the fraction of the  power dissipated in the boundary layer should
correspondently increase.  Spectra 7, 12, 13 and 32 do not follow this
trend; these spectra belong to  the topFB spectra, and in these cases,
we   are   possibly,  underestimating   the   disk  contribution,   or
overestimating the hard flux.

\begin{figure*}
\centering
\begin{tabular}{l l}
\includegraphics[height=6cm, width=4cm,  angle=-90]{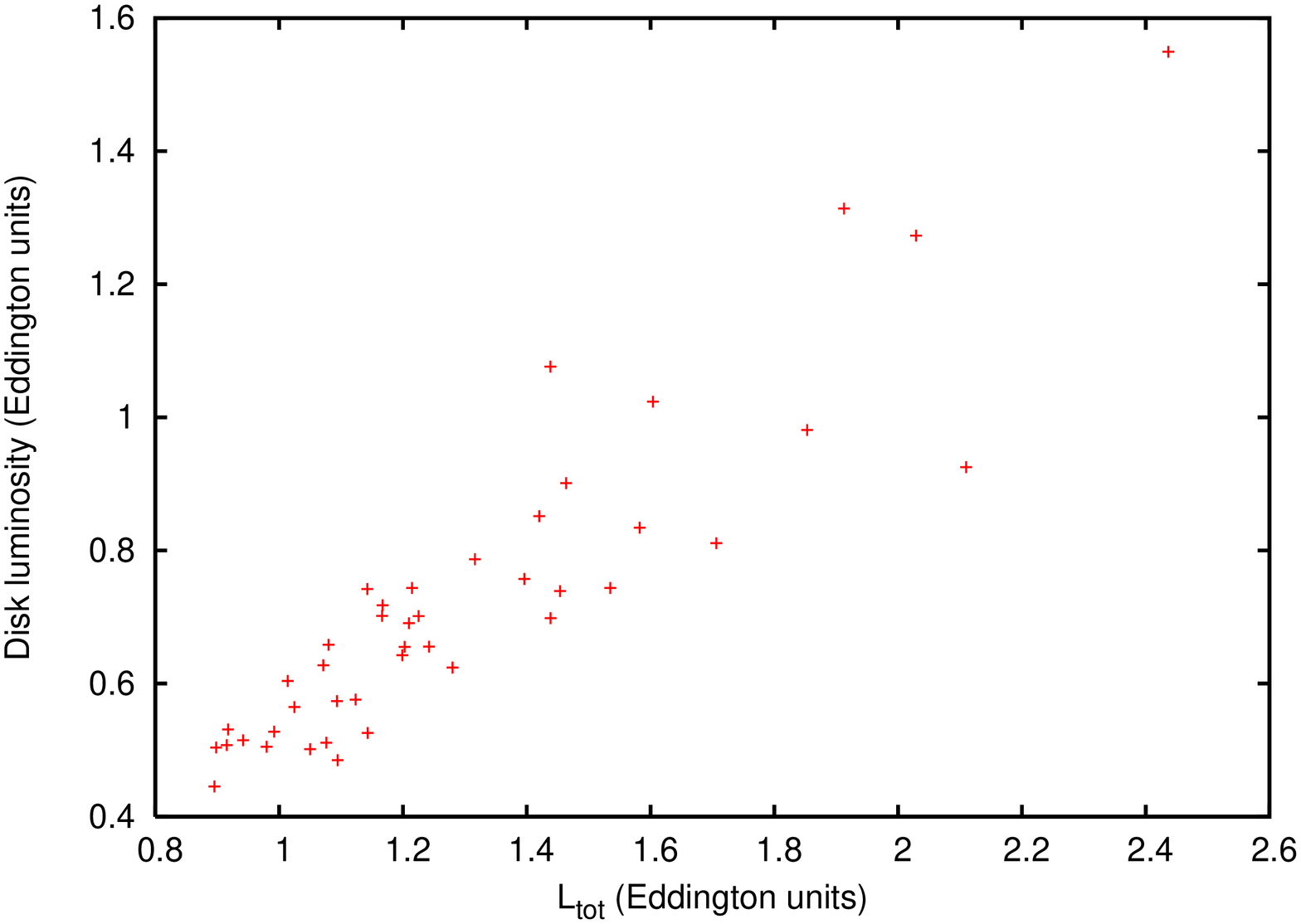} &
\includegraphics[height=6cm, width=4cm,  angle=-90]{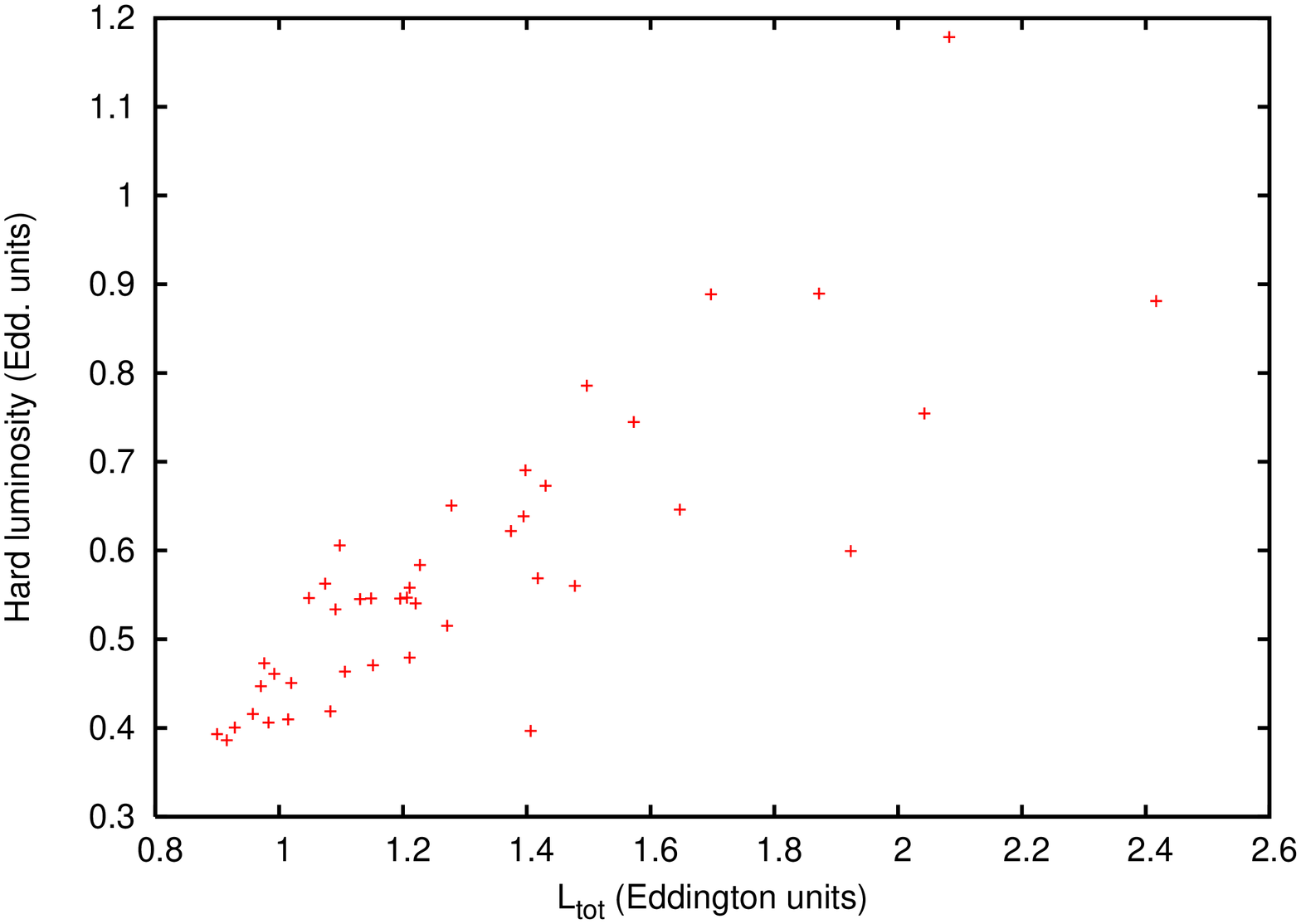} \\
\end{tabular}
\caption{ \small \linespread{1} Left panel: soft (disk) luminosity for
  the extracted  spectra vs.  total (0.1-200 keV)  luminosity.  Right
  panel: hard (Comptonized) luminosity  for the extracted spectra vs.
  total (0.1-200  keV) luminosity.  All luminosities are  in Eddington
  units for a 1.4 \msun~ NS. }
\label{fig2}
\end{figure*}

\begin{figure*}
\centering
\begin{tabular}{l l l l }
\includegraphics[height=4cm, width=2.8cm]{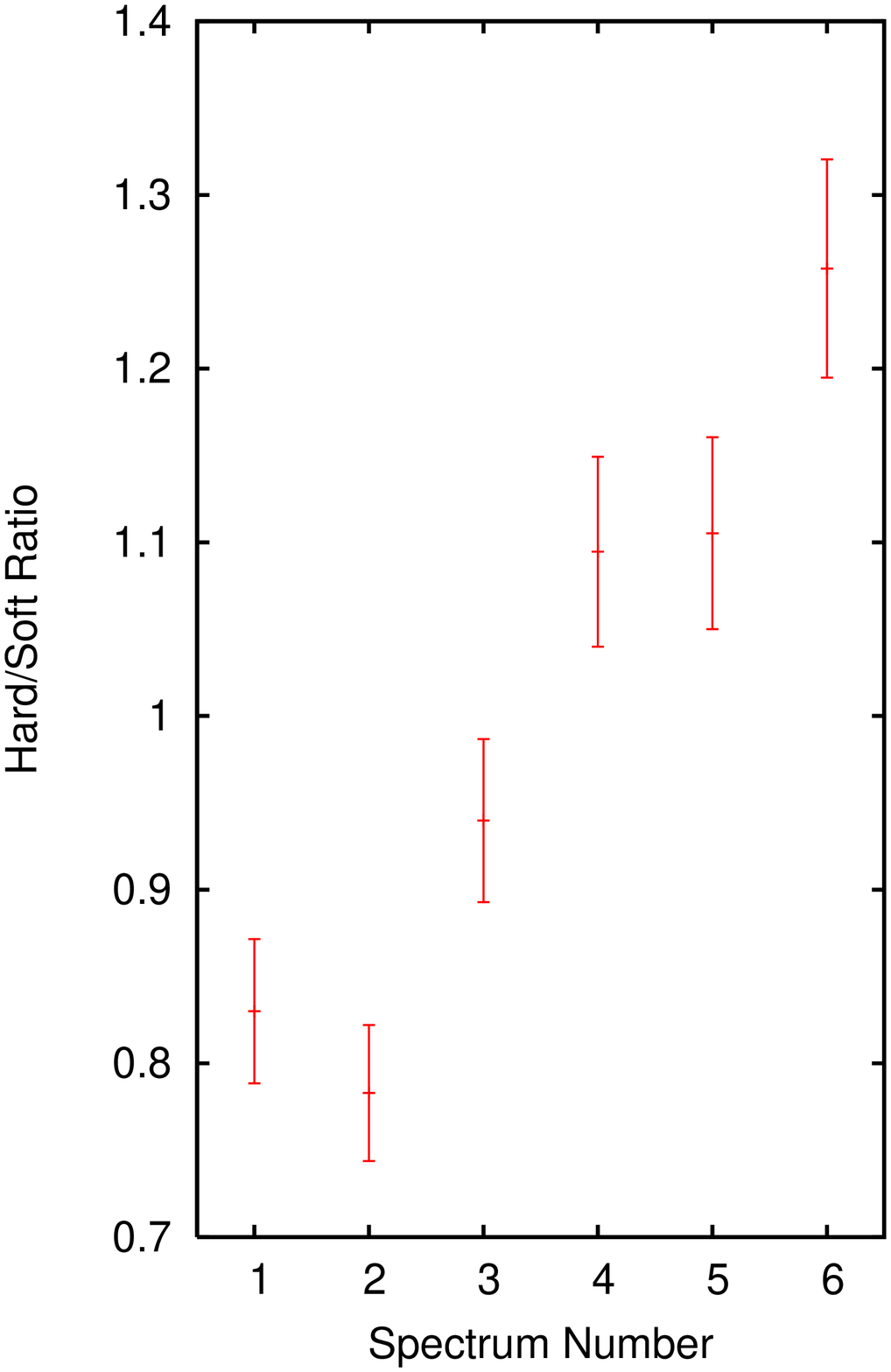} &
\includegraphics[height=4cm, width=2.8cm]{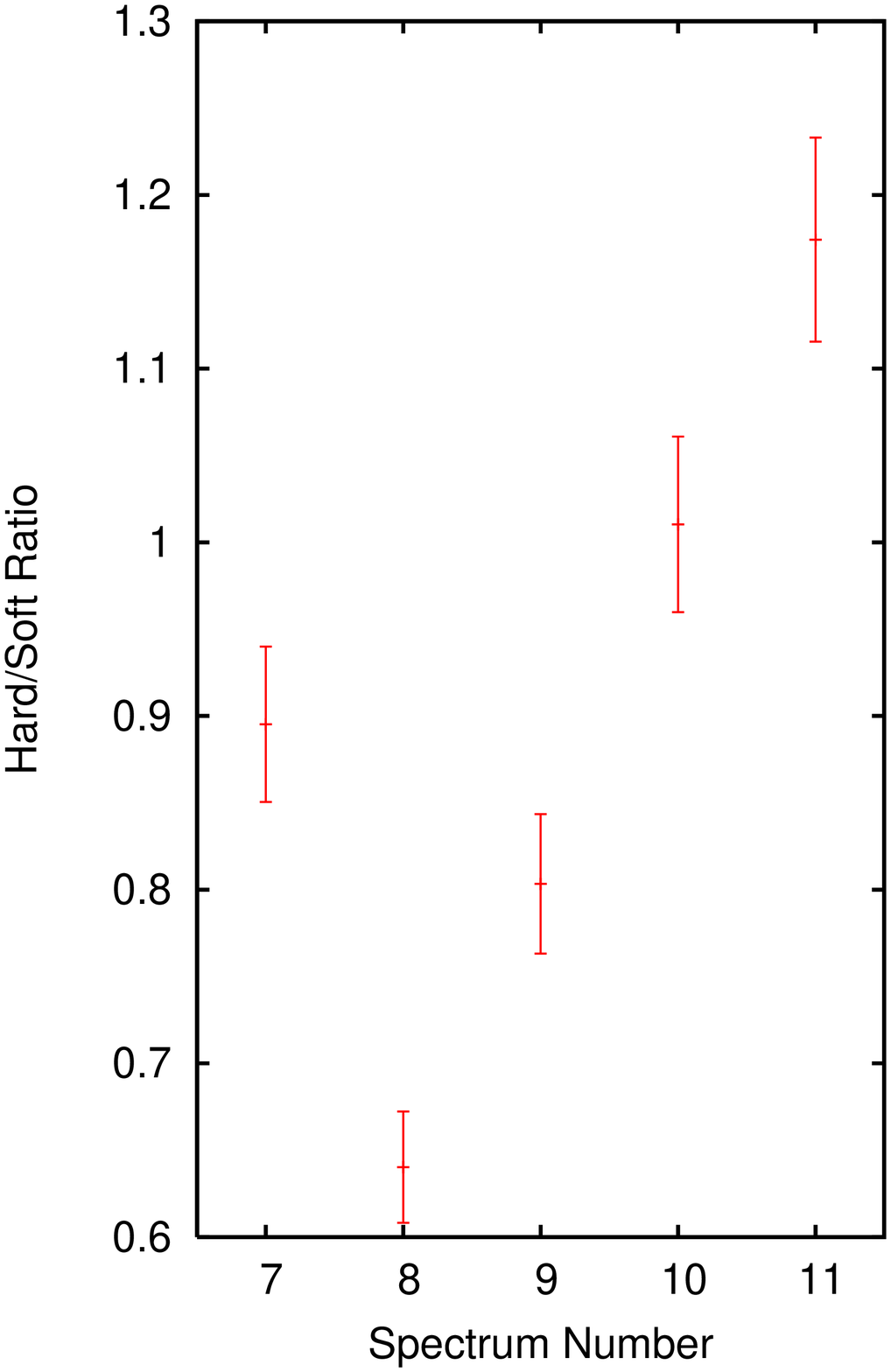} &
\includegraphics[height=4cm, width=2.8cm]{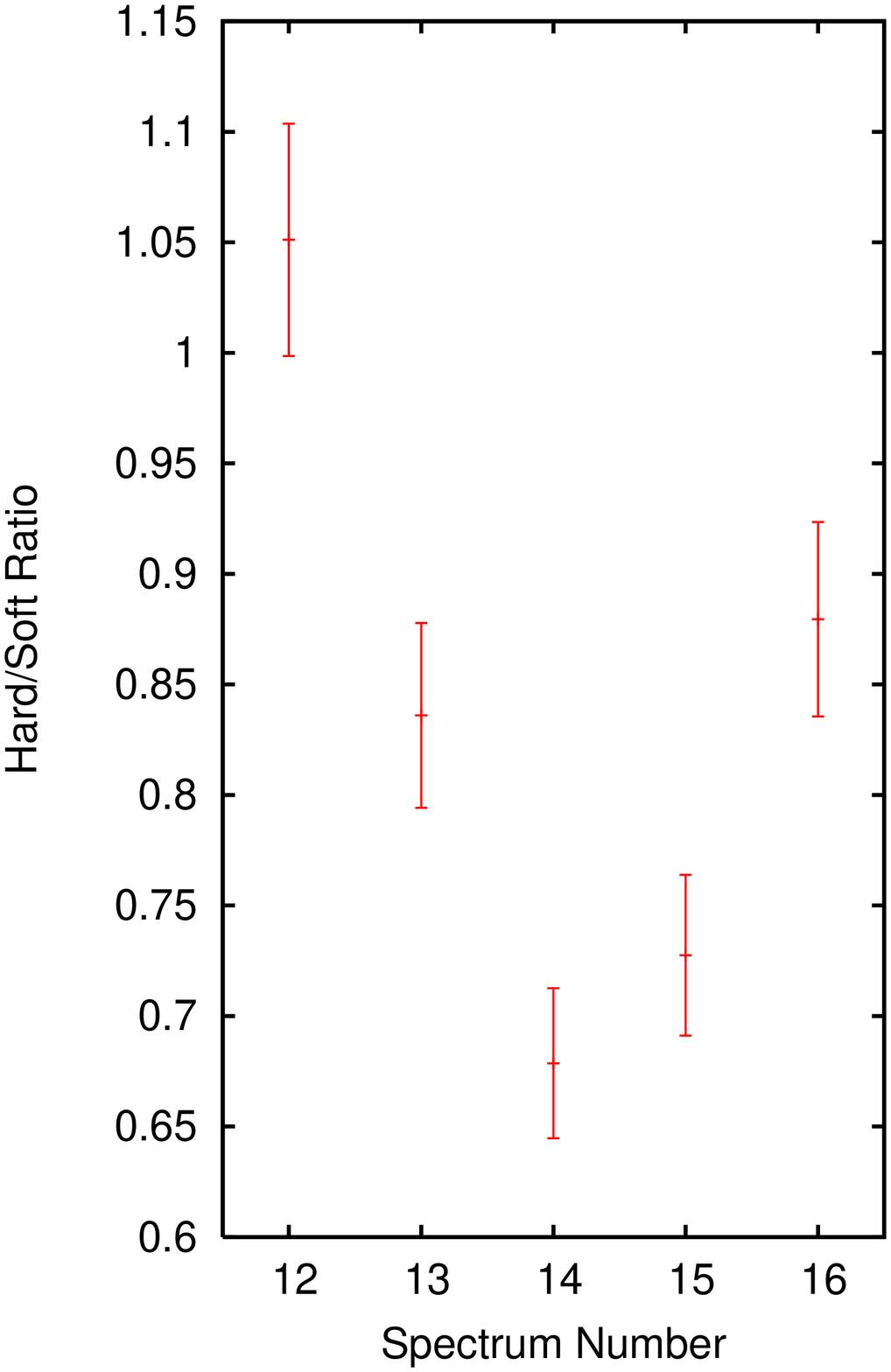} &
\includegraphics[height=4cm, width=2.8cm]{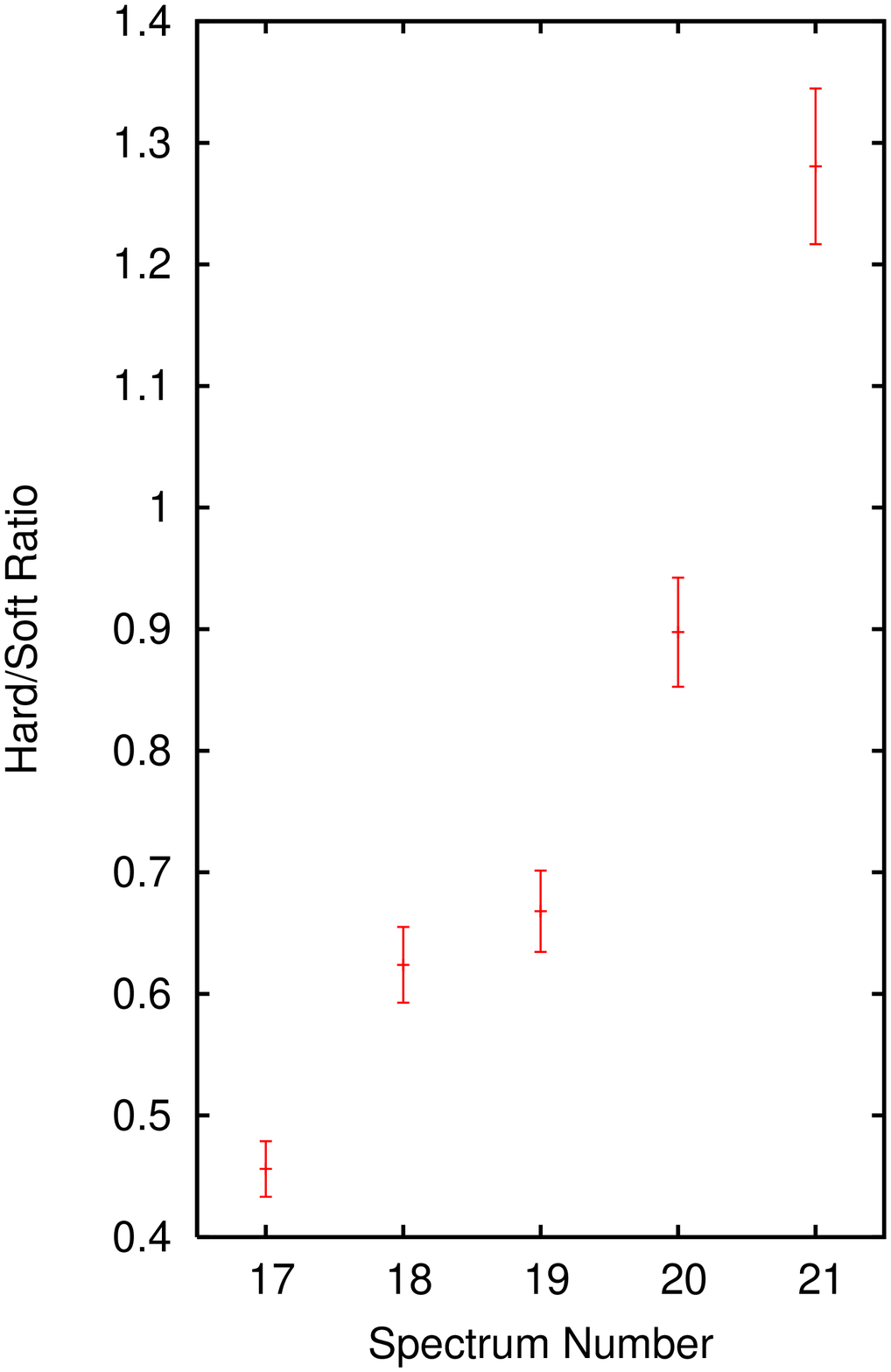} \\
\includegraphics[height=4cm, width=2.8cm]{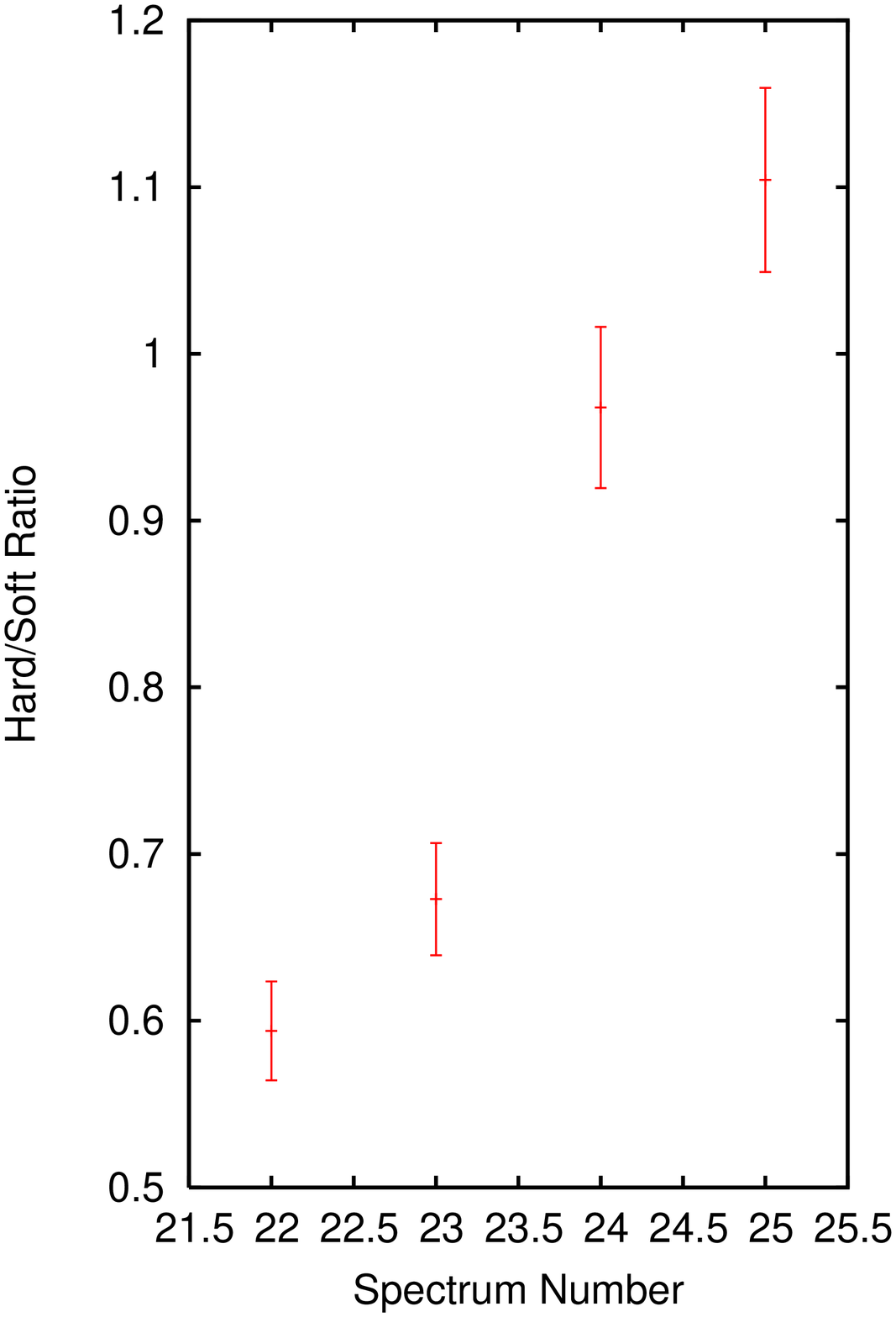} &
\includegraphics[height=4cm, width=2.8cm]{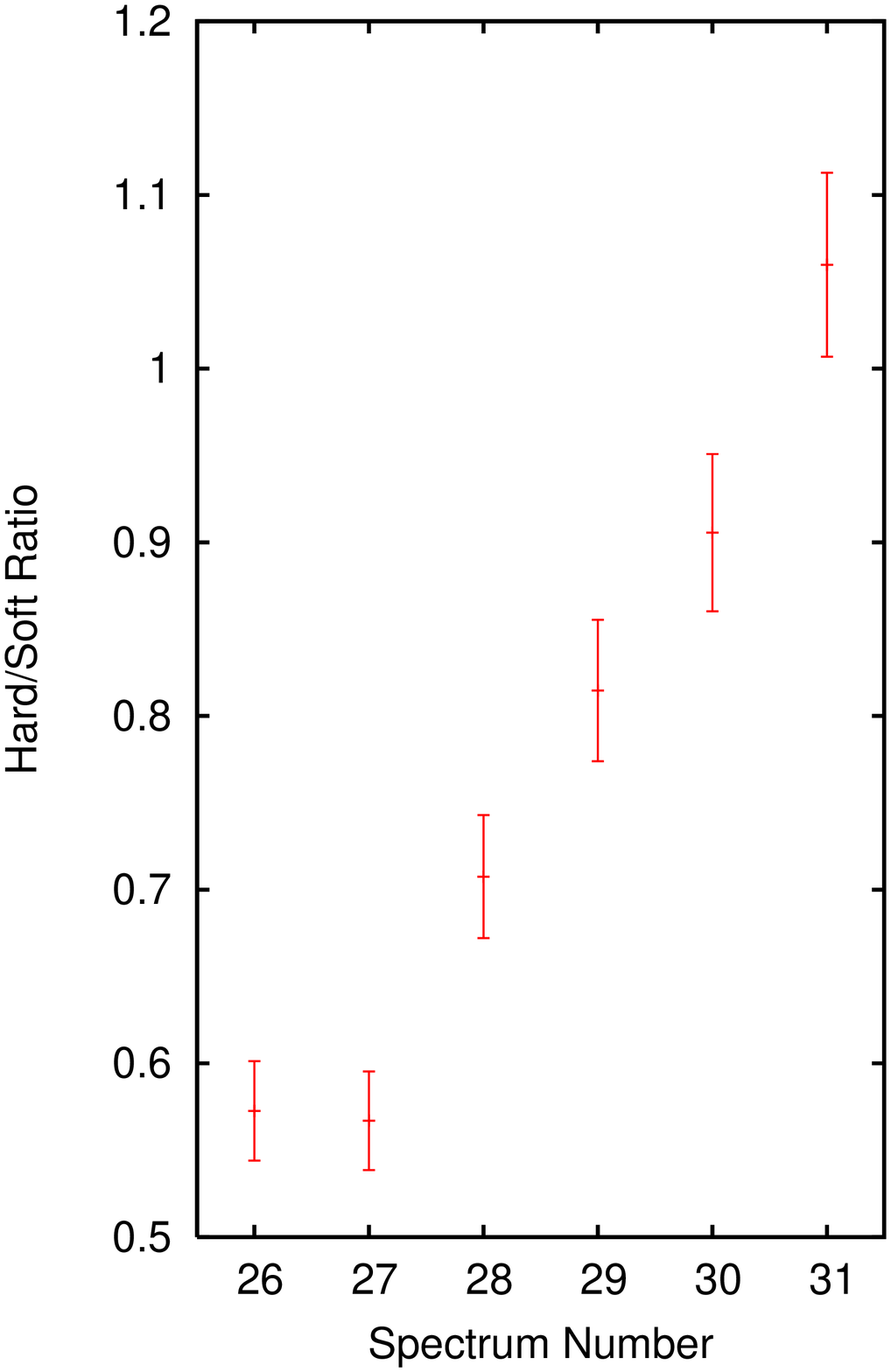} &
\includegraphics[height=4cm, width=2.8cm]{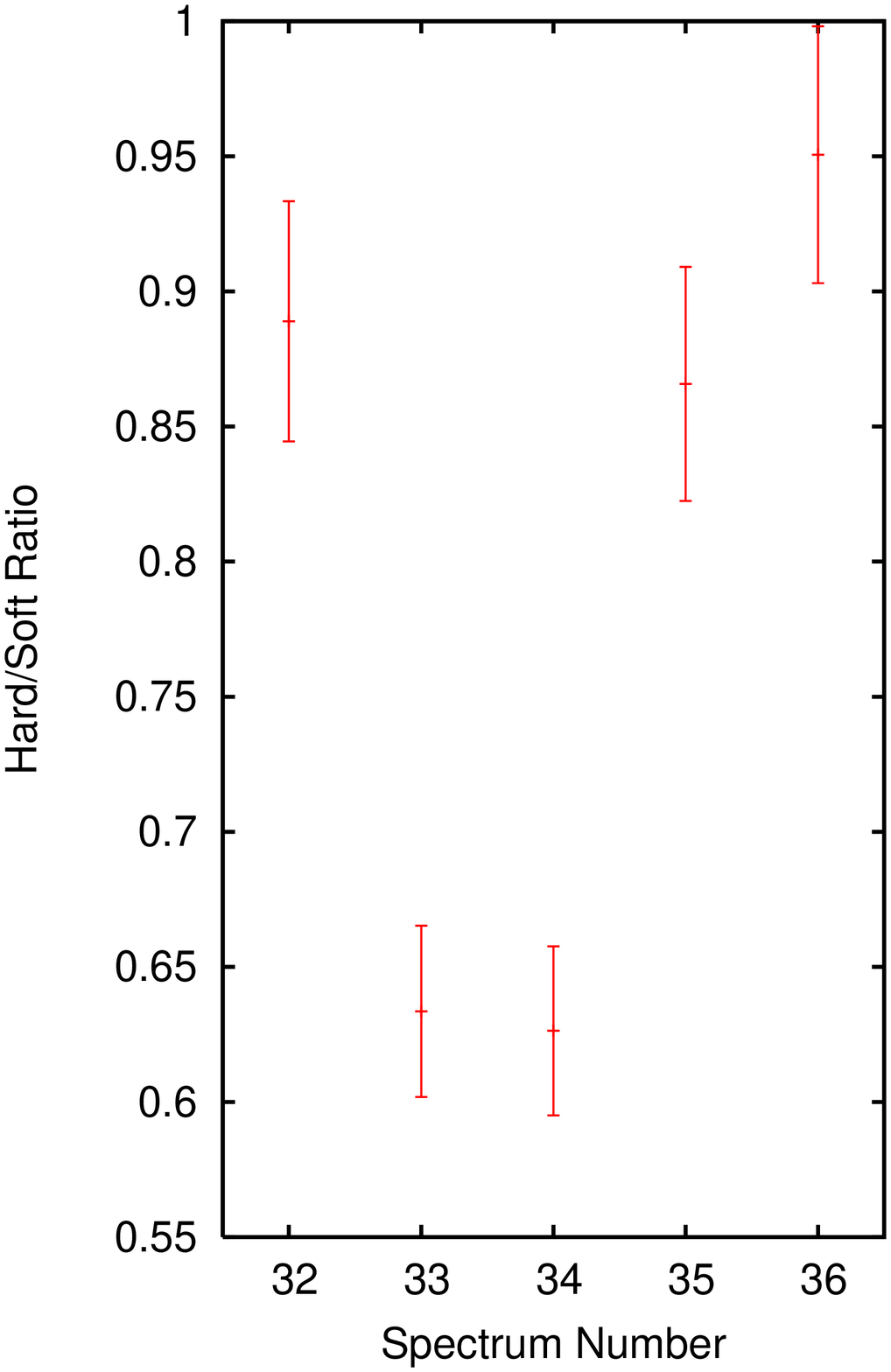} &
\includegraphics[height=4cm, width=2.8cm]{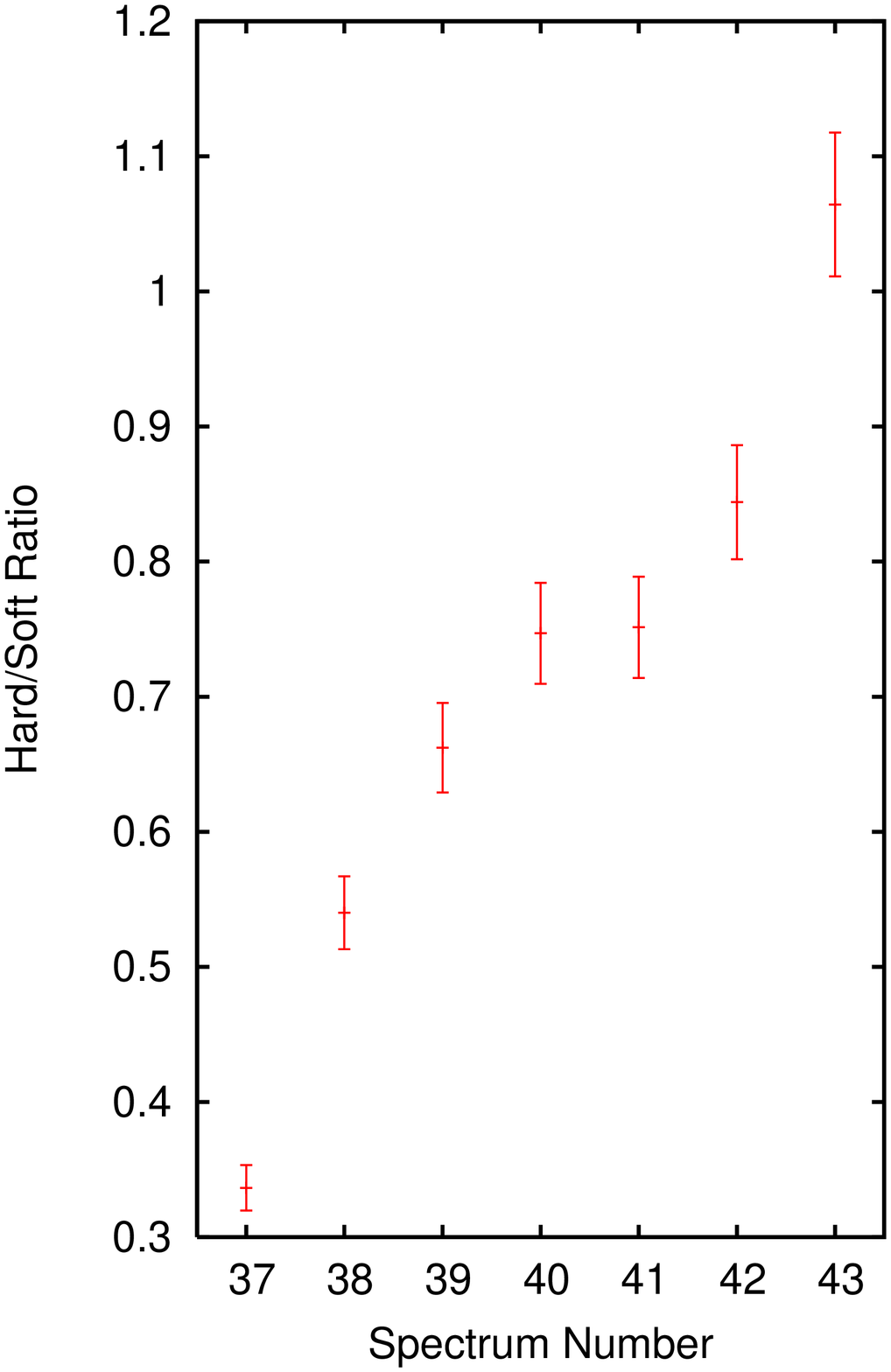} \\
\end{tabular}
\caption{  \small \linespread{1}
Ratio of the hard/soft luminosities for all the CD extracted spectra.
In the x-axis the number refers to the spectrum as indicted in Table ~\ref{tab2}.
We assumed a 5\% error for ratio of the two fluxes.}
\label{fig3}
\end{figure*}

\subsection{Disk emission and thermal Comptonization}
\begin{figure*}
\centering
\begin{tabular}{l l}
\includegraphics[height=6cm, width=4cm,  angle=-90]{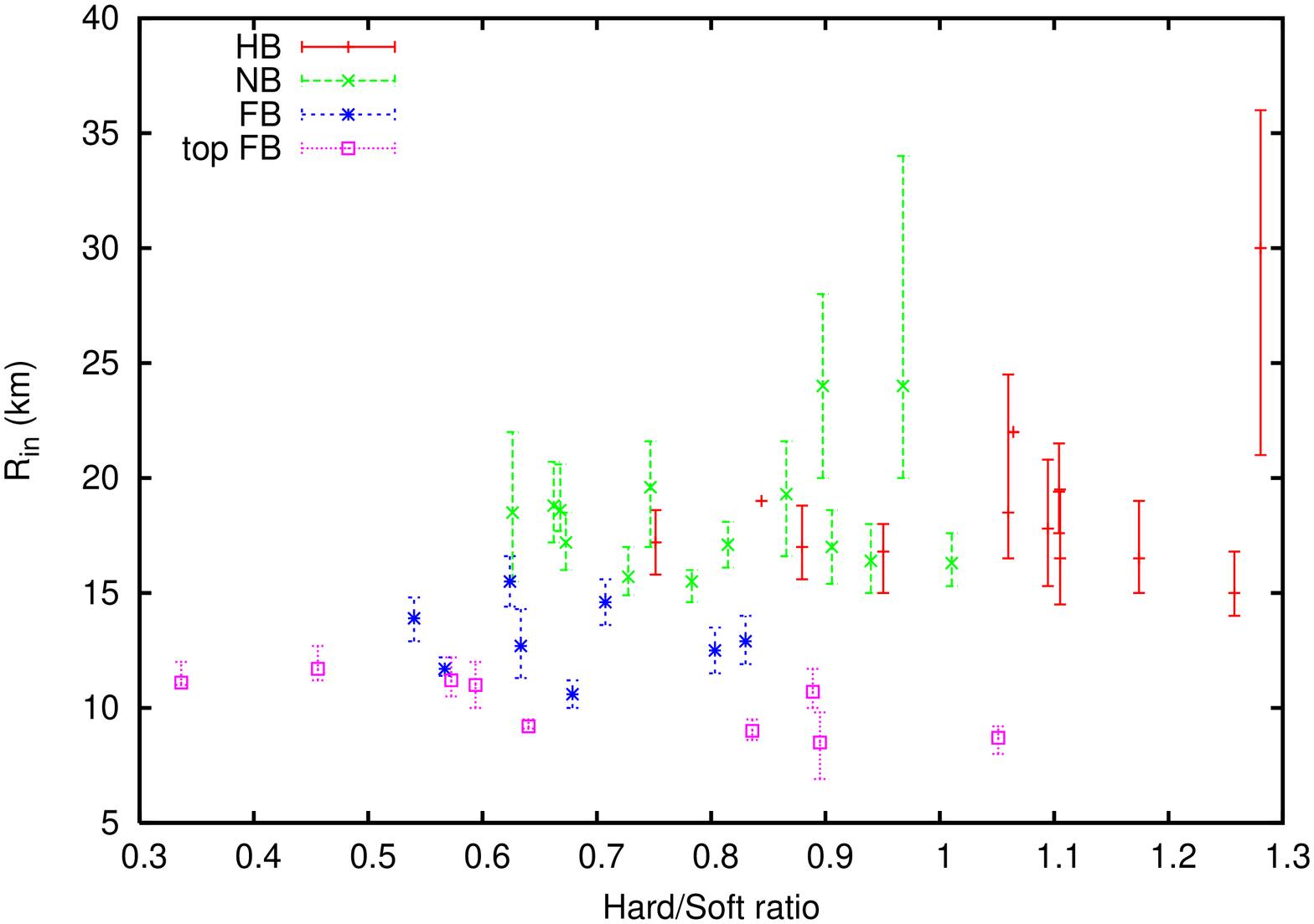} &
\includegraphics[height=6cm, width=4cm,  angle=-90]{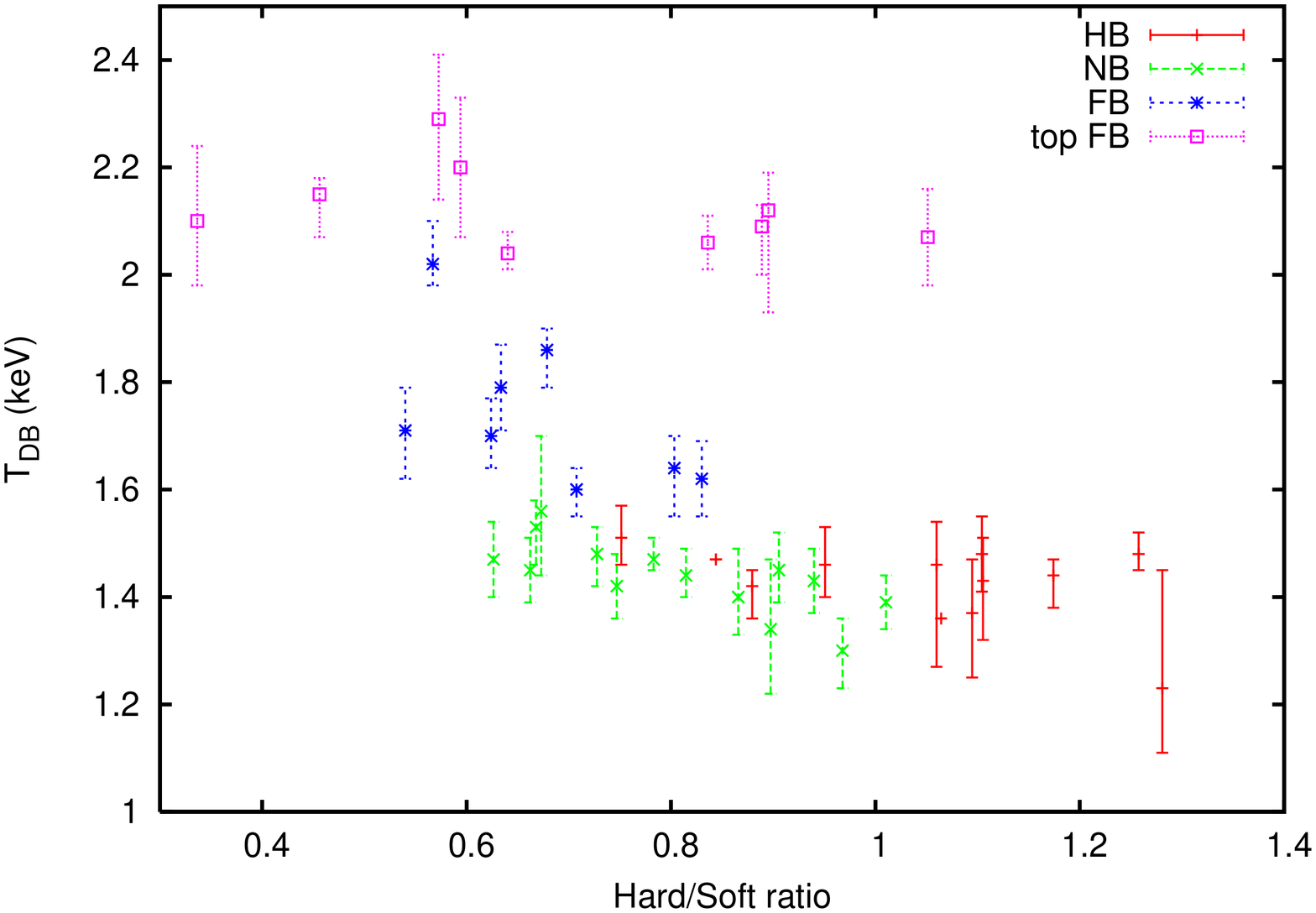} \\
\end{tabular}
\caption{ \small \linespread{1} Left  panel: inner disk radius vs. the
  hard/soft ratio.  Right panel:  temperature at the inner disk radius
  vs.  the  hard/soft  ratio.}
\label{fig4}
\end{figure*}

We  present in  Figure~\ref{fig3}  the dependency  of  the inner  disk
radius  and  of  the  inner  disk  temperature,  plotted  against  the
hard/soft   luminosity   ratio.   It   is   shown,   both   in   Table
~\ref{tab3a},~\ref{tab3b}  and  in  Figure ~\ref{fig4},  the  apparent
radius, derived without  considering important correction factors, due
to  the color/effective temperature  ratio, relativistic  effects, and
the non-zero torque boundary  conditions; moreover, it should be noted
that  these  corrective  factors  can  also  vary,  depending  on  the
accretion state.   The sum of all  these effects could  be a drastical
rescale of the measured inner radii up to almost an order of magnitude
higher \citep{merloni00},  but the general trend  that correlates this
parameter with the accretion state of the source should be
qualitatively preserved.\\
The  apparent inner  disk radius,  for spectra  in the  topFB  has the
lowest values,  $\leq$ 10 km, while  the average value  for NB spectra
and HB spectra is considerably higher (average value: $\sim$ 18 km for
the HB and  NB, $\sim$ 13.5 km for  the FB and $\sim$ 10.2  km for the
topFB).  The inner  disk temperature, on the other  hand, presents the
highest values, as one could  expect, on the FB, while the temperature
diminishes with the accretion rate. A small subset of FB spectra falls
out of the trend, but as  pointed out earlier, in this case, the exact
determination  of the  relative  contributions of  the two  components
could suffer  of systematic  uncertainties linked to  the particularly
luminous state of  the source.  On the FB the  disk temperature has an
average value of 2.12 keV, while on  the lower part of the FB sinks to
1.73 keV. There  is not any appreciable difference  on temperature for
spectra taken on the NB and on the HB with an average value of 1.43
keV.\\

For   the   hard,   thermal   Comptonized,  component   the   accurate
determination of all the  spectral parameters was not always possible.
The soft-seed photon temperature  presents, for each spectra analyzed,
a substantially higher  value with respect to the  disk temperature at
the  inner  radius.   We  hence  propose that  the  soft  photons  are
originated from  the boundary layer/NS  surface, partially thermalized
with the softer photons of inner part of the accretion disk around the
NS. Moreover, calculating the radius of the soft-seed photons emitting
region $R_w$ from:
\begin{equation} 
R_w = 3 \times 10^4 d \sqrt{\frac{f_{bol}}{1+y}}/(kT_0)^2,
\end{equation}
where  $d$ is  the distance  to the  source in  kpc, $f_{bol}$  is the
\COMPTT~  bolometric flux and  $y$ (see  Eq. \ref{ypar}),  the Compton
parameter, we derived for our sample values in the 3--6 km range, that
clearly indicate  a rather small emitting region,  thus supporting our
identification.\\
\begin{figure*}
\centering
\begin{tabular}{l l}
\includegraphics[height=6cm, width=4cm,  angle=-90]{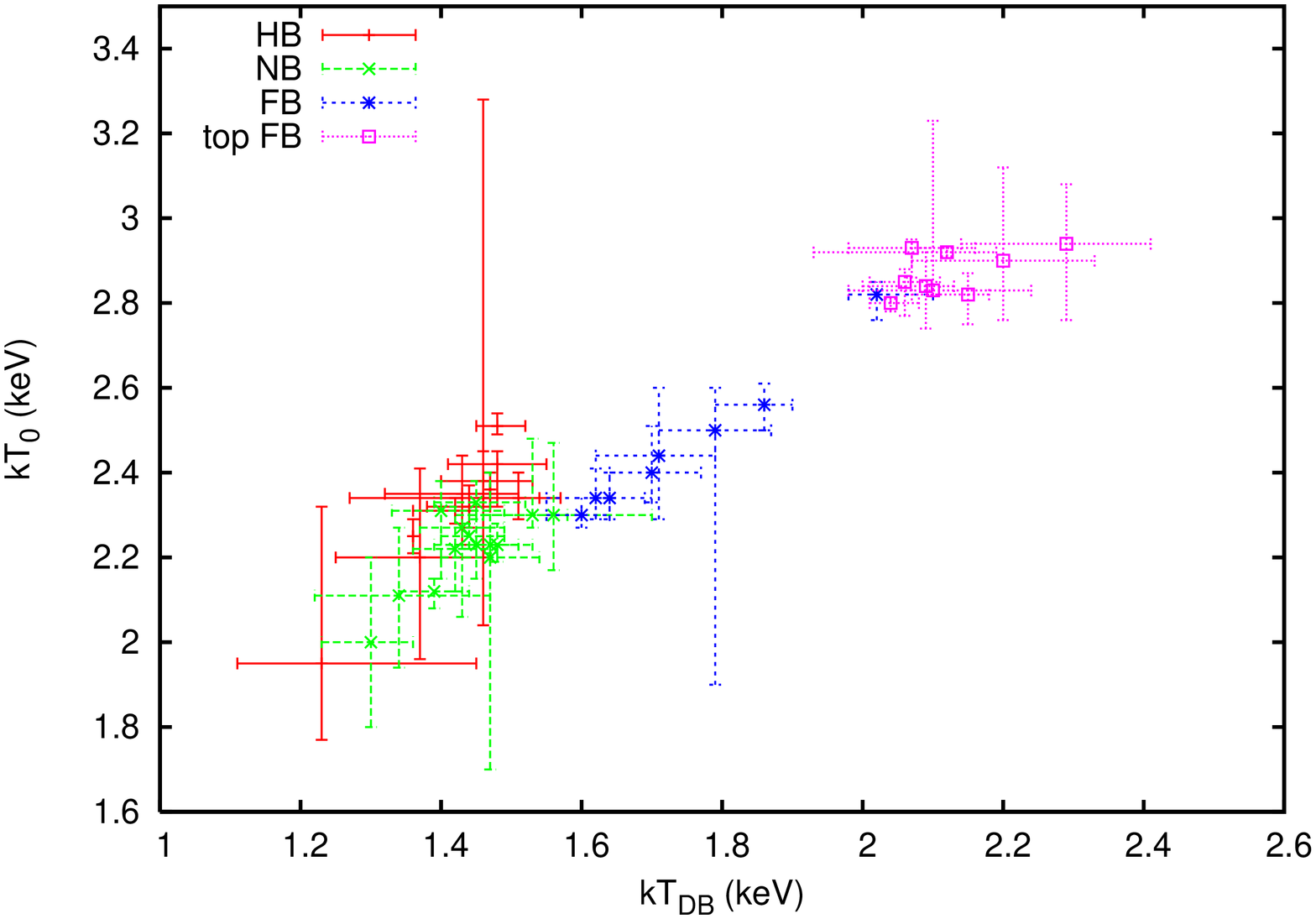} &
\includegraphics[height=6cm, width=4cm,  angle=-90]{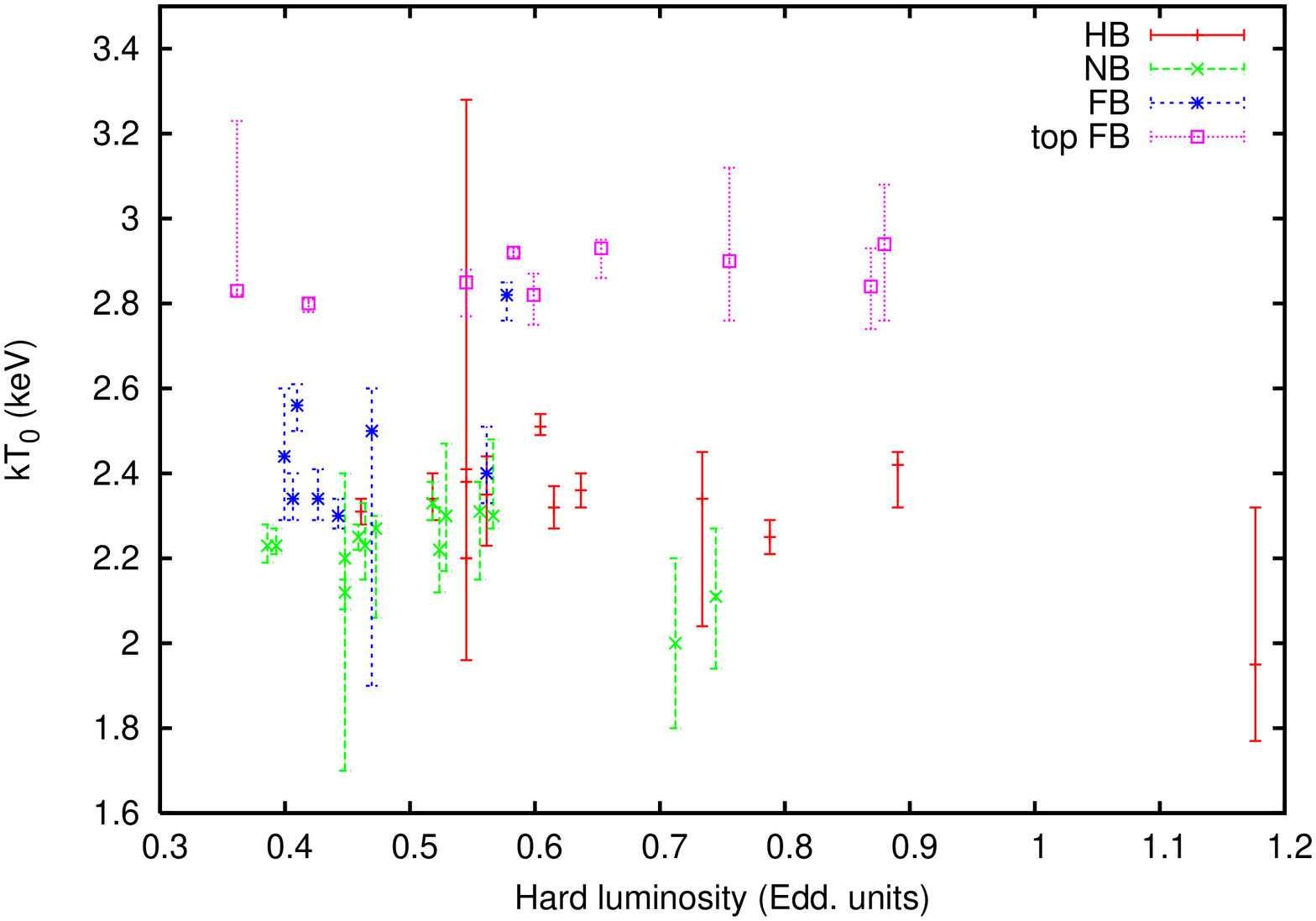} \\
\end{tabular}
\caption{   \small  \linespread{1}   Left   panel:  soft   seed-photon
  temperature  kT$_{0}$  vs.  the  inner  disk temperature  kT$_{DB}$.
  Right  panel:  soft   seed-photon  temperature  kT$_{0}$  vs.   hard
  (\COMPTT+\PEGPWRLW)  luminosity.}
\label{fig5}
\end{figure*}
The CD  correlated changing of the thermal  temperatures is compatible
with  a scenario in  which the  position of  the source  on the  CD is
determined  by the  instantaneous  \mdot, and  higher accretion  rates
correspond to higher seed-photons  temperatures, resulting in a hotter
radiation field.  The soft seed-photon temperature has  a 2.2 keV,
2.4 keV, 2.9 keV average values for spectra on the HB/NB, FB and topFB
respectfully.  When  plotted against each other,  the disk temperature
and the soft seed-photon temperature (see Figure~\ref{fig5}, left panel)
follow each other  quite closely, being these two  parameters the main
driving  physical quantities  related to  the accretion  state  of the
source.  The correlation  between the CD resolved spectra  and the two
temperatures shows, although in a qualitative representation, that the
link between  the spectral evolution  of the source and  its accretion
state is well motivated.\\
\begin{figure*}
\centering
\begin{tabular}{l l}
\includegraphics[height=6cm, width=4cm,  angle=-90]{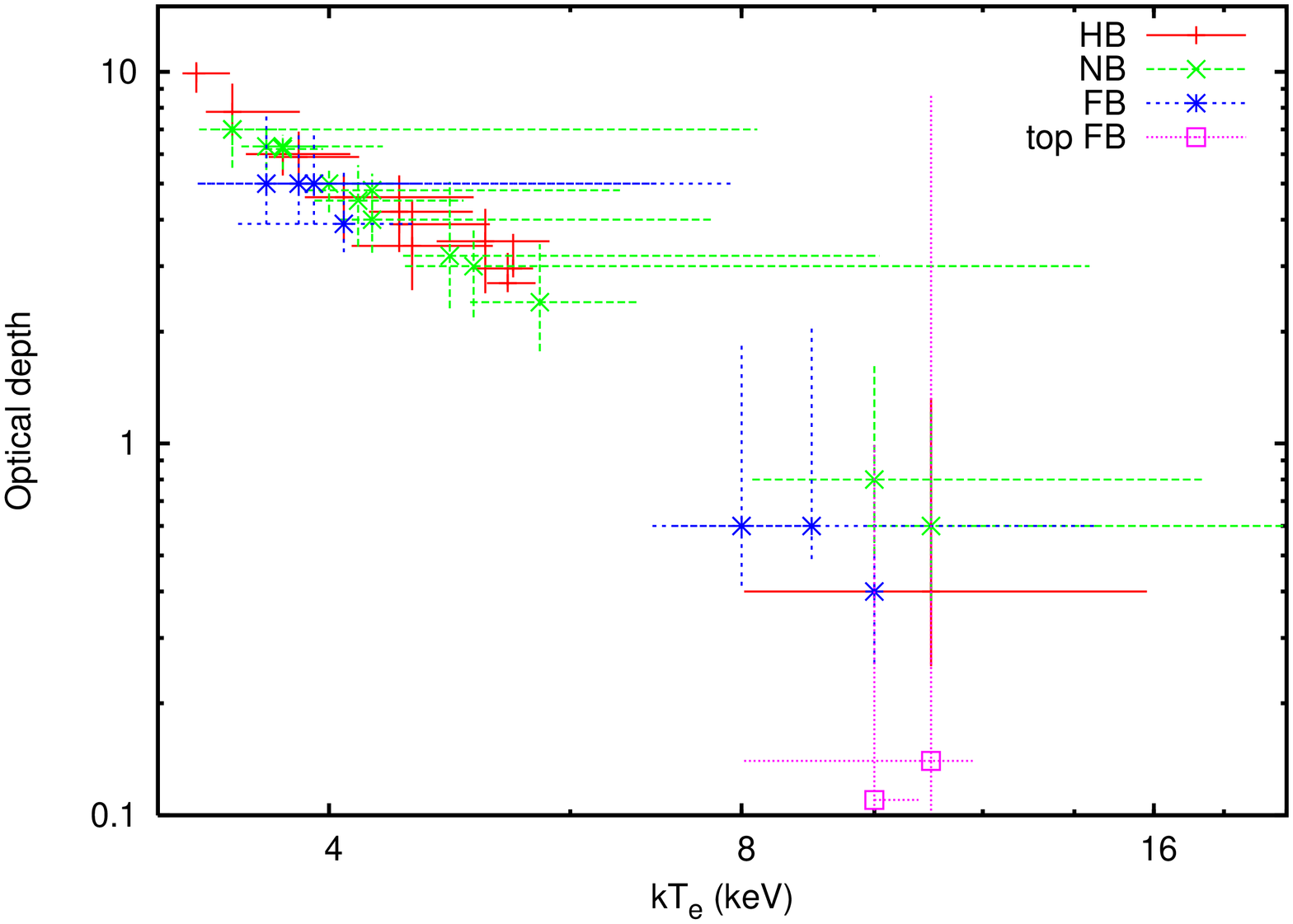} &
\includegraphics[height=6cm, width=4cm,  angle=-90]{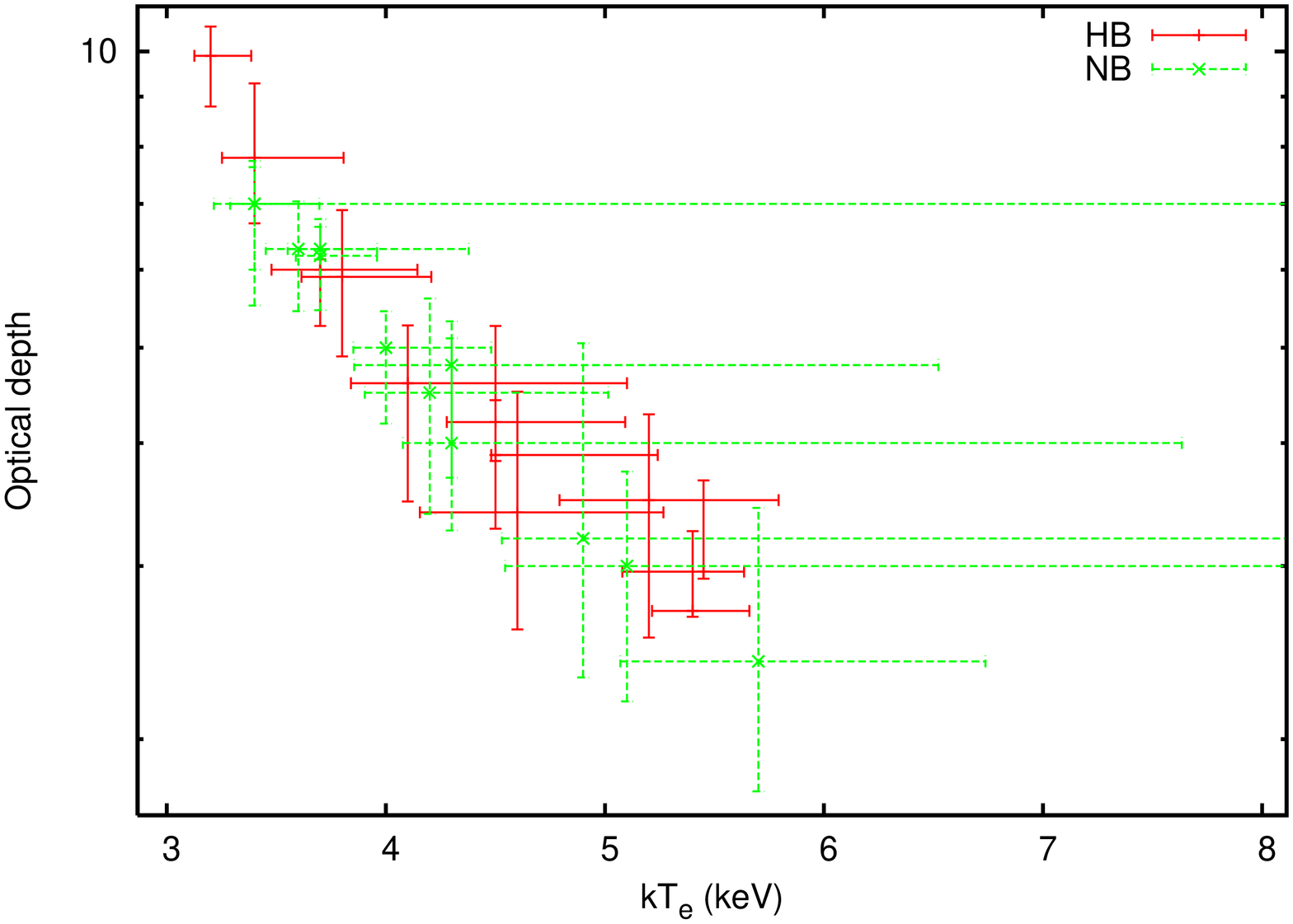} \\
\end{tabular}
\caption{ \small \linespread{1} Left panel: optical depth vs. electron
  temperature (all spectra).  Right  panel: optical depth vs. electron
  temperature (subset  of spectra on  the HB/NB).  Errorbars  show 1.0
  $\sigma$  errors.}
\label{fig6}
\end{figure*}
The  other  two spectral  parameters,  that  describe the  high-energy
curvature of  the Comptonized  component, the electron  temperature of
the Comptonizing  plasma (kT$_e$) and  the optical depth of  the cloud
$\tau$, are not always well constrained  by the fit.  As it is evident
from Figure~\ref{fig6}, for spectra at high accretion rate we obtained
high values for the electron  temperature (kT$_e \geq$ 10 keV) and low
values for the optical depth  ($\tau \leq$ 1).  As the sources resides
on  zones  of  lower   \mdot,  the  Comptonizing  cloud  substantially
thickens, while  the electron temperature  correspondently diminishes.
Plotting the  subset of the spectra in  the NB and HB  helps to better
visualize this  tendency (Figure ~\ref{fig6}, right panel).   It is to
be noted that  the determination of these two  parameters is partially
related with the hard  power-law component; although there are spectra
where the  \PEGPWRLW~ component is  not strictly necessary for  a good
fit,  its introduction  slightly  shifts the  the thermal  Comptonized
component to lower energies, thus giving a noticeable reduction in the
uncertainties of the Compton curvature.\\
In all the  spectra it is required a Gaussian  line at energies $\sim$
6.4 keV. It  mostly appear rather broad, with a  line width that often
reaches our  constraint of 0.8 keV.  The  equivalent widths associated
with the line  are in 80-200 eV  range, and are to be  taken as rather
indicative given the low energy resolution of the PCA and the presence
of systematic residuals in this energy range.

\subsection{The hard tail behavior}
The presence of  a hard X-ray excess in the  extracted spectra of \sco~
is mostly evident  in 18  spectra from  a total of  43 CD selected spectra.
There is  little evidence  of the  presence of   hard X-ray  excess for
spectra lying in the  bottom part of the FB and of  the NB, namely for
spectra taken  near the  apex that connects  the left and  right track
that compose the V pattern.\\
In  all the  examined CD  patterns  we always  detect a  power-law
high-energy  excess,  dominant above  $\sim$  40  keV,  as the  source
resides at the top of the  left track, i.e.~ a portion of the diagram
that  we  tentatively  identified   with  the  HB,  and  which  should
correspond to the  lowest mass accretion rate.  This  group of spectra
are characterized by  the lowest values of inner  disk temperature and
soft seed-photon temperature; the  derived inner disk radii correspond
to  a disk  truncated at  about  10-15 R$_g$,  while the  Comptonizing
optical  cloud is  substantially  thicker with  respect  to any  other
spectrum in  other zones  of the  CD; the fraction  of the  total power
dissipated in the  Comptonizing corona has the highest  values, with a
ratio hard/soft luminosity well above unity.\\
\begin{figure*}
\centering
\begin{tabular}{l l}
\includegraphics[height=6cm, width=4cm,  angle=-90]{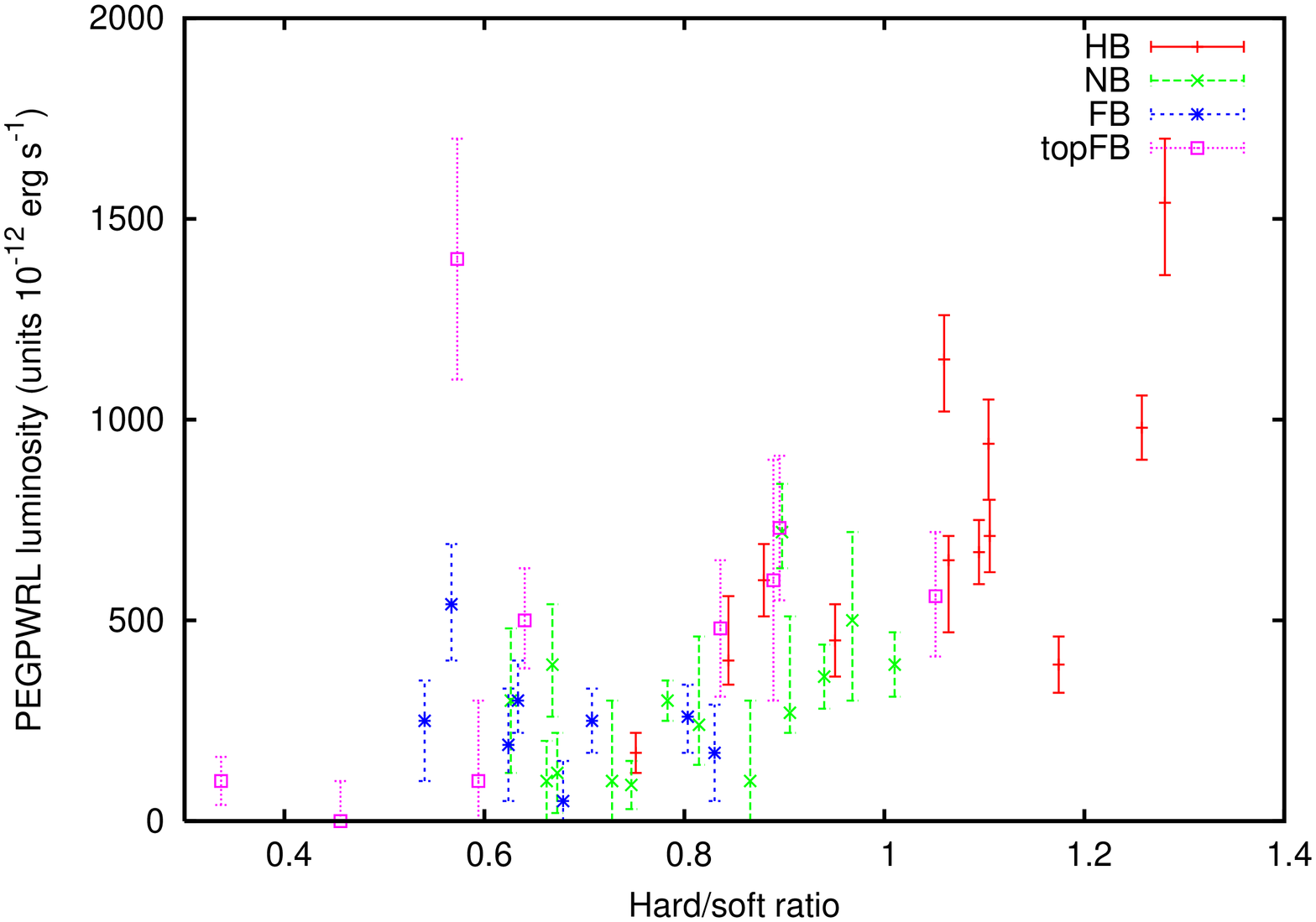} &
\includegraphics[height=6cm, width=4cm,  angle=-90]{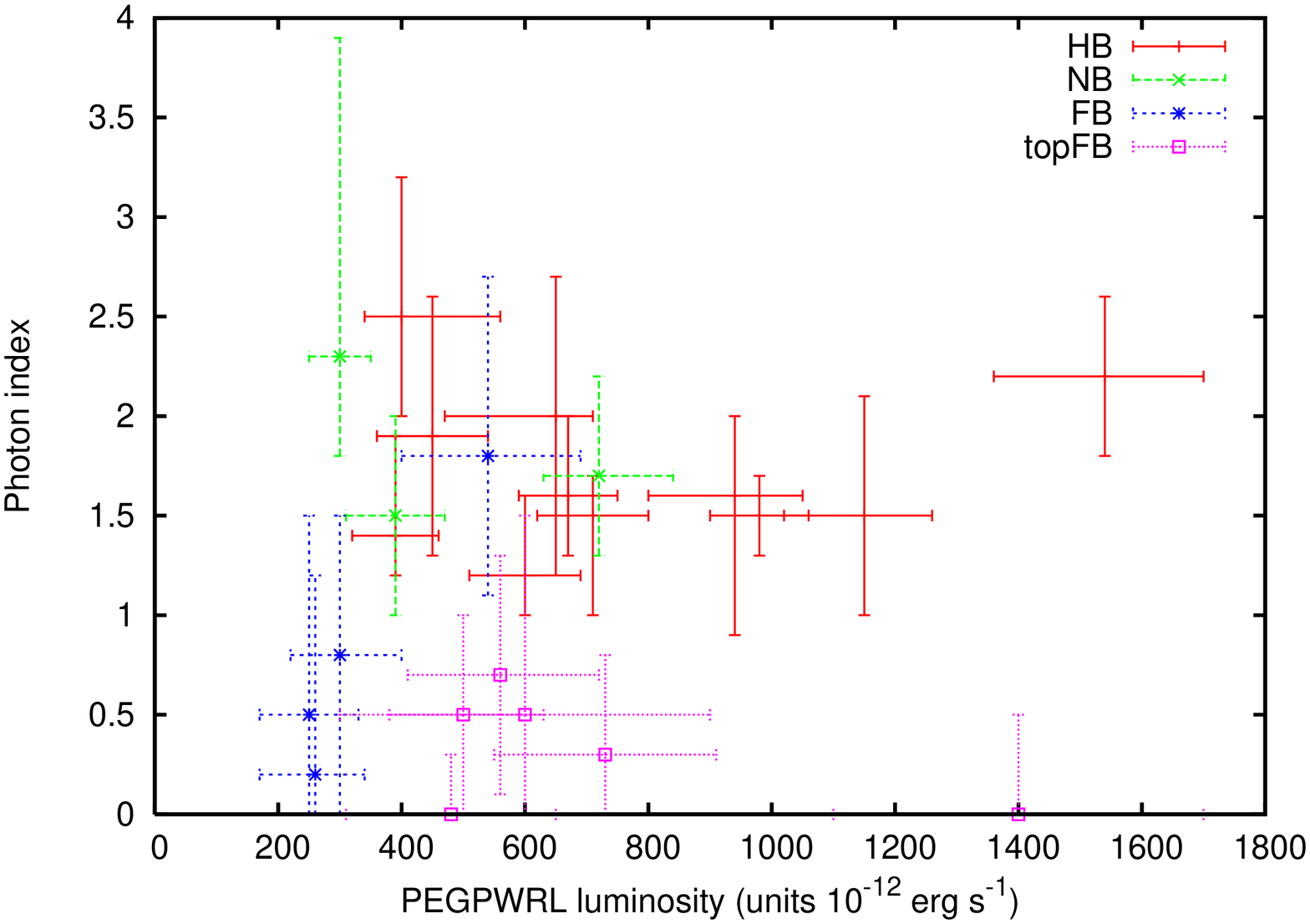} \\
\end{tabular}
\caption{ \small  \linespread{1} Left panel: hard  tail luminosity (in
  the 20-200 keV energy range) vs.  the hard/soft ratio.  Right panel:
  the energy  index of the  \PEGPWRLW~ component vs.  its  20--200 keV
  luminosity (the subset of spectra for which the photon index was not
  costrained by the fit is excluded from this figure).}
\label{fig7}
\end{figure*}
Apart from  this consistent  group of HB/NB  spectra, we also  found a
minor group  of spectra (namely 5 spectra:  7, 8, 12, 13  and 26), for
which the  fit required a third high-energy  component.  These spectra
belong to 3 CDs (20035B, 30036 and 40020) and are spectra taken at the
top of  the FB.   Although these  spectra are located  in the  same CD
region, not all the CD that  display a topFB clearly show this excess.
This could be due to a statistical reason, as in all the other CDs the
HEXTE countrate  (see Table ~\ref{tab2})  drops of about one  order of
magnitude  with respect  to  these  3 CDs;  because  the countrate  is
essentially concentrated  in the 20--35  keV energy band, that  is the
range  that  constrains  the  curvature  of  the  thermal  Comptonized
component,  the  energy channels  in  this  energy  band with  a  high
statistic  will  drive  the  fit,  and  any  small  excess  above  the
exponential tail  of the  Comptonized component will  statistically be
more  accentuated.   However,  we  cannot  exclude  that  an  opposite
reasoning  is true,  i.e.   that we  are underestimating  instrumental
effects  that   appear  only  for  higher   count-rates  as  dead-time
corrections or poor background estimate, so that these particular hard
tails  are  artifacts of  an  incorrect  modelization  of the  thermal
Comptonization  curvature.   As we  have  no  apparent  way to  better
calibrate  our  spectra,  we  discuss  also  the  appearance  of  this
component for  this group of spectra.  Future  observations with other
high-energy satellites, such as INTEGRAL  or SUZAKU, will test if this
particular topFB  state of  the source is  also accompanied by  a high
energy excess or it is an HEXTE faked detection.\\
The hard excess  has no evident relation with  the total luminosity of
the source,  but this could be  expected since the  different zones of
the CD  track are not related  with the total  X-ray luminosity. Using
the pegged power-law, we directly derived from the normalization value
of this  component, the  flux in the  20.0-200.0 keV energy  band.  We
show in Figure~\ref{fig6}, left panel, the hard tails luminosity in the
20--200 keV range  vs.  the hard/soft ratio.  From the  plot it can be
clearly seen  that the hardening of  the spectrum is  reflected in the
increasing  luminosity of  the  hard-tail component;  the CD  resolved
spectra are,  correspondently, disposed on  this plot: the  FB spectra
reside on  the bottom  left of  the diagram with  hard fluxes  $\leq 2
\times 10^{-10}$ erg s$^{-1}$, NB  spectra are harder and half of them
show significant fluxes ($\sim 4 \times 10^{-10}$ erg s$^{-1}$), while
the HB spectra, in the right part of the plot, show the strongest hard
X-ray emission (in the range 6--12 $\times 10^{-10}$ erg s$^{-1}$).\\
As pointed out in the previous section, we found for the topFB sample,
spectra with no  detectable hard X-ray emission, that  occupy the left
bottom  part of  this figure  and that  smoothly join  the correlation
between hard/soft ratio  and hard flux and spectra  that do not follow
this trend  and for  which there is  both in  the \COMPTT~ and  in the
\PEGPWRLW~ component  a hard  flux higher than  we expected.   From the
figure is  also evident  that one  of the most  luminous hard  tail is
found in a particular spectrum on  the topFB (namely spectrum 26 of CD
40020);  the luminosity  of this  component  is related  to the  total
broad-band high  luminous state  of the source,  that reaches  in this
case the highest value of total luminosity (2.6 L$_{Edd}$) observed in
our sample.\\
On  the other  hand, the  values of  the photon  index present  a well
defined bimodal  distribution according to the position  of the source
on the CD. We find quite  flat power-laws, with index values less that
unity for all  the FB and topFB spectra, while for  the other group of
spectra, values  generally range between  1.5 and 2  (Figure~\ref{fig6},
right panel).
\subsection{The hybrid Comptonization model}  
Models of  hybrid Comptonization  have so far  been mostly  adopted to  
explain state transitions both  in black-hole candidate systems (as in  
Cyg  X-1, \citealt{gierlinski99}; GRS  1915+105, \citealt{zdziarski01};  
or  GX  339+4,  \citealt{wardzinski02})  and in  NS  system  \citep[GX  
17+2,~][]{farinelli05}.\\
We  tried to  model the  spectra for  which we  detected a  hard tail,
making  the  hypothesis  that  both  the hard  tail  and  the  thermal
Comptonized component  were related to each other. In this way,  all the
broad-band evolution of the source  is covered in a self-consistent way
by only  two spectral components:  the thermal soft disk  emission and
the hard hybrid Comptonized component.  In table \ref{tab4} we present
the results  of our fits for  the group of  18 spectra, distinguishing
between the HB/NB group (top part of the table), for which we used the
\DISKBB~ component as in the  case of \DBBTTPEG~ modelization, and the
topFB spectra,  for which  we used the  \DISKPN~ component,  under the
assumption that for this state  the disk is not truncated, and reaches
the surface of the NS (we assumed that this happens at a distance of 6
R$_g$).

\begin{figure*}
\centering
\begin{tabular}{l l}
\includegraphics[height=6cm, width=4cm,  angle=-90]{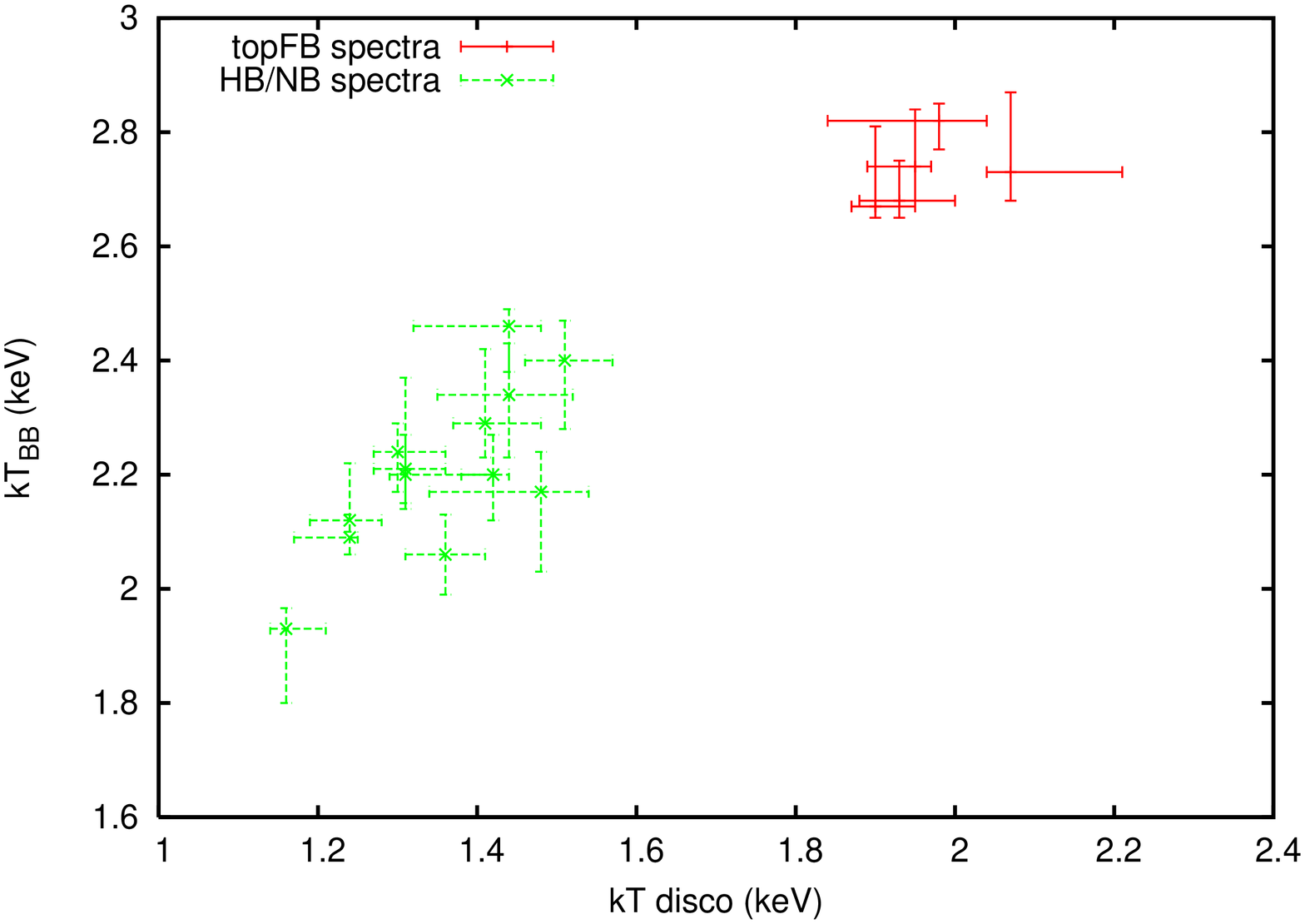} &
\includegraphics[height=6cm, width=4cm,  angle=-90]{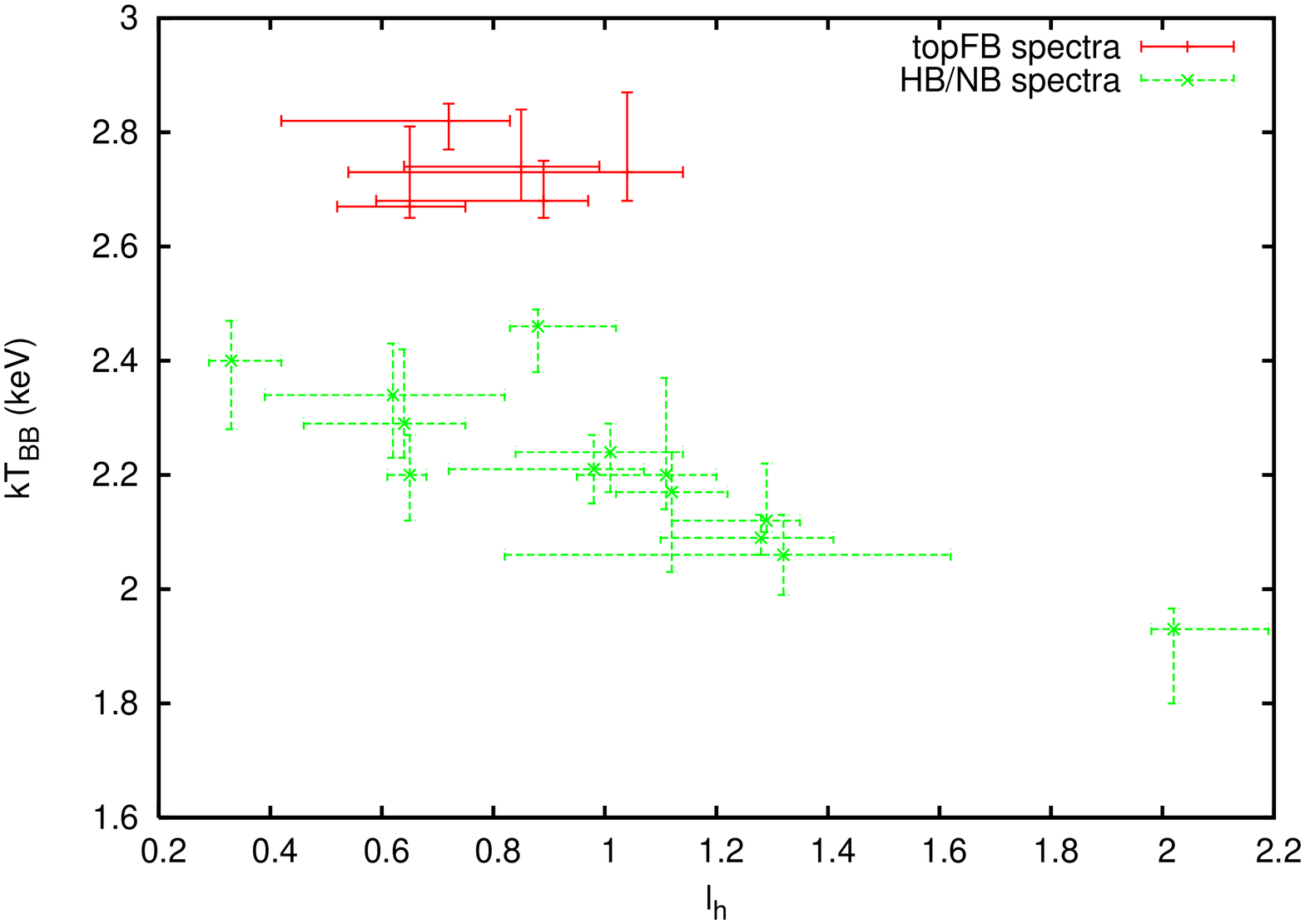} \\
\end{tabular}
\caption{   \small  \linespread{1}   Left   panel:  soft   seed-photon
  temperature kT$_{BB}$ vs.  inner disk temperature kT$_{DISK}$ (which
  corresponds to the kT$_{DB}$ value of the \DISKBB~ component for the
  HB/NB spectra and to the  kT$_{max}$ value of the \DISKPN~ component
  for  the topFB  spectra) for  spectra fitted  using  the \DBBEQPAIR~
  modelization.   Right  panel: the  soft  seed-photon temperature  is
  plotted against the hard compactness parameter.}
\label{fig8}
\end{figure*}

As in  the case  of the previously  adopted modelization,  the thermal
disk  temperature and the  soft-seed photon  temperature are  the main
driving  physical  parameters  that  determine  the  changing  of  the
spectral state of the source. We show in Figure~\ref{fig8} (left panel),
the correlation between the two temperatures: the topFB spectra occupy
the top  right part of  the plot, with  an average $kT_{BB}$  value of
2.73 keV and  inner disk temperature $kT_{DB}$ of  1.97 keV, while the
set of  the HB-NB  spectra are  disposed along a  linear trend  in the
1.2-1.8  keV energy  range for  the disk  temperature and  1.9-2.5 keV
range for  the soft seed-photon temperature. The  plot closely follows
the plot of Figure~\ref{fig5}, right panel.\\
The hard compactness ${\it l_h}$  results for all the examined spectra
only a small fraction of the soft compactness ${\it l_s}$, with values
of hard/soft ratio  ${\it l_h/l_s}$ below 0.2. The  HB/NB spectra span
almost an order of magnitude in the ${\it l_h}$ range, while the topFB
spectra are all grouped in a more narrow range of values, around ${\it
  l_h}$ = 0.6--0.8 (see Figure~\ref{fig8}, right panel).\\
The two groups of spectra  significantly differ both in the $\tau$ and
${\it  l_{nth}/l_h}$ values  (it is  to  be noted,  however, that  the
optical depth reported from the fits has not the same physical meaning
of the classical  Thomson optical depth of the  Comptonizing cloud, as
it  more properly  refers to  the  optical depth  associated with  the
background  photon  radiation  field):  HB spectra  present  generally
higher values of optical depth (with an average value of $\tau_p \sim$
2.2)   and  a  hybrid   electron  distribution   (with  0.3   $<  {\it
  l_{nth}/l_{h}} <  $ 0.8,  and an average  value of 0.55),  while the
topFB spectra  have considerably lower  values of $\tau_p \leq  1$ and
essentially non-thermal  spectra, with ${\it l_{nth}}  \simeq 1$.  The
slope   of  the   high-energy   power-law  of   the  hybrid   electron
distribution, $\Gamma_{inj}$ is consistent with  a value of 2, for the
HB-NB spectra, while  for the topFB spectra its  value is considerably
lower (between 0 and 1).\\

\begin{figure*}
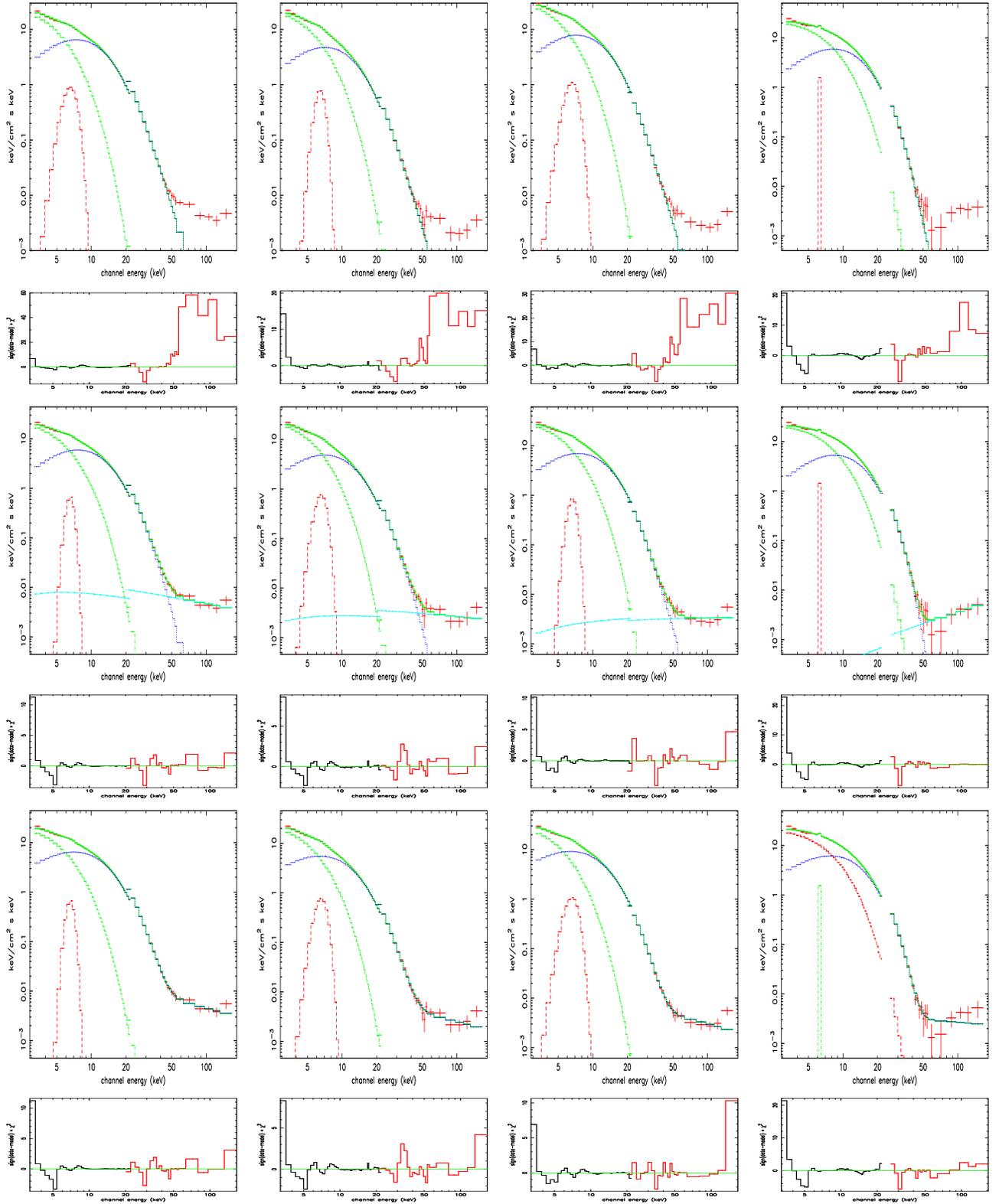

\centering
\begin{tabular}{l l l l}
\includegraphics[height=4cm, width=4.8cm,angle=-90 ]{f9a.eps} &
\includegraphics[height=4cm, width=4.8cm,angle=-90 ]{f9b.eps} &
\includegraphics[height=4cm, width=4.8cm,angle=-90 ]{f9c.eps} &
\includegraphics[height=4cm, width=4.8cm,angle=-90 ]{f9d.eps} \\
\includegraphics[height=4cm, width=1.8cm,angle=-90 ]{f9e.eps} &
\includegraphics[height=4cm, width=1.8cm,angle=-90 ]{f9f.eps} &
\includegraphics[height=4cm, width=1.8cm,angle=-90 ]{f9g.eps} &
\includegraphics[height=4cm, width=1.8cm,angle=-90 ]{f9h.eps} \\
\includegraphics[height=4cm, width=4.8cm,angle=-90 ]{f9j.eps} &
\includegraphics[height=4cm, width=4.8cm,angle=-90 ]{f9k.eps} &
\includegraphics[height=4cm, width=4.8cm,angle=-90 ]{f9l.eps} &
\includegraphics[height=4cm, width=4.8cm,angle=-90 ]{f9m.eps} \\
\includegraphics[height=4cm, width=1.8cm,angle=-90 ]{f9n.eps} &
\includegraphics[height=4cm, width=1.8cm,angle=-90 ]{f9o.eps} &
\includegraphics[height=4cm, width=1.8cm,angle=-90 ]{f9p.eps} &
\includegraphics[height=4cm, width=1.8cm,angle=-90 ]{f9q.eps} \\
\includegraphics[height=4cm, width=4.8cm,angle=-90 ]{f9r.eps} &
\includegraphics[height=4cm, width=4.8cm,angle=-90 ]{f9s.eps} &
\includegraphics[height=4cm, width=4.8cm,angle=-90 ]{f9t.eps} &
\includegraphics[height=4cm, width=4.8cm,angle=-90 ]{f9u.eps} \\
\includegraphics[height=4cm, width=1.8cm,angle=-90 ]{f9v.eps} &
\includegraphics[height=4cm, width=1.8cm,angle=-90 ]{f9w.eps} &
\includegraphics[height=4cm, width=1.8cm,angle=-90 ]{f9x.eps} &
\includegraphics[height=4cm, width=1.8cm,angle=-90 ]{f9y.eps} \\
\end{tabular}
\caption{  \scriptsize \linespread{1}  Deconvolved spectra,  in  the E*f(E)
  representation, and  contributions per  channel to the  $\chi^2$ for
  four  spectra  of our  sample;  first  column:  spectrum 06,  second
  column:  spectrum  16, third  column:  spectrum  31, fourth  column:
  spectrum07.  Deconvolved spectra (upper panels) and contributions to
  the  $\chi^2$ (lower  panels) from  the best-fit  using  the \DBBTT,
  \DBBTTPEG~ and  \DBBEQPAIR~ models for  the first, second  and third
  row  respectively.  In the  spectral  decomposition  the green  line
  component  shows  the  disk   emission,  the  blue  line  shows  the
  Comptonized emission, red line the Gaussian line, the thin blue line
  the \PEGPWRLW~ component.}
\label{fig9}
\end{figure*}

\section{Conclusions}  
We outline  our main conclusions  from an extensive analysis  of \sco~
RXTE  observations  in the  3--200  keV  energy  range: we  observe  a
spectral  evolution of  the source  that is  clearly dependent  on the
position of the source along  its CD track; the spectral decomposition
consists of four different spectral components that we interpret, from
the softest  to the  highest X-rays respectively,  as follows:  a soft
thermal component  from an optically  thick accretion disk,  a thermal
Comptonization from  the boundary layer, a  reflection component which
mainly manifests  in a broad Gaussian  line at 6.4 keV  and a variable
hard excess above 30 keV, mostly present at low \mdot.\\
We have  shown that  the CD correlated  variations of the  two thermal
temperatures (disk temperature  and soft seed-photon temperature), the
trend  in   the  hard/soft  ratio,  the  Compton   thickening  of  the
Comptonizing  cloud  and  the  apparency  of  significant  hard  X-ray
emission on  the HB/NB  are the main  spectral characteristics  of the
source.\\
We paid particular attention to  the modelization of the hardest X-ray
component, which is of primary importance for the understanding of all
the continuum emission, as it  has a major impact on the determination
of  all the other  spectral parameters.   To fit  this hard  excess we
modelled  our  spectra using  a  phenomenological  component, i.e.   a
power-law  with a low  energy exponential  cut-off at  the seed-photon
temperature   and   a  self-consistent   physical   model,  a   hybrid
thermal/non-thermal Comptonization  code.  In  both cases we  found an
adequate description of the spectra for each source state, and the two
modelizations are statistically equivalent.\\
The use  of simple  power-laws to  fit the hard  tails in  the Z-class
sources, but also in  some atolls \citep{tarana07,fiocchi06}, has been
largely used in the past, given  the lack of broad-band coverage up to
the MeV energies, and a  poor understanding of the physical origins of
this component. Contrary to what previously reported \citep{damico01},
the presence of the hard  tail $is$ related to the broad-band spectral
evolution of the source as inferred  from the source position on the Z
track of its CD: the presence  and the values of the photon indexes in
\sco~ are consistent with the results from fits to spectra on the same
zone  of the  CD for  the  other Z  sources \citep{asai94,  disalvo00,
  disalvo01, disalvo06};  at the same  time, the flux  contribution to
the total  energy output of  the sources is  of the same order  and is
anticorrelated to  the inferred mass accretion rate.   This is clearly
shown  in  the  case of  the  CD  20035A,  where  the hard  tail  flux
monotonically increases as the source moves from the bottom to the top
of its NB/HB.\\
The only difference,  between \sco~ and the other  similar sources, is
constituted by rather flat hard tails that we found when the source is
at the top  of the FB and  which has never been observed  in other NSs
sources;   however,  while  for   every  CD   that  we   obtained,  we
systematically found a hard tail as  the source was at the top of left
V track, the detection on the  FB is presumedly dependent on the HEXTE
statistics.   Past  surveys with  the  use  of scintillation  counters
on-board on  balloons reported the unusual flattening  of the spectrum
of  \sco~ above  40  keV  \citep{agrawal71}, as  well  as more  recent
surveys \citep{manchanda06};  it would be  interesting to definitively
assess the existence of such component; from our analysis, we can only
conjecture that  this component, could be  a signature, on  the FB, of
the trespassing  of the Eddington  limit on the interface  between the
accretion  disk and the  board of  the boundary  layer, followed  by a
violent expulsion of part of the accreting matter.\\
Hard tails on  the HB could have  a thermal origin, as in  the case of
the thermal Comptonized component that dominates the spectrum at lower
energies (less  than 20 keV); in  this case the power-law,  as long as
the optical depth is low and  the electron temperature is not too high,
is  a  good  approximation  of  the Comptonized  spectrum,  whose  $y$
parameter
\begin{equation} \label{ypar}
y = \frac{4 k T_e}{m_e c^2} \times (\tau + \tau^2)
\end{equation}
is related to the photon index of the power-law by the following relation:
\begin{equation}
\alpha = - \frac{1}{2} + \sqrt{\frac{9}{4}+\frac{\pi^2}{3y}}.
\end{equation}
In the case  of \sco~ spectra on the HB/NB,  we derived photon indexes
in  the 1.8--2.2  range, that  imply a  $y$ Compton  parameter  in the
0.6--1.1 range; a $\tau = 1$  thick plasma would then require a $\sim$
50 keV thermal  plasma, while a $\tau = 2$, on  the contrary, a $\sim$
15  keV  electron  temperature.   We  did not  find  any  evidence  of
high-energy cut-off,  obtaining only lower  limits in the  50--100 keV
energy range,  and this constraints the  optical depth to  be less than
one while the electron temperature  would be greater than 50 keV.  The
formation of hot zones, or blobs, of very hot plasma, possibly powered
by  episodes of magnetic  reconnection above  the disk,  constitutes a
plausible scenario \citep{haardt94}; this model has been also proposed
in the case of the  hard/low states of BH candidates \citep{malzac04},
which share with  the bright Z-sources systems, from  a spectral point
of  view,  similar values  of  the hard  photon  indexes  and the  low
accretion state, while they  naturally differ from the strong boundary
layer emission present in the NSs sources and absent in the BH case.\\
A thermal origin, on the other hand, for the topFB hard tails seems to
be  excluded, as  it would  require  unrealistic large  values of  the
electron temperature.\\
Our analysis has shown  that this scenario is statistically equivalent
to a  hybrid Comptonization model,  where the non-thermal  fraction of
the power  injected in the electron  heating counts up to   half of
the total  injected power.  The non-thermal fraction  is significantly
higher with respect  to the value found by  \citet{farinelli05} in the
case of GX 17+2.  The  non-thermal fraction for the topFB spectra must
be distinguished from the non-thermal spectra of the BH sources, as in
the case of \sco, the hard  luminosity is a small fraction of the soft
luminosity, that dominates the 10--30 keV energy spectrum. Most of the
soft photons do not Compton  interact with the electron cloud, because
of the low optical depth values.\\
Another  possible  physical  mechanism   to  explain  the  hard  X-ray
variability      is      the      bulk      motion      Comptonization
\citep[e.g.][]{psaltis01},  present in  systems with  a  high velocity
radial accretion flows.  This radiation mechanism, used to explain the
hard  power-law component  in BH  systems spectra  in  their soft/high
states, is,  however, not able to  account for the the  absence of any
cut-off up  to energies  of 0.5  MeV and for  the hard  photon indexes
observed ($\Gamma  \leq$ 2.0)  \citep{niedzwiecki06}.  In the  case of
bright NS systems,  the high radiation pressure of  the inner boundary
layer constitutes  a strong barrier  for any incoming  convergent bulk
motion flow.   On the other side,  because jets have  been observed in
the radio from \sco, bulk  motion Comptonization inside the jet can be
an  important  production mechanisms  also  of  hard X-ray  radiation.
Theoretical  spectra  and  energetic  contribution  according  to  the
relevant physical parameters involved in the case of a strict coupling
with  inflow  (i.e.   accretion  to  the compact  object  through  the
formation  of  an accretion  disk)  and  outflow  (i.e. jets),  are  a
promising  way to cover  in a  self-consistent way  all the  phases of
accretion.  Details about the geometry of the scattering media, amount
of reflection on the cold disk and dependence from the accretion state
of the  source are yet to  be fully explored.  Moreover,  in this case
synchrotron  emission  of  soft  photons  by a  beamed  population  of
relativistic electrons  becomes a  competitive source of  photons with
respect to the thermal soft  photon emission of the boundary layer and
accretion disk.  An attempt to explicitly compute a jet spectrum, from
radio to hard X-rays, has been recently proposed by \citet{markoff05},
where  the jet  base  subsumes  the role  of  the static  Comptonizing
corona; spectral fits in the case of BH systems (namely Cyg X-1 and GX
339-4) in hard  states are consistent with this  scenario, but BH soft
states  and NS  systems spectra  need  yet to  be tested  in order  to
understand the limits of validity of the jet model.\\
Questions to be further addressed in future observations are: the exact  
shape and  contribution of the soft  component below the  3 keV range,  
the investigation with good spectral resolution in the 6--10 keV range  
of  the  reflection features,  the  extension up  to  0.5  MeV of  the  
spectral  coverage  in  order  to constrain  the  physical  parameters  
characterizing   the   non-thermal   electron  distribution   of   the  
Comptonizing plasma. The latter  point is well within the capabilities  
of the recently launched SUZAKU satellite, so that future observations  
in this direction can be a stringent test to our conclusions.  
\bibliographystyle{apj}
\bibliography{references}
\clearpage
\begin{landscape}
\begin{deluxetable}{lrrrrrrrrrrrr}
\tablewidth{0pt}
\tablecolumns{13}
\scriptsize
\tablecaption{FIT RESULTS: THERMAL COMPTONIZATION MODEL  \label{tab3a}}
\tablehead{
\colhead{Parameter}     
& \colhead{kT$_{DB}$} 
& \colhead{R$_{DB}$} 
& \colhead{$E_{gaus}$} 
& \colhead{$\sigma_{gaus}$} 
& \colhead{N$_{gaus}$} 
& \colhead{$T_0$}   
& \colhead{$T_e$} 
& \colhead{$\tau$}  
& \colhead{Index} 
& \colhead{$F_{PEG}$} 
& \colhead{$F_{TT}$} \\
\colhead{Units}     
& \colhead{keV}    
& \colhead{km}            
& \colhead{keV}    
& \colhead{keV} 
& \colhead{ph cm$^{-2}$ s$^{-1}$}              
& \colhead{keV}    
& \colhead{keV}    
& 
&       
& \colhead{erg cm$^{-2}$ s$^{-1}$}
& \colhead{erg cm$^{-2}$ s$^{-1}$} \\
}
\startdata
     1 &   $1.62_{-0.07}^{+0.07}$ &  $12.9_{-1.0}^{+1.1}$    &  $6.45_{-0.05}^{+0.16}$ &  $0.42_{-0.4}^{+0.3}$  &  $0.18_{-0.06}^{+0.08}$ &  $2.34_{-0.05}^{+0.07}$ &  $3.8_{-1.6}^{+11}$ 	        & 	  $5_{-3}^{+5}$ &  $2$  			&  $170_{-120}^{+120}$ 			&  7.90  \\
     2 &   $1.47_{-0.02}^{+0.04}$ &  $15.5_{-0.9}^{+0.5}$    &  $6.49_{-0.09}^{+0.13}$  &  $0.54_{-0.25}^{+0.24}$     &  $0.17_{-0.05}^{+0.04}$ &  $2.23_{-0.02}^{+0.04}$ &  $3.4_{-0.3}^{+0.8}$ & $7_{-2.7}^{+1.7}$   	&   $2.3_{-0.5}^{1.6}$ 	&   $300_{-50}^{+50}$ 	&  7.28  \\
     3 &   $1.43_{-0.06}^{+0.06}$ &  $16.4_{-1.4}^{+1.6}$    &  $6.50_{-0.13}^{+0.21}$  &  $0.66_{-0.3}^{+0.14}$ 	&  $0.19_{-0.08}^{+0.08}$ &  $2.27_{-0.21}^{+0.03}$  &  $4.3_{-0.6}^{+9}$ 		& 	  $4_{-2}^{+3}$ &   $2$			&  $360_{-80}^{+80} $			&  8.76  \\
     4 &   $1.37_{-0.12}^{+0.10}$ &  $17.8_{-2.5}^{+3}$   &  $6.48_{-0.08}^{+0.23}$  &  $0.7_{-0.5}^{+0.07}$ 	&  $0.21_{-0.11}^{+0.11}$ &  $2.20_{-0.24}^{+0.21}$ &  $3.4_{-0.4}^{+1.1}$ 	& 	  $7.8_{-3}^{+4}$ 	&   $1.6_{-0.3}^{+0.4}$ &   $670_{-80}^{+80} $ 	&  10.1  \\
     5 &   $1.43_{-0.11}^{+0.08}$ &  $16.5_{-2.0}^{+3.0}$   &  $6.50_{-0.10}^{+0.26}$  &  $0.64_{-0.6}^{+0.16}$ 	&  $0.16_{-0.08}^{+0.11}$ &  $2.35_{-0.12}^{+0.09}$ &  $3.8_{-0.5}^{+1.1}$ 	& 	  $5.9_{-2.7}^{+2.7}$ 	&   $1.5_{-0.5}^{+0.2}$ 	&   $710_{-90}^{+90}$ 	&  10.4  \\
     6 &   $1.48_{-0.03}^{+0.04}$ &  $15.0_{-1.0}^{+1.8}$    &  $6.51_{-0.11}^{+0.23}$  &  $0.45_{-0.4}^{+0.4}$ 	&  $0.12_{-0.05}^{+0.08}$ &  $2.51_{-0.02}^{+0.03}$ &  $5.45_{-1.0}^{+0.5}$ 	&	  $2.96_{-0.13}^{+1.9}$ &   $1.5_{-0.2}^{+0.2}$ &   $980_{-80}^{+80}$ 	&  11.2  \\	 
\hline
     7 &   $2.12_{-0.19}^{+0.07}$ &  $8.5_{-1.6}^{+1.3}$ &  $6.40_{}^{+0.17}$       &  $0_{}^{+0.5}$ 		&  $0.08_{-0.04}^{+0.03}$ &  $2.92_{-0.01}^{+0.01}$ &  $10_{}^{}$ 		& 	  $0.02_{-0.01}^{+0.06}$ &   $0.3_{-0.3}^{+0.5}$  &   $730_{-180}^{+180}$ &  10.8  \\
     8 &   $2.04_{-0.03}^{+0.04}$ &  $9.2_{-0.1}^{+0.3}$ &  $6.40_{}^{+0.18}$ 	  &  $0_{}^{+0.4}$ 		&  $0.09_{-0.03}^{+0.04}$ &  $2.80_{-0.02}^{+0.01}$ &  $10_{}^{}$ 		& 	  $0.01_{-0}^{+0.18}$ 	&   $0.5_{-0.5}^{+0.5}$ &   $500_{-120}^{+130}$ &  7.76  \\
     9 &   $1.64_{-0.09}^{+0.06}$ &  $12.5_{-1.0}^{+1.0}$    &  $6.44_{-0.17}^{+0.16}$  &  $0.3_{-0.3}^{+0.3}$ 	&  $0.14_{-0.05}^{+0.07}$ &  $2.34_{-0.05}^{+0.06}$ &  $3.9_{-0.6}^{+2}$ 	& 	  $5_{-3}^{+5}$ 	&   $0.2_{-0.2}^{+1.0}$   &  $260_{-90}^{+80}$    &  7.53  \\
    10 &   $1.39_{-0.05}^{+0.05}$ &  $16.3_{-1.0}^{+1.3}$    &  $6.49_{-0.17}^{+0.15}$  &  $0.5_{-0.3}^{+0.3}$ 	&  $0.20_{-0.06}^{+0.08}$ &  $2.12_{-0.04}^{+0.03}$ &  $3.6_{-0.4}^{+2.1}$		&  $6.3_{-2.3}^{+2.0}$ 	&   $1.5_{-0.5}^{+0.5}$  &  $390_{-80}^{+80}$			&  8.3  \\
    11 &   $1.44_{-0.06}^{+0.03}$ &  $16.5_{-1.5}^{+2.5}$   &  $6.52_{-0.12}^{+0.18}$  &  $0.57_{-0.3}^{+0.2}$ 	&  $0.17_{-0.06}^{+0.08}$ &  $2.32_{-0.05}^{+0.05}$ &  $4.50_{-0.05}^{+2.0}$ & $3.89_{-0.15}^{+3.7}$ &   $1.4_{-0.2}^{+0.5}$ &   $390_{-70}^{+70}$ 	&  11.4  \\
\hline
    12&   $2.07_{-0.09}^{+0.09}$ &  $8.7_{-0.7}^{+0.5}$     &  $6.40_{}^{+0.18}$ 	  &  $0_{}^{+0.5}$ 		&  $0.09_{-0.05}^{+0.04}$ &  $2.93_{-0.07}^{+0.02}$ &  $10_{}^{}$ 		& 	  $0.01_{}^{+0.16}$ 	&   $0.7_{-0.6}^{+0.6}$ &   $560_{-150}^{+160}$ &  12.1  \\
    13&   $2.06_{-0.05}^{+0.05}$ &  $9.0_{-0.4}^{+0.5}$     &  $6.40_{}^{+0.21}$ 	  &  $0_{}^{+0.5}$ 		&  $0.09_{-0.05}^{+0.04}$ &  $2.85_{-0.08}^{+0.03}$ &  $10_{}^{}$		& 	  $0.01_{}^{+0.17}$ 	&   $0_{}^{+0.3}$	&   $480_{-170}^{+170}$ &  10.1  \\
    14&   $1.86_{-0.07}^{+0.04}$ &  $10.6_{-0.6}^{+0.6}$    &  $6.45_{-0.05}^{+0.13}$  &  $0_{}^{+0.5}$ 		&  $0.11_{-0.04}^{+0.06}$ &  $2.56_{-0.06}^{+0.05}$ &  $9_{-5}^{+15}$ & $0.6_{-0.3}^{+4}$ 	&   	2		&  $50_{-50}^{+100}$			&  7.59  \\
    15&   $1.48_{-0.06}^{+0.05}$ &  $15.7_{-0.8}^{+1.3}$    &  $6.48_{-0.0.8}^{+0.15}$ &  $0.5_{-0.3}^{+0.3}$ &  $0.20_{-0.07}^{+0.08}$ &  $2.23_{-0.04}^{+0.05}$ &  $4.9_{-1}^{+14}$ 		& 	  $3.2_{-2.4}^{+5}$ 	&   2			&  $100_{-100}^{+200}$			&  7.15  \\
    16&   $1.42_{-0.06}^{+0.03}$ &  $17.0_{-1.4}^{+1.8}$    &  $6.52_{-0.12}^{+0.19}$  &  $0.65_{-0.3}^{+0.15}$ 	&  $0.20_{-0.07}^{+0.08}$ &  $2.31_{-0.03}^{+0.03}$ &  $5.4_{-0.5}^{+0.7}$ 	& 	  $2.7_{-0.1}^{+1.5}$ 	&   $1.20_{-0.2}^{+0.4}$&   $600_{-90}^{+90}$ 	&  8.54  \\
\hline
    17&   $2.15_{-0.08}^{+0.03}$ &  $11.7_{-0.5}^{+1.0}$    &  $6.40_{}^{+0.26}$ 	  &  $0_{}^{+0.5}$ 		&  $0.18_{-0.12}^{+0.08}$ &  $2.82_{-0.07}^{+0.05}$ &  $10_{}^{}$ 		& 	  $0.01_{}^{+0.19}$ 	&   	2		&   $0_{}^{+100}$			&  11.1  \\
    18&   $1.70_{-0.06}^{+0.07}$ &  $15.5_{-1.1}^{+1.1}$    &  $6.48_{-0.0.8}^{+0.16}$ &  $0.3_{-0.3}^{+0.3}$ 	&  $0.24_{-0.08}^{+0.11}$ &  $2.40_{-0.07}^{+0.11}$ &  $3.6_{-1}^{+9}$ 	& 	  $5_{-3}^{+7}$ 	&   2			&    $190_{-140}^{+140}$ 			&  10.4  \\
    19&   $1.53_{-0.07}^{+0.05}$ &  $18.6_{-0.9}^{+2.0}$    &  $6.51_{-0.11}^{+0.15}$  &  $0.5_{-0.3}^{+0.3}$ 	&  $0.26_{-0.08}^{+0.14}$ &  $2.30_{-0.03}^{+0.18}$ &  $4.3_{-1.2}^{+6}$ 		&  $4.8_{-3}^{+1.4}$	&   2			&      $390_{-130}^{+150}$ 			&  10.5  \\
    20&   $1.34_{-0.12}^{+0.13}$ &  $24_{-4}^{+4}$  &  $6.42_{-0.02}^{+0.25}$  &  $0.8_{-0.4}^{}$ 		&  $0.41_{-0.19}^{+0.14}$ &  $2.11_{-0.17}^{+0.16}$ &  $3.7_{-0.3}^{+0.7}$ 	& 	  $6.2_{-2}^{+1.5}$ 	&   $1.7_{-0.4}^{+0.5}$ &   $720_{-90}^{+120}$ 	&  13.8  \\
    21&   $1.23_{-0.12}^{+0.22}$ &  $30_{-9}^{+6}$  &  $6.42_{-0.02}^{+0.25}$  &  $0.8_{-0.4}^{}$ 		&  $0.52_{-0.29}^{+0.16}$ &  $1.95_{-0.18}^{+0.37}$ &  $3.2_{-0.2}^{+0.5}$ 	& 	  $9.9_{-3}^{+1.9}$ 	&   $2.2_{-0.4}^{+0.4}$ &   $1540_{-180}^{+160}$ &  21.8  \\
\enddata
\end{deluxetable}
\clearpage 
\end{landscape}

\begin{landscape}
\begin{deluxetable}{lrrrrrrrrrrrr}
\tablewidth{0pt}
\tablecolumns{13}
\scriptsize
\tablecaption{FIT RESULTS: THERMAL COMPTONIZATION MODEL  \label{tab3b}}
\tablehead{
\colhead{Parameter}     
& \colhead{kT$_{DB}$} 
& \colhead{R$_{DB}$} 
& \colhead{$E_{gaus}$} 
& \colhead{$\sigma_{gaus}$} 
& \colhead{N$_{gaus}$} 
& \colhead{$T_0$}   
& \colhead{$T_e$} 
& \colhead{$\tau$}  
& \colhead{Index} 
& \colhead{$F_{PEG}$} 
& \colhead{$F_{TT}$} \\
\colhead{Units}     
& \colhead{keV}    
& \colhead{km}            
& \colhead{keV}    
& \colhead{keV} 
& \colhead{ph cm$^{-2}$ s$^{-1}$}              
& \colhead{keV}    
& \colhead{keV}    
& 
&       
& \colhead{erg cm$^{-2}$ s$^{-1}$}
& \colhead{erg cm$^{-2}$ s$^{-1}$} \\
}
\startdata
    22&   $2.20_{-0.13}^{+0.13}$ &  $11_{-1.0}^{+1.0}$    &  $6.40_{}^{+0.3}$ 	 &   $0_{}^{+0.5}$ 		&  $0.20_{-0.04}^{+0.12}$ &  $2.90_{-0.14}^{+0.22}$ &  $11_{-8}^{+20}$ 		& 	  $0.14_{-0.13}^{+23}$ 	&   2			&    $100_{-100}^{+200}$ 			&  14.0  \\
    23&   $1.56_{-0.12}^{+0.14}$ &  $17.2_{-1.2}^{+1.3}$  &  $6.43_{-0.03}^{+0.19}$  &  $0.75_{-0.4}^{+0.05}$ 	&  $0.45_{-0.19}^{+0.14}$ &  $2.30_{-0.13}^{+0.17}$ &  $4.0_{-0.4}^{+1.3}$ 	& 	  $5.0_{-2.2}^{+1.2}$ 	&   2			&   $120_{-100}^{+100}$ 			&  9.8  \\
    24&   $1.30_{-0.07}^{+0.06}$ &  $24_{-4}^{+10}$   &  $6.49_{-0.09}^{+0.20}$  &  $0.68_{-0.3}^{+0.12}$ 	&  $0.30_{-0.11}^{+0.14}$ &  $2.00_{-0.2}^{+0.2}$ &  $3.7_{-0.4}^{+0.7}$ 		& 	  $6.3_{-0.4}^{+0.9}$ 	&   	2		&    $500_{-200}^{+220}$ 			&  13.2  \\
    25&   $1.48_{-0.07}^{+0.07}$ &  $19.4_{-1.8}^{+2.1}$    &  $6.54_{-0.14}^{+0.24}$  &  $0.65_{-0.4}^{+0.15}$ 	&  $0.26_{-0.11}^{+0.15}$ &  $2.42_{-0.10}^{+0.03}$ &  $5.2_{-1.1}^{+1.6}$ 	& 	  $3.5_{-2.6}^{+2.1}$ 	&   $1.6_{-0.7}^{+0.4}$ &   $940_{-140}^{+110}$ &  16.5  \\
\hline
    26&   $2.29_{-0.15}^{+0.12}$ &  $11.2_{-0.7}^{+1.0}$    &  $6.46_{-0.06}^{+0.21}$  &  $0.24_{-0.24}^{+0.3}$ 	&  $0.27_{-0.10}^{+0.12}$ &  $2.94_{-0.18}^{+0.14}$ &  $11_{-7}^{+9}$ & 	  $0.06_{-0.05}^{+8}$ 	&   $0_{}^{+0.5}$ 	&   $1400_{-300}^{+300}$  &  16.3  \\
    27&   $2.02_{-0.04}^{+0.08}$ &  $11.7_{-0.3}^{+0.5}$    &  $6.54_{-0.14}^{+0.10}$  &  $0.48_{-0.5}^{+0.3}$ 	&  $0.29_{-0.11}^{+0.16}$ &  $2.82_{-0.06}^{+0.03}$ &  $8_{-3}^{+3}$ &  $0.6_{-0.5}^{+3.4}$ 	&    $1.8_{-0.7}^{+0.9}$			&  $540_{-140}^{+150}$ 			&  10.7  \\
    28&   $1.60_{-0.05}^{+0.04}$ &  $14.6_{-1.0}^{+1.0}$    &  $6.50_{-0.10}^{+0.13}$  &  $0.61_{-0.2}^{+0.19}$ 	&  $0.29_{-0.07}^{+0.10}$ &  $2.30_{-0.03}^{+0.04}$ &  $17_{-10}^{+13}$ 	& 	  $0.06_{-0.05}^{+0.6}$ &   $0.5_{-0.5}^{+1.0}$			&   $250_{-80}^{+80}$ 			&  8.2  \\
    29&   $1.44_{-0.04}^{+0.05}$ &  $17.1_{-1.0}^{+1.0}$    &  $6.50_{-0.10}^{+0.13}$  &  $0.68_{-0.2}^{+0.12}$ 	&  $0.26_{-0.07}^{+0.08}$ &  $2.25_{-0.03}^{+0.03}$ &  $4.2_{-0.8}^{+2.2}$ & 	  $4.5_{-3}^{+3}$ 	&  	2		&   $240_{-100}^{+220}$ 			&  8.5  \\
    30&   $1.45_{-0.06}^{+0.07}$ &  $17.0_{-1.6}^{+1.6}$    &  $6.49_{-0.09}^{+0.21}$  &  $0.80_{-0.3}^{}$ 		&  $0.28_{-0.11}^{+0.07}$ &  $2.33_{-0.04}^{+0.05}$ &  $5.7_{-1.7}^{+2.8}$ 	& 	  $2.4_{-1.7}^{+2.8}$ 	&  	2		&   $270_{-50}^{+240}$ 			&  9.6  \\
    31&   $1.46_{-0.19}^{+0.08}$ &  $18.5_{-2.0}^{+6}$    &  $6.52_{-0.12}^{+0.21}$  &  $0.5_{-0.5}^{+0.3}$ 	&  $0.25_{-0.07}^{+0.10}$ &  $2.34_{-0.3}^{+0.11}$ &  $3.7_{-0.6}^{+1.2}$ 	& 	  $6.0_{-2.0}^{+0.4}$&   $1.5_{-0.5}^{+0.6}$ &   $1150_{-130}^{+110}$ &  13.6  \\
\hline
    32&   $2.09_{-0.09}^{+0.04}$ &  $10.7_{-0.7}^{+1.0}$    &  $6.40_{}^{+0.29}$ 	  &  $0.3_{-0.3}^{+0.5}$ 	&  $0.22_{-0.12}^{+0.18}$ &  $2.84_{-0.10}^{+0.09}$ &  $10$ 	& 	  $0.11_{-0.10}^{+2.4}$ 	&    $0.5_{-0.5}^{+1.0}$			&  $600_{-300}^{+300}$	&  16.1  \\
    33&   $1.79_{-0.08}^{+0.08}$ &  $12.7_{-1.4}^{+1.6}$    &  $6.40_{}^{+0.24}$ 	  &  $0.4_{-0.4}^{+0.4}$ 	&  $0.23_{-0.10}^{+0.17}$ &  $2.5_{-0.6}^{+0.1}$ &  $4.1_{-1.8}^{+1.4}$ 		&  $3.9_{-1.7}^{+4}$ 	&   $0.8_{-0.8}^{+0.7}$	&  $300_{-80}^{+100}$		&  8.7  \\
    34&   $1.47_{-0.07}^{+0.07}$ &  $18.5_{-3}^{+3.5}$    &  $6.40_{}^{+0.27}$ 	  &  $0.68_{-0.5}^{+0.12}$ 	&  $0.20_{-0.10}^{+0.11}$ &  $2.2_{-0.5}^{+0.2}$ &  $3.4_{-0.5}^{+13}$ 	& 	  $7_{-4}^{+2}$ 	&   2 &   $300_{-180}^{+180}$ 	&  8.3  \\
    35&   $1.40_{-0.07}^{+0.09}$ &  $19.3_{-2.7}^{+2.3}$    &  $6.40_{}^{+0.24}$ 	  &  $0.62_{-0.4}^{+0.18}$ 	&  $0.22_{-0.09}^{+0.13}$ &  $2.31_{-0.16}^{+0.07}$ &  $5.1_{-1.5}^{+25}$ 		& 	  $3.0_{-2.2}^{+2.0}$ 	&  2			&   $100_{-100}^{+200}$			&  10.3  \\
    36&   $1.46_{-0.06}^{+0.07}$ &  $16.8_{-1.8}^{+1.2}$    &  $6.40_{}^{+0.3}$ 	          &  $0.8_{-0.4}^{}$ 		&  $0.29_{-0.14}^{+0.09}$ &  $2.38_{-0.18}^{+0.9}$ &  $4.1_{-0.7}^{+2.7}$ 		& 	  $4.6_{-3}^{+1.8}$ 	&   $1.9_{-0.6}^{+0.7}$			&   $450_{-90}^{+90}$			&  10.1  \\
\hline
    37&   $2.10_{-0.12}^{+0.14}$ &  $11.1_{-0.1}^{+0.9}$      &  $6.47_{-0.07}^{+0.26}$       &  $0.6_{-0.6}^{+0.2}$ 	&  $0.28_{-0.13}^{+0.17}$    &  $2.83_{-0.01}^{+0.4}$ &  $10$		& 	  $0.01_{-0}^{+0.4}$ 	&  	2		&    $100_{-60}^{+60}$			&  6.7  \\
    38&   $1.71_{-0.09}^{+0.08}$ &  $13.9_{-1.0}^{+0.9}$      &  $6.40_{}^{+0.22}$ 	  &  $0.8_{-0.2}^{}$ 		&  $0.38_{-0.13}^{+0.12}$    &  $2.44_{-0.15}^{+0.16}$ &  $10$ 	& 	  $0.4_{-0.39}^{+0.3}$ & 2  &   $250_{-150}^{+100}$			&  7.4    \\
    39&   $1.45_{-0.06}^{+0.06}$ &  $18.8_{-1.6}^{+1.9}$      &  $6.40_{}^{+0.20}$ 	  &  $0.8_{-0.2}^{}$ 		&  $0.40_{-0.12}^{+0.08}$    &  $2.23_{-0.08}^{+0.10}$ &  $11_{-3}^{+30}$ 	& 	  $0.6_{-0.6}^{+1.7}$ &  2			&   $100_{-100}^{+100}$			&  8.6 \\
    40&   $1.42_{-0.06}^{+0.06}$ &  $19.6_{-2.6}^{+2.0}$     &  $6.42_{-0.02}^{+0.21}$       &  $0.8_{-0.2}^{}$          &   $0.35_{-0.10}^{+0.10}$   &  $2.22_{-0.10}^{+0.10}$ &  $10_{-5}^{+20}$ 	        & 	  $0.8_{-0.8}^{+2.2}$ &  2			&   $90_{-60}^{+60}$			&  9.7  \\
    41&   $1.51_{-0.05}^{+0.06}$ &  $17.2_{-1.4}^{+1.4}$      &  $6.44_{-0.04}^{+0.23}$       &  $0.8_{-0.2}^{}$ 	        &  $0.33_{-0.13}^{+0.07}$   &  $2.34_{-0.05}^{+0.06}$ &  $11_{-8}^{+13}$ 		& 	  $0.4_{-0.4}^{+2.5}$ 	&  	2		&    $170_{-50}^{+50}$			&  9.6  \\
    42&   $1.47_{}^{}$ 	       &   $19_{}^{}$         &  $6.47_{-0.07}^{+0.28}$      &  $0.8_{-0.4}^{}$ 	        &  $0.28_{-0.10}^{+0.05}$   &  $2.36_{-0.04}^{+0.04}$ &  $4.6_{-1.2}^{+1.8}$ 	& 	  $3.4_{-2.2}^{+3}$ 	     &   $2.5_{-0.5}^{+0.7}$  &   $400_{-60}^{+160}$ 	&  11.8  \\
    43&   $1.36_{}^{}$         &   $22_{}^{}$         &  $6.40_{}^{+0.24}$ 	  &  $0.8_{-0.14}^{}$ 		&  $0.40_{-0.10}^{+0.13}$   &  $2.25_{-0.04}^{+0.04}$ &  $4.5_{-0.6}^{+1.6}$ 	&	  $4.2_{-2.5}^{+0.6}$ 	     &   $2.0_{-0.8}^{+0.7}$    &   $650_{-180}^{+60}$ 	&  14.6  \\
\hline
\hline
\enddata 
\tablecomments{Parameters values  and associated  errors for  all  the CDs
  resolved spectra  adopting the  \DBBTT~ or the \DBBTTPEG~  model. 
  The column  $F_{PEG}$ gives the power-law flux in units of 10$^{-12}$
  erg cm$^{-2}$ s$^{-1}$ calculated in the 20-200 keV energy range.
  The last column $F_{TT}$ gives the flux of the thermal Comptonization
  component in units of 10$^{-8}$ erg cm$^{-2}$ s$^{-1}$ in the 0.1--200 keV energy range.
  See Section  \ref{spectralmodels} for an
  explanation  of   the  parameters  involved  in   the  fits. All  the
  uncertainties  are calculated  at 90\%  confidence level.}
\end{deluxetable}
\clearpage 
\end{landscape}

\begin{landscape}
\begin{deluxetable}{lrrrrrrrrrrr}
\tablewidth{0pt}
\tablecolumns{11}
\tablecaption{FIT RESULTS: THE HYBRID COMPTONIZATION MODEL  \label{tab4}  }
\tablehead{
\colhead{Parameter}                                   
& \colhead{kT$_{DB}$}    
& \colhead{$R_{in}$ (K)}
& \colhead{$LineE$}    
& \colhead{$\sigma$}          
& \colhead{N$_{Gaus}$}   
& \colhead{$l_h$}               
& \colhead{$kT_{BB}$}     
& \colhead{$l_{nth}/l_{h}$}        
& \colhead{$\tau_p$}        
& \colhead{$G_{inj}$}
& \colhead{Flux$_{eqp}$} \\
\colhead{Units}        
& \colhead{keV} 
& \colhead{km}                       
& \colhead{keV}                        
& \colhead{keV}                
& \colhead{ph cm$^{-2}$ s$^{-1}$}           
&                         
& \colhead{keV}                 
&       
& 
& 
&      \\ 
}
\startdata
\hline
\hline
02 
& $1.42_{-0.04}^{+0.02}$ 
&   $16_{-4}^{+7}$   
&  $6.50_{-0.1}^{+0.13}$  
&  $0.58_{-0.23}^{+0.22}$ 
&  $0.18_{-0.04}^{+0.05}$   
&  $0.65_{-0.04}^{+0.03}$ 
&  $2.20_{-0.08}^{+0.07}$ 
&   $38_{-5}^{+9}$ 
&  $0.81_{-0.13}^{+1.07}$  
& 2 
& 8.59   \\
04 & $1.31_{-0.04}^{+0.05}$ &   $19_{-7}^{+8}$   &  $6.45_{-0.05}^{+0.25}$ &  $0.8_{-0.5}^{}$        &  $0.26_{-0.15}^{+0.04}$   &  $0.98_{-0.26}^{+0.09}$ &  $2.21_{-0.06}^{+0.06}$ &   $62_{-4}^{+3}$ &  $1.7_{-0.7}^{+0.7}$    & 2 & 11.52  \\
05 & $1.30_{-0.03}^{+0.06}$ &   $19_{-9}^{+8}$   &  $6.50_{-0.1}^{+0.20}$  &  $0.8_{-0.4}^{}$        &  $0.22_{-0.1}^{+0.06}$    &  $1.01_{-0.17}^{+0.13}$ &  $2.24_{-0.07}^{+0.05}$ &   $63_{-4}^{+3}$ &  $1.8_{-0.7}^{+0.3}$    & 2 & 12.33  \\
06 & $1.44_{-0.12}^{+0.04}$ &   $16_{-6}^{+10}$  &  $6.50_{-0.1}^{+0.20}$  &  $0.46_{-0.11}^{+0.09}$ &  $0.12_{-0.04}^{+0.04}$   &  $0.88_{-0.05}^{+0.14}$ &  $2.46_{-0.08}^{+0.03}$ &   $73_{-4}^{+2}$ &  $0.64_{-0.09}^{+0.6}$  & 2 & 12.46  \\
11 & $1.48_{-0.14}^{+0.06}$ &   $18_{-6}^{+13}$  &  $6.40_{}^{+0.16}$      &  $0.6_{-0.3}^{+0.2}$    &  $0.25_{-0.07}^{+0.11}$   &  $1.12_{-0.10}^{+0.10}$ &  $2.17_{-0.14}^{+0.07}$ &   $53_{-6}^{+11}$&  $2.81_{-0.16}^{+0.14}$ & 2  & 13.25  \\
16 & $1.16_{-0.02}^{+0.05}$ &   $25_{-10}^{+12}$ &  $6.40_{}^{+0.20}$      &  $0.8_{-0.3}^{}$        &  $0.38_{-0.07}^{+0.03}$   &  $2.02_{-0.04}^{+0.17}$ &  $1.93_{-0.13}^{+0.036}$&   $75_{-3}^{+7}$ &  $7.3_{-0.3}^{+0.3}$    & 2  & 11.39  \\
20 & $1.36_{-0.05}^{+0.05}$ &   $23_{-7}^{+8}$   &  $6.50_{-0.1}^{+0.20}$  &  $0.6_{-0.2}^{+0.2}$    &  $0.31_{-0.09}^{+0.2}$    &  $1.32_{-0.5}^{+0.3}$   &  $2.06_{-0.07}^{+0.07}$ &   $54_{-1}^{+3}$ &  $4.0_{-3}^{+0.8}$      & 2  & 14.99  \\
21 & $1.24_{-0.05}^{+0.04}$ &   $28_{-14}^{+10}$ &  $6.46_{-0.06}^{+0.19}$ &  $0.8_{-0.4}^{}$        &  $0.43_{-0.13}^{+0.13}$   &  $1.29_{-0.17}^{+0.06}$ &  $2.12_{-0.02}^{+0.1}$  &   $52_{-2}^{+3}$ &  $2.8_{-0.9}^{+1.6}$    & 2  & 12.31  \\
25 & $1.24_{-0.07}^{+0.01}$ &   $26_{-6}^{+10}$  &  $6.42_{-0.02}^{+0.18}$ &  $0.8_{-0.2}^{}$        &  $0.41_{-0.09}^{+0.06}$   &  $1.28_{-0.18}^{+0.13}$ &  $2.09_{-0.03}^{+0.04}$ &   $47_{-2}^{+3}$ &  $2.8_{-0.2}^{+0.6}$    & 2 & 11.14  \\
31 & $1.31_{-0.02}^{+0.11}$ &   $22_{-9}^{+8}$   &  $6.48_{-0.08}^{+0.22}$ &  $0.8_{-0.2}^{}$        &  $0.36_{-0.14}^{+0.06}$   &  $1.11_{-0.16}^{+0.09}$ &  $2.20_{-0.06}^{+0.17}$ &   $67_{-1}^{+8}$ &  $1.9_{-0.7}^{+0.4}$    & 2 & 16.52  \\
36 & $1.44_{-0.09}^{+0.08}$ &   $17_{-8}^{+9}$   &  $6.40_{}^{+0.27}$      &  $0.7_{-0.5}^{+0.1}$    &  $0.21_{-0.12}^{+0.12}$   &  $0.62_{-0.23}^{+0.2}$  &  $2.34_{-0.11}^{+0.09}$ &   $50_{-8}^{+10}$&  $0.6_{-0.4}^{+0.8}$    & 2 & 2.47  \\
42 & $1.51_{-0.05}^{+0.06}$ &   $11_{-3}^{+5}$   &  $6.40_{}^{+0.3}$       &  $0.8_{-0.5}^{}$        &  $0.11_{-0.05}^{+0.03}$   &  $0.33_{-0.04}^{+0.09}$ &  $2.40_{-0.12}^{+0.07}$ &   $43_{-8}^{+7}$ &  $0.20_{-0.15}^{+0.6}$  & 2  & 12.1   \\
43 & $1.41_{-0.04}^{+0.07}$ &   $12_{-5}^{+5}$   &  $6.42_{-0.02}^{+0.26}$ &  $0.8_{-0.3}^{}$        &  $0.125_{-0.05}^{+0.008}$ &  $0.64_{-0.18}^{+0.11}$ &  $2.29_{-0.06}^{+0.13}$ &   $41_{-6}^{+4}$ &  $0.6_{-0.4}^{+0.6}$    &  2  & 1.4.3  \\
\hline
\hline
07 & $1.95_{-0.06}^{+0.02}$ &  $1.03_{-0.06}^{+0.09}$ &  $6.4_{}^{+0.17}$     &  $0.0_{}^{+0.5}$     &  $0.09_{-0.03}^{+0.021}$  &  $0.85_{-0.21}^{+0.14}$  &  $2.74_{-0.06}^{+0.10}$ &   $100_{-1}^{}$   &        $0.60_{-0.24}^{+0.25}$ &  $0.9_{-0.9}^{+0.8}$    & 12.62  \\   
08 & $1.93_{-0.05}^{+0.07}$ &  $1.17_{-0.07}^{+0.05}$ &  $6.4_{}^{+0.2}$      &  $0.0_{}^{+0.5}$     &  $0.10_{-0.05}^{+0.03}$   &  $0.89_{-0.3}^{+0.08}$   &  $2.68_{-0.03}^{+0.07}$ &   $100_{-4}^{}$   &	 $0.47_{-0.5}^{+0.3}$   &  $0.8_{-0.8}^{+1}$      & 8.62   \\
12 & $1.98_{-0.14}^{+0.06}$ &  $1.00_{-0.01}^{+0.01}$ &  $6.4_{}^{+0.2}$      &  $0.0_{}^{+0.4}$     &  $0.09_{-0.05}^{+0.03}$   &  $0.72_{-0.3}^{+0.11}$   &  $2.82_{-0.05}^{+0.03}$ &   $100_{-6}^{}$   &	 $0.32_{-0.3}^{+0.4}$   &  $0.2_{-0.2}^{+1.7}$    & 13.08  \\
13 & $1.90_{-0.03}^{+0.05}$ &  $1.20_{-0.1}^{+0.05}$  &  $6.4_{}^{+0.2}$      &  $0.0_{}^{+0.5}$     &  $0.09_{-0.04}^{+0.02}$   &  $0.65_{-0.13}^{+1.0}$   &  $2.67_{-0.02}^{+0.14}$ &   $92_{-17}^{+4}$ &	 $0.71_{-0.13}^{+0.7}$  &  $0.86_{-0.86}^{+0.54}$ & 10.68  \\
26 & $2.07_{-0.03}^{+0.14}$ &  $1.97_{-0.3}^{+0.08}$  &  $6.5_{-0.1}^{+0.25}$  &  $0.3_{-0.3}^{+0.5}$ &  $0.30_{-0.13}^{+0.2}$    &  $1.04_{-0.5}^{+0.1}$    &  $2.73_{-0.05}^{+0.14}$ &   $100_{-3}^{}$   &	 $0.70_{-0.7}^{+0.3}$   &  $0.0_{}^{+1.6}$        & 3.023  \\
\hline
\hline
\enddata
\tablecomments{ Parameters values  and associated errors, adopting the
  model \DBBEQPAIR~ for the selected  spectra on the HB/NB (first part
  of the table) and  for spectra on the top of the  FB (second part of
  the table).  The model for the  five FB spectra differs from the one
  adopted for  the HB/NB spectra, as  it uses a  more appropriate disk
  emission  model   (\DISKPN,  in  XSPEC)  instead   of  the  \DISKBB~
  component.  For the \DISKBB~  component we list the derived apparent
  inner  disk radii  in  km, while  for  the spectra  fitted with  the
  \DISKPN~ component, the inner disk  radius has been fixed to 6 R$_g$
  and  the second column  shows the  value of  the normalizion  of the
  component in units of  \msun$^2$ cos(i)/(D$^2 \beta^4 \times 10^2$),
  where the distance  is given in kpc, $\beta$  is the color/effective
  temperature  and $i$  is  the  inclination angle  of  the disk.  The
  unabsorbed flux of the \EQPAIR~ component has been calculated in the
  0.1--200  keV energy  range  and it  is  in units  of $10^{-8}$  erg
  cm$^{2}$  s$^{-1}$.    See  Section  ~\ref{spectralmodels}   for  an
  explanation of  all the other  parameters involved in the  fits. All
  the uncertainties are calculated at 90\% confidence level.}
\end{deluxetable}
\clearpage
\end{landscape}

\end{document}